\documentstyle[amsfonts,12pt,twoside,cite,epsf,setspace,a4]{thesis}
\begin{document}

\doublespacing
\parskip=4pt
\parindent 0pt
\baselineskip 17pt

\newcommand{\vb}[1]{{\mathbf{#1}}}
\newcommand{\be}{\begin{equation}}
\newcommand{\ee}{\end{equation}}
\newcommand{\bea}{\begin{eqnarray}}
\newcommand{\eea}{\end{eqnarray}}
\newcommand{\ba}[1]{\begin{array}{#1}}
\newcommand{\ea}{\end{array}}
\newcommand{\real}{{\rm l}\! {\rm R}}
\newcommand{\ra}{\rightarrow}
\newcommand{\tr}{\mbox{ tr }}
\newcommand{\Tr}{\mbox{ Tr }}
\newcommand{\al}{\alpha}
\newcommand{\bt}[1]{\begin{table}[h]\centering\begin{tabular}{#1}}
\newcommand{\et}[1]{\end{tabular}\caption{\small#1}\end{table}}
\newcommand{\bz}{{\bar{z}}}
\newcommand{\r}[1]{\ref{#1}}
\newcommand{\lb}[1]{\label{#1}}
\newcommand{\del}{\delta}
\newcommand{\Del}{\Delta}
\newcommand{\der}{\partial}
\newcommand{\Th}{\Theta}
\newcommand{\bj}{{\bar{j}}}
\newcommand{\lra}{\leftrightarrow}
\newcommand{\td}{\tilde{\del}}
\newcommand{\g}{\gamma}
\newcommand{\m}{\mu}
\newcommand{\n}{\nu}
\newcommand{\bchi}{\bbox{\chi}}
\newcommand{\nn}{\nonumber}
\newcommand{\fig}[3]{\begin{figure}[ht]\epsfxsize=140mm\epsfbox{#1}\caption{\small #2 \label{#3}}\end{figure}}
\newcommand{\figh}[3]{\begin{figure}[h]\epsfxsize=140mm\epsfbox{#1}\caption{\small #2 \label{#3}}\end{figure}}
\newcommand{\mod}{\,{\mathrm{mod}}\,}
\newcommand{\im}{\,{\mathrm{Im}}\,}


\thispagestyle{empty}
\begin{center}
\vspace{1cm}

{\Large\bf Heterotic, Open and Unoriented String Theories\vspace{.5cm}\\
from Topological Membrane}
\vspace{4.5cm}

{\large Pedro Castelo Caetano Ferreira\\
\vspace{1.3cm}

Department of Physics\vspace{-.4cm}\\
Keble College\vspace{-.4cm}\\
University of Oxford
\vspace{1cm}

{\font\oxcrest=oxcrest40
\oxcrest\char'01}
\vfill
Thesis submitted for the degree of Doctor of Philosophy\\
in the University of Oxford\\[10mm] Supervisor: Ian I. Kogan}

5th October 2001, Michaelmas term
\end{center}

\newpage
\thispagestyle{empty}
\ \\

\newpage
\thispagestyle{empty}
\pagenumbering{Roman}
\begin{center}
{\large\bf Heterotic, Open and Unoriented String Theories\\
from Topological Membrane}
\vspace{.3cm}

Pedro Castelo Caetano Ferreira\\
\vspace{.3cm}

Department of Physics\\
Keble College\\
University of Oxford
\vspace{.4cm}
\vfill
{\large\sc Abstract}
\end{center}

In this work I consider several topics in the Topological Membrane
(TM) approach to string theory. In TM the string worldsheets emerge as
boundary theories of the 3D membrane with topology
$\Sigma\times[0,1]$. The string dynamics is generated in this way from
the bulk physics, namely from the $3D$ Topologically Massive Gauge
Theory (TMGT) and Topologically Massive Gravity (TMG).

Both (equivalent) path integral and canonical methods of quantizing
TMGT are considered. When this theory is defined on a manifold with
two disconnected boundaries there are induced chiral Conformal Field
Theories (CFT's) on the boundaries which can be interpreted as the
left and right sectors of closed strings.

A detailed study of the charge spectrum of $3D$ Abelian TMGT is
given. It is shown that Narain constraints on toroidal
compactification (integer, even, self-dual momentum lattice) have a
natural interpretation in purely three dimensional terms.  This is an
important result which is necessary to construct toroidal
compactification and the heterotic string from TM(GT).  The block
structure of $c=1$ Rational Conformal Field Theory (RCFT) from the
point of view of three dimensional field theory is also derived.

Open and unoriented strings in TM(GT) theory are also studied through
orbifolds of the bulk $3D$ space. This is achieved by gauging discrete
symmetries of the theory. Open and unoriented strings can be obtained
from all possible realizations of $C$, $P$ and $T$ symmetries. The
important role of $C$ symmetry to distinguish between Dirichlet and
Neumann boundary conditions is discussed in detail.

Future directions of research in this field are also suggested and
discussed.

\vfill
\begin{center}
Thesis submitted for the degree of Doctor of Philosophy\\
in the University of Oxford

5th October 2001, Michaelmas term
\end{center}

\newpage
\thispagestyle{empty}
\ \\

\newpage
\thispagestyle{empty}
\begin{center}
\ \vspace{.2cm}

{\large\sc Acknowledgments}
\end{center}
\vspace{.8cm}

Firstly I would like to thank Ian Kogan for his guidance in
physics and his personal support over the last four years.

I would like to thank my collaborator Bayram Tekin.

Also I thank Alex Kovner and Andr\'e Lukas
for many discussions, suggestions and their patience.

Thanks also to Martin Schvellinger and Arshad Momen for several discussions.

I thank my examiners John Weather and Richard Szabo as well for their
suggestions and comments.

Further I would like to thank my colleagues M\'ario Santos, Nuno Reis,
Florian Merz, Emiliano Papa, Adrian Campbell Smith, Dave Skinner, Alex
Nichols, Gl\'oria Luzon, Jos\'e Nat\'ario, Francesco Mancini, Fermin
Viniegra, Sanjay and specially to Guilherme Milhano and Jo\~{a}o
Correia for they company and for several discussions.

Thanks to my friends in Oxford, Rob Wallis, Jos\'e Fernandez-Calvo,
Charlie Crichton, Mike Hall, Rebecca Smith, Enda Leaney, Margaret
Jackson, Bob Cowan, James Beresford, Pierre Courtois, Andr\'e Gouveia,
Ana Lopes, Patricia Castro, Yasser Omar, Ruy and Lory
Ribeiro for their friendship and specially to Luis Gomes for his
ability for taking me out of troubles.

And finally thanks to my friends faraway Ricardo Alves, Paulo Aur\'elio Azevedo Moniz,
Silvia Ribeiro, Tiago Perdig\~{a}o, Jorge Maria and Jo\~ao Jorge. Also
to C. for showing me that flowers grow in the desert.

Thanks to my family, Ant\'onio,
Irene, Frederico, Rodrigo and Zara for their help and support over the
last four troubled years and specially to Beatriz for always finding
good use for my equations: tearing them apart. Also to her mother,
Maria for the year we have been together in Oxford.

I acknowledge Keble College and the Theoretical Physics Group for
receiving me and for the travel grants, in particular to Janet Betts and
Gill Dancey for their efficiency,

This thesis was supported by PRAXIS XXI/BD/11461/97 grant from FCT (Portugal).

\newpage
\thispagestyle{empty}
\ \\

\newpage
\thispagestyle{empty}
\begin{center}
\ \vspace{5cm}\\
to Beatriz\\
\end{center}

\newpage
\thispagestyle{empty}
\ \\

\newpage
\thispagestyle{empty}
\begin{flushright}
\ \vspace{5cm}\\
Als das Kind Kind war, wu\ss te es nicht, da\ss\ es Kind war,\\
alles war ihm beseelt, und alle Seelen waren eins... 

\ \\
Als das Kind Kind war,\\
war das die Zeit der folgenden Fragen:\\
Warum bin ich Ich, und warum nicht Du?\\
Warum bin ich hier und warum nicht dort?\\
Wann begann die Zeit und wo endet der Raum?\\
Ist das Leben unter der Sonne nicht blo\ss\ ein Traum?

\ \\
\textit{in} ''Der Himmel \"Uber Berlin'' (1986) by Wim Wenders
\end{flushright}

\newpage
\thispagestyle{empty}
\ \\

\newpage
\tableofcontents
\newpage
\pagenumbering{arabic}
\chapter{Introduction}

This chapter gives an historical overview of $3D$ Chern-Simons theory
as well as a short review of the Topological Membrane (TM) approach to
string theory. Not many details are given, only a general overview of
past work on the subject. Some fundamental concepts will be introduced
and a description at the level of the action, justifying the several
terms that constitute the full action of TM.  In the remaining of this
work only Abelian Topological Massive Gauge Theories (TMGT) will be
addressed in detail.

In Chapter~\r{ch.quant} the quantization of TMGT is addressed, both from
a path integral and canonical perspective, and some fundamental issues 
studied and introduced.

In Chapter~\r{ch.tor} the correct toroidal compactification spectrum
of bosonic closed string theories and the heterotic string will be built
purely from $3D$ TM(GT) physics. The Narain lattice spectrum
conditions (Lorentzian even, integer, self dual lattice) will also be
derived. The fusion rules will as well be rederived from the point of
view of the bulk theory. A Heterotic string theory background encoding
the $E_8\times E_8$ group will be rederived in the light of these results.

In Chapter~\r{ch.opun}, open and unoriented string theories are built by
orbifolding TM(GT) and gauging the $3D$ QFT symmetries $PT$ and $PCT$. A
preliminary discussion on modular invariance and T-duality is carried out.

Finally Chapter~\r{ch.conc} summarizes the accomplishments of this
thesis and further directions of research in these topics are suggested.

\newpage
\section{$3D$ Chern-Simons Theories}

In mathematical terms the Chern-Simons (CS) term~\cite{CS_0} is a
topological invariant. It  comes from the Chern class of $4D$
manifolds. The introduction in physics of this term
in $3D$ was originally suggested by Schwarz~\cite{CS_1} and Singer
in unpublished work on its Abelian version
\be
S_{CS,a}[A]=k\int_Md^3xA\wedge F
\ee
where $A$ is some gauge connection field on a $3D$ manifold $M$ and
$F$ its curvature. Several other
authors~\cite{CS_2,CS_3,CS_4,CS_5} studied independently the CS term
together with Maxwell-Yang-Mills in several different contexts. For a
brief historical review and applications of CS in physics
see~\cite{CS_6} and references therein.

Together with the Maxwell term the CS term introduces many new
properties in the gauge theory. Although being a topological invariant
affecting the long-range behavior of the free field Maxwell theory,
namely large gauge transformations depend on the CS term coefficient,
it is a quadratic term and also leads to finite-range effects, namely
it gives a mass to the photon~\cite{CS_4} (an alternative to the Higgs
mechanism of mass generation).  It also changes the charge spectrum of
the theory and is $P$ and $T$ violating changing the allowed discrete
symmetries of the theory.

In its non-Abelian version the CS action is
\be
S_{CS,na}[A]=\frac{k}{8\pi}\int_Md^3x\tr\left[A\wedge dA+\frac{2}{3}A\wedge A\wedge A\right]
\ee
and the CS coefficient has to be quantized in order for the quantum theory
to be well defined~\cite{CS_5} due to the compacteness of the gauge group.

In $3D$ there is also a gravitational CS term
\be
S_{CS,g}=-\frac{\kappa}{4}\tr\int_Md^3x\left[R\wedge\omega+\frac{2}{3}\omega\wedge\omega\wedge\omega\right]
\ee
where $R$ is the Riemann tensor and $\omega$ is the spin connection.
Note that $\kappa$ is not quantized since the Lorentz group is not
compact. In $3D$ there is a direct mapping between $3D$ CS-Einstein gravity
and non-Abelian CS-Maxwell gauge theory which will be explored below.

The connection between $3D$ CS gauge theory and Conformal Field Theory
(CFT) was first set forward by Witten~\cite{W_0}. Basically a pure CS
gauge theory is exactly soluble. Further, as long as the manifold $M$
admits a local topology of the form $M=\Sigma\times R$ ($R$ is some
compact interval associated with time coordinate), it induces chiral
Wess-Zumino-Witten-Novikov (WZWN) actions on fixed slices $\Sigma$
(see figure~\r{fig.witten1}). These WZWN models are simply chiral
CFT's. In this way a $3D$ gauge theory, which is a
thickening of a $2D$ CFT, is obtained.

\fig{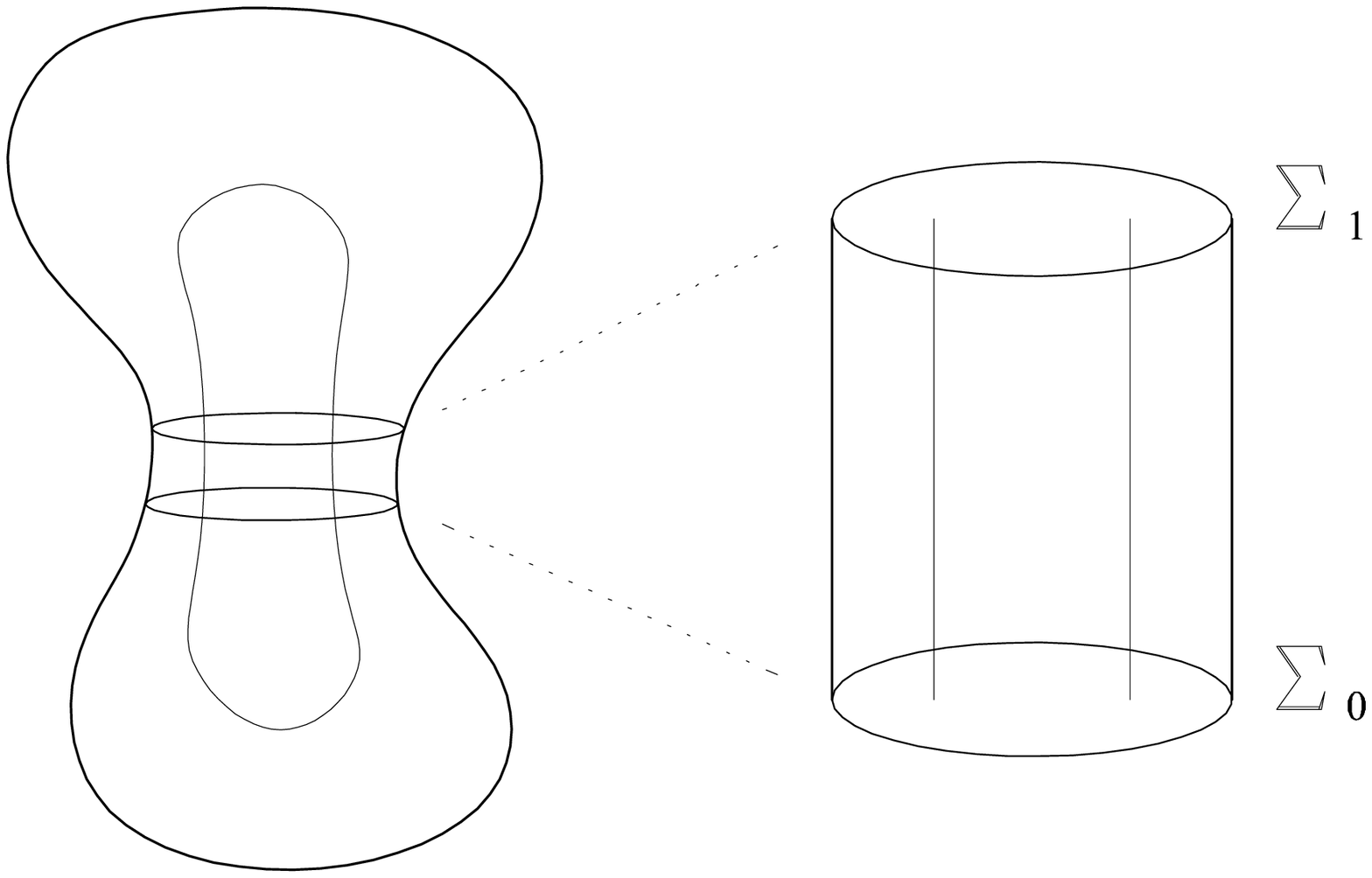}{Heegard splitting of a closed manifold. Locally it
is obtained from an open manifold with topology $\Sigma\times R$ (with $R$
being some compact interval).}{fig.witten1}

The CFT partition functions and operator content can in this way be
obtained from the $3D$ gauge theory giving a new insight and meaning
to the $2D$ conformal theory. These constructions were originally formalized
in~\cite{MS_1,MS_2,BN_1,LR_1,O_1}.

One further important remark is that in $3D$ gauge theories both the
Maxwell and CS terms are present. Starting from either of the pure
cases (Maxwell or CS) the other term is induced by quantum radiative
corrections once fermions are introduced. One can, as well, think of
the Maxwell term as a regulator of the pure CS theory. Then the CS
theory is the infra-red limit of the Maxwell-CS
theory. See~\cite{GD_1} for a review of $3D$ CS theories.

The main point to stress is that neither Maxwell nor CS theories exist
as a self consistent quantum gauge theories. Further, it will be argued next
that gravity also needs to be taken into account to describe a
full-fledged $3D$ theory and, hence, a fully-fledged boundary CFT.

\newpage
\section{TMGT - Topologically Massive Gauge Theory}

The action for an Abelian TMGT with gauge group $U(1)^N$ is
\be
S_{TMGT,N}=\int_M\left[-\frac{\sqrt{-g}}{4\gamma}F^I_{\mu\nu}F_I^{\mu\nu}+\frac{K_{IJ}}{8\pi}\epsilon^{\mu\nu\lambda}A^I_\mu 
\partial_\mu A^J_\lambda\right]
\lb{STMGTN}
\ee
where $A_I$ are the $N$ gauge connection fields ($I=1,\ldots,N$) and
$F^I$ the corresponding curvatures. $K_{IJ}=G_{IJ}+B_{IJ}$ are the
couplings between the several $U(1)$ gauge fields. $G$ stands for the
symmetric part of the matrix $K$ and $B$ for its antisymmetric
part. The manifold where the theory is defined is considered to have
the topology $M=\Sigma\times[0,1]$ with $t\in [0,1]$. Therefore $M$
has two boundaries: $\partial M=\Sigma_{t=0}\cup\Sigma_{t=1}$.

The $F^2/\gamma$ term can either be thought as being introduced
as a regulator, in this the pure topological CS is retrieved in
the limit $\gamma\to\infty$, or it can be thought as being in the
action necessarily, in which case the pure CS case is simply the
low-energy limit of the theory (i.e. the ground state of the
Maxwell-CS theory is the same as the pure CS theory). Anyhow in the
quantum theory both terms will be present due to radiative corrections.

Note that since $M$ has boundaries the theory is not gauge invariant,
nor has it any classical extrema (due to boundary induced terms under
infinitesimal transformations of the fields). Depending on the
boundary conditions, new degrees of freedom will emerge on the boundary
corresponding to the gauge transformation parameters. The induced
actions will be WZWN actions of the form
\be
\ba{ll}
I(g)=&\displaystyle\frac{k}{8\pi}\int_\Sigma d^2z\tr\left(g^{-1}dg.g^{-1}dg
\right)+\vspace{.2cm}\\
&\displaystyle\frac{k}{24\pi}\int_Md^3x \tr\left(g^{-1}dg\wedge g^{-1}dg\wedge
g^{-1}dg\right)
\ea
\ee

These degrees of freedom will correspond to bosonic fields of chiral
CFT's on both boundaries $\Sigma_0$ and $\Sigma_1$.  Furthermore, at least
one of the gauge fields $A_i$ ($i=z,\bz$) will necessarily be fixed at
the boundary in order to ensure that the theory has a classical
extremum. The boundary conditions are implemented by the insertion of
boundary actions or, equivalently, by inserting wave functions at the
boundaries. Note that there is a one to one correspondence between the
ground state wave functions and the conformal blocks of the boundary
CFT.  It is also important to stress that the relative boundary
conditions (between both boundaries) are selecting the kind of
relative chirality of the two CFT's, remember that the final aim here
is to obtain string theory and all the aspects inherent to it related
to CFT's. In the case of a full non-chiral CFT, the left movers will
live on one boundary (say $\Sigma_0$), while the right movers will
live in the other one ($\Sigma_1$).

The Maxwell term turns out to be fundamental in this construction: In
the pure CS case there are only two canonical coordinates,
$\pi^z=\epsilon^{z\bz}A_\bz$ is the canonical conjugate momenta to
$A_z$. The way to impose boundary conditions in pure CS is to choose a
proper polarizations in each boundary, which means that $A_z$ is chosen
as spatial coordinate in one boundary while $A_\bz$ is chosen on the
other one. In the Maxwell-CS theory there are four canonical
coordinates, the canonical conjugate momenta to $A_i$ is
$\pi^i=F^{0i}/\gamma+k\epsilon^{ij}A_j/8\pi$.  Fixing $A_z$ only,
$\pi^\bz$ is fixed as well due to the commutation relation
$[A_z,\pi^\bz]=0$. In this way it will indeed be possible to have
(anti)holomorphic bosonic degrees of freedom on the boundaries holding
two boundary chiral CFT's (one in each boundary) or alternatively
killing all the degrees of freedom on only one of the boundaries
allowing the construction of the heterotic string.

The pure CS theory is geometry independent (i.e. does not depend on
the metric). The CS term only  depends on the antisymmetric tensor,
being a topological invariant. The metric on the boundary is induced
by the antisymmetric tensor $h^{(2)}_{12}=\epsilon^{012}$. Depending
on the choice of boundary conditions (anti-holomorphic chiral CFT's)
two inequivalent classes of metrics will be selected on the
boundaries. Basically one is selecting a polarization of the theory,
that is which of the $A_i$'s is the momenta and which is the position
in the phase space.
By introducing the Maxwell term the theory is made geometry dependent
since $F^2/\sqrt{g}$ depends explicitly on the metric.
The induced metric on the boundaries is in this case
$h^{(2)}_{ij}=\epsilon^{0ij}/\sqrt{g^{(3)}}$ for $i<j$.

\newpage
\section{Topological Membrane Approach to String Theory}

At the heart of the five known string theories (I, IIA, IIB and
Heterotic $SO(32)$ and $E_8\times E_8$) is a $2D$ CFT, a sigma
model. Together with $11D$ supergravity, these five theories are
connected by a web of dualities. It is also believed that these five
theories are effective limits of a wider $11D$ theory,
M-theory~\cite{W_M,T_M}.

Therefore it can be considered that such a theory is described by a $3D$
gauge theory defined on a manifold $M$ with some $2D$ boundary where
the corresponding string CFT's are induced. Although the description
of the 11th dimension has not so far been clarified in terms of a $3D$
gauge theory one fact is sure: the principles inherent to the $3D$
gauge theory are fewer than the principles inherent to all the
existing string theories, which is something desirable in any unifying
theory!

Although in a different formalism, this kind of approach was
originally suggested for open strings by Witten~\cite{W_1}
(see also~\cite{W_0}). But these first works took into account
only the pure topological CS theory such that in the bulk $3D$ gauge
theory and only topological degrees of freedom would survive.

Kogan~\cite{TM_00} suggested that closed string theories be described
by a Topological Massive Gauge Theory (TMGT). The Maxwell term was
introduced as a regulator which would allow a description of off-shell
closed string theory on the $2D$ boundary of the membrane. The
on-shell string theory would be the infra-red limit. Since then much
work has been done in this
subject~\cite{TM_01,TM_02,TM_03,TM_04,TM_04a,TM_05,TM_06,TM_07,TM_08,TM_09,TM_10,TM_11,TM_12,TM_13,TM_14,TM_15,TM_16,TM_last}.

Although the off-shell description was not fully
developed, it becomes apparent that the Maxwell term is fundamental in
the description of string theory in order to have the correct
left and right spectrum on opposite boundaries, to have the
correct degrees of freedom on the boundary chiral CFTs and
allowing, as well, the construction of the heterotic string.

Further to the TMGT it is necessary to introduce Topological Massive
Gravity (TMG). This is due to the Maxwell term which depends explicitly
on the metric. Hence it is necessary to consider the path integral
over metrics averaged by the Einstein action. Similar to the pure TMGT
case, a gravitational CS term will also be induced. In terms of
the boundary string theory the TMG sector will induce the integration
over moduli (holding the partition function modular invariant) and, if
necessary (as it will be explained), a Liouville field (maintaining
Weyl invariance).

Generic WZWN models can be built by considering appropriate bulk gauge
groups. In particular coset constructions correspond, from the point of
view of the bulk, to summing several CS actions (or more generally,
TMGT actions).

Superstring theories can be built by considering supersymmetric
membranes. This can be achieved in two ways; either by considering
Majorana fermions in the bulk such that the $3D$ supersymmetry
transformations induce fermionic edge states (in a similar way to the
gauge transformations in TMGT), or considering an appropriate coset
construction such that, on the boundary, one of the several gauge
fields that are present correspond to the bosonized fermions while the
other ones correspond to the boson fields.

\newpage
\section{TMG - Topologically Massive Gravity\lb{ch.int:sec.tmg}}

Since our action is explicitly dependent on the metric it is necessary to 
consider as well a gravitational term~\cite{TM_02,TM_03} and integrate 
over metrics in the path integral.
Note that, if this term is not present, it will be induced
by quantum corrections. The obvious candidate is the $3D$ Einstein action
\be
S_E=\kappa\int_Md^3x\sqrt{g}R=\kappa\int_Md^3x\epsilon^{\mu\nu\lambda}\epsilon_{abc}e_\mu^a\left(\partial_\nu w_\lambda^{bc}-\partial_\lambda
w_\nu^{bc}+[w_\nu,w_\lambda]^{bc}\right)
\ee
where $R$ is the curvature
and both a vierbein $e$ and a spin connection $w$ were introduced
($g_{\mu\nu}=e_\mu^ae_\nu^b\eta_{ab}$).

There also exists a gravitational CS term which can be thought of
as induced by quantum corrections, $\frac{k'}{4\pi}
\left(\eta_{ad}\eta_{bc}w_\mu^{ab}\partial_\nu w_\lambda^{cd}\\ +
\frac{2}{3}\eta_{af}\eta_{bc}\eta_{de}w_\mu^{ab}w_\nu^{cd}w_\lambda^{ef}\right)$.

The spin connection $w$ is a function of the vierbein $e$.
They can be made independent by introducing 
a Lagrange multiplier term~\cite{TM_02,TMG_1}.

The full TMG action is then given by
\be
\ba{rcl}
S_{TMG}=\int_Mdtd^2z&\epsilon^{\mu\nu\lambda} &\left[\displaystyle\kappa\epsilon_{abc}e_\mu^a
\left(\partial_\nu w_\lambda^{bc}-\partial_\lambda w_\nu^{bc}+[w_\nu,w_\lambda]^{bc}\right)\right.\vspace{.1cm}\\
&+&\displaystyle\frac{k'}{4\pi}\left(\eta_{ad}\eta_{bc}w_\mu^{ab}\partial_\nu w_\lambda^{cd}+\frac{2}{3}\eta_{af}\eta_{bc}\eta_{de}w_\mu^{ab}w_\nu^{cd}w_\lambda^{ef}\right)\vspace{.1cm}\\
&+&\displaystyle\left.\lambda_\mu^{a}\left(\eta_{ab}\partial_\nu e_\lambda^b+\eta_{ab}\eta_{cd}w_\mu^{bc}e_\lambda^d\right)\right]\vspace{.1cm}
\ea
\lb{STMG}
\ee
where $\lambda$ in the last
term is a Lagrange multiplier such that $w$ and $e$ are
independent and there are four canonical variables: $w_z$,
its conjugate $w_\bz$, $e$ and its canonical conjugate $
\beta_\nu^a=\lambda_\nu^a+\frac{1}{\tau^2}\eta_{bd}\eta_{ce}\epsilon^{abc}w_\nu^{de}$.

In $3D$ there is a one to one map between
non-Abelian gauge theory and gravity~\cite{TM_02,TMG_2,TMG_3}.
Define the gauge field as
\be
A_\nu=e_\nu^aP_a+\eta^{cd}\epsilon_{abc}w_\nu^{ab}J_d
\ee
where $J^a\equiv\frac{1}{2}\epsilon^{abc}J_{bc}$, $J^{ab}$ are the
Lorentz generators and $P^a$ the translation generators.
The gauge group corresponding to the action~(\r{STMG})
is $SO(2,1)$ with quadratic invariants
$\left<J_a,J_b\right>=\left<P_a,P_b\right>=\delta_{ab}$ and
$\left<J_a,P_b\right>=0$.

Following a similar treatment to that of the gauge theory,
(gravitational) WZWN actions 
with gauge group
\be
SO(2,1)\cong PSL(2,{\mathbb{C}})\cong SL(2,{\mathbb{R}})\times SL(2,{\mathbb{R}})
\ee
emerge on the boundaries of the manifold $M$.

It is interesting to note that similar to TMGT one can impose boundary
conditions such that (e.g. for the sphere) one of the $SL$'s is on one
boundary and the other one on the other. This means that these groups
generate conformal transformations on the boundaries. Remember that
only holomorphic degrees of freedom live on one boundary and
antiholomorphic on the other one such that conformal transformations
act in $z$ in one boundary and $\bz$ in the other one.

Note that $2D$ ghosts do not emerge due to gauge fixing the
gravitational diffeomorphism since the Faddeev-Popov determinant is
absorbed by the constraints of the theory (see~\cite{TM_last} and
references therein for details).

Fixing and integrating over moduli in string theory will be
accomplished by considering boundary actions similarly to the method
described below in section~\r{ch.quant:sec.path} or, equivalently, by 
considering the insertion of boundary wave functions as described
below in section~\r{ch.quant:sec.can} for the gauge sector.

The gravitational wave function was computed in~\cite{TM_04a} for the
case of the torus (see also~\cite{TM_last} and references
therein). Taking two commuting holonomies of $SO(2,1)$ $\exp\{\mu
J_2\}$ and $\exp\{\nu J_2\}$ and the extrinsic curvature $K$ of
$\Sigma$ it is
\be
\psi_{\mathrm{grav}}[\mu,\nu,K]=\int_{{\mathcal{F}}}\frac{d^2\tau}{(\im\tau)^2}\ \frac{\mu-\tau\nu}{\pi K\sqrt{\im\tau}}\ \exp\left\{-\frac{i|\mu-\nu\tau|^2}{K\im\tau}\right\}\chi(\tau,\bar{\tau})
\ee
where the integration is over a fundamental region $\mathcal{F}$ of
the modular group of the torus and $\tau$ is the modular parameter and 
$\chi$ is some mass form. In particular the ground state is given by
\be
\chi^{(0)}(\tau,\bar{\tau})=\sqrt{\im\tau}|\eta(\tau)|^2
\ee
with $\eta(\tau)$ the usual Dedekind eta function
\be
\eta(\tau)=\exp\left\{2\pi
i\tau/24\right\}\prod_{r=1}^{\infty}\left(1-\exp\left\{2\pi i 
r\tau\right\}\right)
\lb{dede}
\ee

Computing the ground state to ground state transition amplitude one
finds
\be
\left<\psi^{(0)}_{\mathrm grav}|\psi^{(0)}_{\mathrm grav}\right>=\int_{{\mathcal{F}}}\frac{d^2\tau}{(\im\tau)^2}|\eta(\tau)|^4
\ee

This is exactly the diffeomorphism ghost contribution for the
partition function of string theory when the moduli are gauge fixed
(for the torus).

These computations have not been performed for higher genus Riemann
surfaces but a similar picture should hold.

What remains is to explain how the Liouville field emerges from TMG. To 
do so consider the gravitational WZWN model induced at the boundaries
due to the diffeomorphisms
\be
\ba{ll}
I(g,w)=&\displaystyle\frac{k'}{4\pi}\int_\Sigma d^2z\tr\left(g^{-1}dg.g^{-1}dg
\right)+\vspace{.2cm}\\
&\displaystyle\frac{k}{12\pi}\int_Md^3x \tr\left(g^{-1}dg\wedge g^{-1}dg\wedge
g^{-1}dg\right)+\vspace{.2cm}\\
&\displaystyle\frac{k'}{4\pi}\int_\Sigma
d^2z\tr\left(2w.g^{-1}dg-2M w^a.e_a+w.w\right)
\ea
\ee
where $M=2\pi\kappa/k'$ stands for the mass of the $3D$ bulk graviton.

The previous discussion generating the ghost concerns the ground state
of the theory, or equivalently, the topological limit of the
theory. This topological limit corresponds to the vacuum expectation
values of the driebeins being zero, $\left<e^a_\nu\right>=0$. In the case
that this value is taken to be non-zero
\be
\left<e^a_\nu\right>=\delta^a_\nu
\ee
the bulk graviton becomes massive and the boundary theories are not
diffeomorphism invariant any longer.

The full computation is done in~\cite{TM_last} and is not going to be
carried out here. It is enough to consider the Gauss decompositions for
the $SL(2,{\mathbb{R}})$ group introduced in~\cite{TMG_4}
\be
g=\left(\ba{cc}1&0\\\gamma&1\ea\right)\left(\ba{cc}e^\phi&0\\0&e^{-\phi}\ea\right)\left(\ba{cc}1&\gamma'\\0&1\ea\right)
\ee
After an appropriate redefinition of $\phi$ one gets the Liouville
action on the $2D$ boundary with a cosmological constant $\mu$
\be
S_{\mathrm{Liouv}}=\int_\Sigma d^2z\left(
\partial_\bz\phi\partial_z\phi+\lambda R^{(2)}\phi+\mu\exp\left\{\beta\phi\right\}\right)
\ee
where $\beta$ and $\lambda$ are appropriate constants used in the
redefinition of $\phi$.

Furthermore the $2D$ cosmological constant turns out to be exactly the
square of the graviton mass
\be
\mu=M^2
\ee

The interesting conclusion to draw from the previous discussion is
that the bulk theory, in the topological limit of TMG reproduces
Weyl and conformal invariance and gives us the appropriate ghosts and
integration over moduli space. Once one considers excited states the
boundary theory is not Weyl invariant anymore; instead a new field,
identified with the boundary Liouville field, emerges such that the
total boundary action is again Weyl invariant. Note that this
mechanism is exactly the one in string theory which gives rise to a
new target space coordinate, the Liouville field (see for
instance~\cite{POL_1} for this discussion), but is here obtained
from TMG arguments! This is actually one way of getting an extra 11th
dimension. 

Note that this argument is very similar to the discussion concerning TMGT
when considering strings off-shell. In the topological limit (ground
state) one obtains the usual on-shell string CFT on the boundaries,
while when considering excited states one obtains a non-conformal 
theory on the boundaries corresponding to off-shell string theories.

To finalize this discussion note that the mechanism just described is
simply forcing the boundary theory to be critical. This means the total
central charge is zero and one obtains critical string theory.

\newpage
\section{Supersymmetry}

In order to generate superconformal boundary actions one may take two
different approaches. Either consider a bulk fermionic sector such
that the $3D$ theory is supersymmetric and the supersymmetry
transformations induce on the boundaries the correct superconformal
degrees of freedom~\cite{TM_05,TM_09}, or it may be considered as a coset
model such that one of the gauge fields represent bosonized fermions
and the other fields are responsible for the flux between different
sectors (i.e. between the boundaries), say Ramond and Neveu-Schwarz
for example~\cite{TM_11}.

Consider the $3D$ supersymmetric massive gauge action
\be
S=\int_M\left[-\frac{1}{4\gamma}F_{\mu\nu}F^{\mu\nu}+\frac{k}{8\pi}\epsilon^{\mu\nu\lambda}A_\mu\partial_\nu 
A_\lambda+\frac{1}{2\gamma}\bar{\Psi}\gamma^\mu\partial_\mu\Psi-\frac{k}{8\pi}\bar{\Psi}\Psi\right]
\ee
with $\Psi=(\psi,\bar{\psi})$ being a two-component
Majorana-Weyl fermion.

As for the TMGT case, due to the boundaries, the action does not have an
extremum and is not invariant under
the bulk $N=1$ supersymmetry transformations
\be
\ba{rcl}
\delta
A_\mu&=&\bar{\eta}\left(\partial_\mu\lambda-\gamma_\mu\psi\right)\vspace{.1 cm}\\
\delta \psi&=&\epsilon^{\mu\nu\lambda}\partial_\nu A_\lambda\gamma_\mu\eta
\ea
\ee
where $\eta$ is a global Grassmann parameter. Note that the first term
in the transformation of $A$ can be thought as a gauge transformation
with parameter $\bar{\eta}\lambda$.

In each boundary one can impose one of the components of $\Psi$ to be
fixed such that a chiral action is generated. On each boundary
only $N=1/2$ supersymmetry will survive such that upon identifying both
boundaries the resulting non-chiral world-sheet action will correspond to $N=1$
supersymmetry.

As an example the NSR string action
\be
S_{NSR}=\frac{k}{8\pi}\int_{\Sigma}d^2x\left[\partial_z X\partial_\bz X-i\bar{\psi}\gamma^{\mu}\partial_\mu\psi\right]
\ee
can be obtained by taking $\bar{\psi}$ fixed in $\Sigma_0$ and $\psi$
fixed in $\Sigma_1$ and adding the boundary
action
$k/8\pi\left(\int_{\Sigma_0}\psi\partial_\bz\psi-\int_{\Sigma_1}\bar{\psi}\partial_z\bar{\psi}\right)$.
For a more detailed calculation see for instance the similar computation
carried out for the TMGT sector in section~\r{ch.quant:sec.path}~\cite{TM_05}.

Consider now the second proposed approach using coset constructions.
To generate the supersymmetric Virasoro algebra on the $2D$ boundary
WZWN model corresponding to the $N=2$ superconformal minimal model, the
coset  to be considered is~\cite{TM_11}
\be
M_k=SU(2)_{k+2}/U(1)_{k+2}\equiv SU(2)_k\times SO(2)_2/U(1)_{k+2}
\ee
In terms of the bulk theory this corresponds to the action:
\be
S_{M_k}=k I_{SU(2)} + 2 I_{SO(2)} - (k+2) I_{U(1)}
\ee
The central charge of this minimal model is
\be
c_k=\frac{3k}{k+2}
\ee

The TMGT corresponding to the Abelian gauge group $SO(2)$ will induce
on the boundary a $2D$ action which, upon fermionization, is the
worldsheet fermionic sector of the superstring. Its ground state
solution is degenerate. At level $l$ there are $l/(2\eta^2)^g$
solutions, where $g$ is the genus of the Riemann surface
$\Sigma^g$. The matter is introduced in the theory through closed
Wilson lines carrying half integer charges $Q_\eta=1/2$ corresponding
to Fermi statistics, the wave function for a gauge field $B$ is
\be
\Psi_{r_l}=\exp\left[-\frac{1}{16\pi}b^l\tau_{lm}(\bar{b}-b)^m\right]\Theta^{g}\left[\ba{c}r/4\\ 0\ea\right]\left(\left.\frac{b}{\pi}\right|\frac{8\tau}{\pi}\right)
\ee
with the decomposition $B=b+d\phi+*d\chi$, where
$b=(\bar{b}^l\omega_l-b^l\bar{\omega}_l)$. The indices run
$l,m=1,\ldots,g$ and $r_l=1,2,3,4$($\leq l/2\eta^2$).  $\tau$ is a
$g\times g$ matrix which encodes the modular structure of $\Sigma^g$
and $\omega_l=\tilde{b}^\alpha_l+\tau_{lm}\tilde{b}^\beta_m$ is
parameterized by the harmonic forms $\tilde{b}^\alpha$ and
$\tilde{b}^\beta$ normalized as the Poincar\'e duals
$\oint_{\alpha_m}\tilde{b}^\alpha_l=\oint_{\beta_m}\tilde{b}^\beta_l=\delta_{ml}$
and
$\oint_{\alpha_m}\tilde{b}^\beta_l=\oint_{\beta_m}\tilde{b}^\alpha_l=0$.
The closed contours $\alpha$ and $\beta$ stand for a basis of the
canonical homology cycles of the Riemann surface $\Sigma^g$.

The boundary states of the $3D$ theory corresponding to
the NS and R sectors are obtained as quantum superimpositions
of the 4 possible ground states $\left|r_l\right>$, which is to say
that the correct basis of states will be chosen.
The periodicity on both worldsheet coordinates can be
\textit{measured} in the TM framework using the Wilson loop
operators $W_C[B]=\exp\left[iQ_\eta\oint_C B\right]$ by taking $C$ to be a
fixed time path coinciding with the holonomy cycles of $\Sigma^g$
\be
W_\alpha^l=\exp[i(\bar{b}-b)^l]\ \ \ \ \ \ \ W_\beta^l=\exp[i(\tau_{lm}\bar{b}^m-\bar{\tau}_{lm}b^m)]
\ee
The averages of these operators are $\left<W_i\right>=\pm 1$ ($i=\alpha,\beta$)
corresponding to periodic ($+1$) and antiperiodic ($-1$) fermions.
For genus 1 the results are summarized in table~\r{tab:R-NS}.
\bt{|c|c|c|l|}
\hline
$\Psi$&$\left<W_\alpha\right>$&$\left<W_\beta\right>$&type\\\hline
$\left|1\right>+\left|3\right>$&$+$&$-$&R\\
$\left|1\right>-\left|3\right>$&$-$&$-$&NS\\
$\left|2\right>+\left|4\right>$&$+$&$+$&R\\
$\left|2\right>-\left|4\right>$&$-$&$+$&NS\\\hline
\et{\lb{tab:R-NS}Ground state superpositions holding NS and R
periodicities for genus 1.}

The GSO projections emerge in this way as the sum over these ground
states. This is equivalent to projecting the trace in the
partition function onto states with eigenvalue $+1$ of the Klein
operator $(-1)^F$ (states with an even number of fermions). This is
actually the mechanism which ensures that string theory is modular invariant
and that the $10D$ target space theory of string theory is
supersymmetric.

To end this Chapter note again that the full supersymmetric TM induces
$10D$ critical string theories. But if one considers the full TM,
including TMG as well, the $11D$ may be generated through the
Liouville field and/or the superghosts~\cite{KP_1}.

\chapter{Quantization of TMGT}
\lb{ch.quant}

\section{Path Integral Formalism\lb{ch.quant:sec.path}}

There are several ways to derive CFT from CS theory. The path integral
approach was first suggested by Ogura~\cite{O_1} (see
also~\cite{TM_04}).  In this section we review some features of TMGT
defined on a three dimensional flat manifold with a boundary.  In
order to clarify, some arguments derived from canonical formalism are
present although the full canonical quantization will only be carried
in the next section. A list of the induced chiral boundary conformal
field theories is presented and a careful analysis of the Gauss law
structure is considered such that the charge spectrum of the theory is 
built.

Two kind boundary conditions on the gauge fields will be considered in
the remaining of this work. Either the fields of the theory are
constraint (are compact) in such a way that large gauge
transformations are allowed or not. They will be labeled by
$\mathcal{B}$ and $\tilde{\mathcal{B}}$:
\be
\ba{ll}
\mathcal{B}:&\mathrm{Gauge\ transformations\ are\ allowed}\\
\tilde{\mathcal{B}}:&\mathrm{Gauge\ transformations\ are\ not\ allowed}
\ea\nonumber
\ee

Similarly to the action~(\r{STMGTN}), the TMGT action for a 
single $U(1)$ gauge group obeying $\mathcal{B}$ boundary conditions is
\be
S_{TMGT,1}=\int_{M} d^2z\,dt\left[-\frac{\sqrt{-g_{(3)}}}{4\gamma}F^{\mu\nu}F_{\mu\nu}+
\frac{k}{8\pi}\epsilon^{\mu\nu\lambda}A_\mu\partial_\nu A_\lambda\right]
\lb{STMGT1}
\ee
where ${M}=\Sigma\times[0,1]$ and $\Sigma$ is a $2D$
compact Euclidean manifold with a complex structure denoted by
$(z,\bz)$.  The time-like coordinate takes values in the compact
domain $t\in[0,1]$. The indices run over $\mu=0,i$ with $i=z,\bz$.

Under an infinitesimal variation of the fields $A\rightarrow~A+\delta
A$ the action changes by
\be
\delta S = \int_{M}\left(\frac{\sqrt{-g_{(3)}}}{\gamma}\partial_{\mu}F^{\mu\nu}+\frac{k}{4\pi}\epsilon^{\mu\lambda\nu}\partial_\mu A_\lambda\right)\delta A_{\nu}-\left[\int_\Sigma \Pi^i\delta A_i\right]^{t=1}_{t=0}
\lb{dS}
\ee
where
\be
\Pi^i= \frac{1}{\gamma}F^{0i}-\frac{k}{8\pi}\tilde{\epsilon}^{ij}A_j
\lb{mom_i}
\ee
is the canonical momentum conjugate to $A_i$.  Note that the $2D$
antisymmetric tensor $\tilde{\epsilon}^{ij}$ is induced by the $3D$
antisymmetric tensor and metric
\be
\tilde{\epsilon}^{ij}=\frac{-\epsilon^{0ij}}{\sqrt{-g_{(3)}}}
\ee
When referring to the usual $2D$ antisymmetric tensor the
notation $\epsilon^{ij}$ (without the tilde) is used~\cite{AFC_1}.

In order for the theory to have a classical extremum it is necessary to
impose suitable \textit{boundary conditions} for which the second
term in the variation of the action vanishes.
Let us assume that the boundary of $M$ has two 
pieces, which are $\Sigma_{(t=0)}=\Sigma_0$ and $\Sigma_{(t=1)}=\Sigma_1$. 
On each of the boundaries, up to gauge transformations, one can fix 
one or both fields $A_z$ and $A_\bz$. In doing so it is necessary to add an
appropriate boundary action $S_B=S_{B0}+S_{B1}$ such that the new
action $S+S_B$ has no boundary variation, and hence well defined
classical extrema. Note that upon canonical quantization
it is necessary to impose the corresponding equal time commutation
relations of $\Pi$ and $A$
\be
[\Pi^i(\vb{z}),A^\bj(\vb{z'})] = g^{i\bj}\delta^{(2)}(\vb{z}-\vb{z'}) 
\lb{com_pa}
\ee
The convention for the metric used here is $g^{z\bz} = g^{\bz z} = 2$.
So upon fixing $A^z$, $\Pi^z$ is fixed as well. The same holds for the $A^\bz$
and $\Pi^\bz$ components.

Then, on each component of the boundary ($\partial M=\Sigma_0\cup\Sigma_1$)
the possible choices of boundary conditions and
boundary actions are
\be
\ba{lcccc}
            &\mathrm\small boundary\ conditions&\Sigma_1\mathrm\small\ bound.\ action&\Sigma_0\mathrm\small\ bound.\ action\\
            & & & \\
N.          &\delta A_z=\delta A_\bz=0&S_{B1}=0&S_{B0}=0\\
            & & & \\
C.          &\delta A_\bz=0&\displaystyle S_{B1}=\int_{\Sigma_1}\Pi^z A_z&\displaystyle S_{B0}=-\int_{\Sigma_0}\Pi^zA_z\\
            & & & \\
\bar{C}.\ \ &\delta A_z=0&\displaystyle S_{B1}=\int_{\Sigma_1}\Pi^\bz A_\bz&\displaystyle S_{B0}=-\int_{\Sigma_0}\Pi^\bz A_\bz
\ea
\lb{bc}
\ee
There are nine allowed choices: $NN$, $NC$, $CC$, $C\bar{C}$, and so
on. The first letter denotes the type of boundary conditions on
$\Sigma_0$ and the second one on $\Sigma_1$.  The $N$ boundary
condition stands for \textit{Non-Conformal} or \textit{Non-Dynamical},
$C$ stands for \textit{Conformal} and $\bar{C}$ for
\textit{anti-Conformal} and are related to the kind of CFT which are
obtained on the boundaries when each of them are chosen, basically if
the bosonic fields are holomorphic or antiholomorphic.  Note the
importance of the $F^2$ term, it gives the theory four independent
canonical coordinates, as opposed to the two of the pure Chern-Simons
theory (where one of the $A$'s is canonically conjugate to the
other). This fact allows us to fix both of the $A$'s and corresponding
$\Pi$ in the same boundary allowing the construction of the heterotic
string. For further details we refer the reader
to~\cite{TM_05,TM_09}. This topic will be addressed again in
section~\r{ch.tor:sec.bc}.

But note that even with the addition of these boundary actions the full 
action is not invariant under gauge transformations. Considering the
transformation $A\to A+d\Lambda$, the bulk action
transforms as
\be
S\to S-\frac{k}{8\pi}\left[\int_\Sigma\epsilon^{ij}\partial_i\Lambda A_j\right]_{t=0}^{t=1}
\lb{gauge1}
\ee
while the boundary actions transform as
\be
\Pi^zA_z\to\Pi^zA_z+\Pi^z\partial_z\Lambda-\frac{k}{8\pi}\epsilon^{z\bz}\partial_\bz\Lambda A_z-\frac{k}{8\pi}\epsilon^{z\bz}\partial_\bz\Lambda\partial_z\Lambda 
\ee

This fact is actually what makes possible the construction of
effective $2D$ boundary theories out of $3D$ ones. The gauge
parameters will constitute the degrees of freedom of those $2D$
theories.  To see it explicitly consider gauge fixing the path integral
by Faddeev-Popov procedure such that
$A_\mu=\bar{A}_\mu+\partial_\mu\Lambda$.  The action and path integral
factorize as
\be
Z=\int{\mathcal{D}}\bar{\vb{A}}\Delta_{FP}\delta\left(F(\bar{\vb{A}})\right)e^{i\bar{S}[\bar{\vb{A}}]}\int{\mathcal{D}}\lambda{\mathcal{D}}\chi e^{i\bar{S}_{B}[\chi,\lambda]}
\lb{Z.int}
\ee
where on the boundaries the gauge parameters
$\Lambda(t=0)=\chi$ and $\Lambda(t=1)=\lambda$
become dynamical degrees of freedom which decouple from the bulk theory.

As an example take $\bar{C}C$ boundary conditions. After gauge fixing
the boundary action becomes
\be
\ba{rl}
S_{B,\bar{C}C}=&\displaystyle\int_{\Sigma_1}\left[\bar{\Pi}^z\bar{A}_z+\bar{\Pi}^z\partial_z\lambda-\frac{k}{8\pi}\epsilon^{z\bz}\partial_z\lambda \bar{A}_\bz-\frac{k}{8\pi}\epsilon^{z\bz}\partial_\bz\lambda\partial_z\lambda\right]\vspace{.2cm}\\
-&\displaystyle\int_{\Sigma_0}\left[\bar{\Pi}^\bz \bar{A}_\bz+\bar{\Pi}^\bz\partial_\bz\chi+\frac{k}{8\pi}\epsilon^{z\bz}\partial_\bz\chi \bar{A}_z+\frac{k}{8\pi}\epsilon^{z\bz}\partial_\bz\chi\partial_z\chi\right]
\ea
\ee

Gluing both boundaries using the identification $z\cong\bz$ and
$A_z\cong A_\bz$ the previous action can be rewritten in the simpler form
\be
\ba{rcl}
S_{B,\bar{C}C}=&\displaystyle\left.\int_\Sigma\right[ &\displaystyle(\bar{\Pi}^z-\frac{k}{8\pi}\epsilon^{\bz
z}\partial_\bz\lambda)(\bar{A}_z+\partial_z\chi)-(\bar{\Pi}^\bz-\frac{k}{8\pi}\epsilon^{z
\bz}\partial_z\chi)(\bar{A}_\bz-\partial_\bz\lambda)\vspace{.2 cm}\\
&-&\displaystyle\left.\frac{k}{8\pi}\epsilon^{z\bz}\partial_z(\chi-\lambda)\partial_\bz(\chi-\lambda)\right]
\ea
\ee
where $\Sigma$ stands for the \textit{identified}
$\Sigma_0\cong\Sigma_1$.

In the path integral there is still a
dependence on the boundary values of $A$'s. Assuming that the measure can be
factorized into a bulk integration times a boundary integration
\be
\int{\mathcal{D}}\bar{A}_z{\mathcal{D}}\bar{A}_\bz=\int{\mathcal{D}}\bar{A}_{z,\mathrm{bulk}}{\mathcal{D}}\bar{A}_{\bz,\mathrm{bulk}}\int{\mathcal{D}}\bar{A}_{z,\mathrm{bound}}{\mathcal{D}}\bar{A}_{\bz,\mathrm{bound}}
\ee

and performing the Gaussian integration on the boundary $A$'s one gets
the action 
\be
\bar{S}_{B}=\frac{k}{4\pi}\int g^{ij}\partial_i(\chi-\lambda)\partial_j(\chi-\lambda)
\ee
presented in the second factor of~(\r{Z.int}).  The metric is identified as
$g^{z\bz}=-i2\epsilon^{z\bz}$. Note that $\epsilon^{z\bz}=i$.

This action can be recognized as the
$d=1$ free boson action $\int \partial X \bar{\partial}X$ with
$X=\chi-\lambda$. $\chi$ stands for the holomorphic part of $X$ and
$\lambda$ for the antiholomorphic part. The point to stress is that
it completely decouples from the bulk and the path integral indeed
factorizes. In this way it is proved that there is actually an
effective $2D$ boundary CFT.

Note that the effective boundary action depends on the chosen boundary
conditions and the identifications used in the gluing procedure.  This
issue will be dealt with in more detail in section~\r{ch.tor:sec.bc}.

Now let us find the allowed charges in this theory. The boundary
conditions of the gauge fields $A$ has to be taken into account and it
turns out to be of fundamental importance due to the existence of
large gauge transformations (for compact $A$) around the holonomies of
the gauge connections. Define the electric and magnetic fields as
\be
\ba{rcl}
E^i&=&\displaystyle 
\frac{1}{\gamma}F^{0i}=\Pi^i+\frac{k}{8\pi}\tilde{\epsilon}^{ij}A_j\vspace{.2 cm}\\
B&=&\displaystyle\epsilon^{ij}\partial_iA_{j}
\ea
\lb{EB}
\ee

The commutation relations follow directly from the Poisson bracket
and read
\be
\ba{rcl}
\left[E^i(\vb{z}),E^j(\vb{z'})\right]&=&-i{k\over{4\pi}}\epsilon^{ij}\delta^{(2)}(\vb{z}-\vb{z'})\vspace{.2 cm}\\
\left[E^i(\vb{z}),B(\vb{z'})\right]&=&-i\epsilon^{ij}\partial_j\delta^{(2)}(\vb{z}-\vb{z'})
\ea
\lb{com_EB}
\ee
If there is an external charge $\rho_0$ (coupled to $A_0$ in the
action), the Gauss law is imposed by integrating the field component
$A_0$ in the path integral and reads simply
\be
\partial_i E^i+{k\over{4\pi}}B=\rho_0
\lb{path.gauss}
\ee
In the quantum theory this equation needs to be satisfied by the physical
states. So, following~\cite{TM_07}, the generator of time independent
gauge transformations $U$ can easily be defined as
\be
U=\exp\left\{i \int_\Sigma\Lambda(\vb{z})\left(\partial_i E^i+{k\over{4\pi}} B-\rho_0\right)\right\}
\lb{U}
\ee
where $\Sigma$ stands for a generic fixed time slice of $M$.
Since the gauge fields are compact, $\Lambda$ is identified with
(mapped to) an angle in the complex plane such that
\be
\ba{rcl}
\ln(z)&=&\ln|z|+i(\Lambda(z)+2\pi n)\vspace{.2 cm}\\
\partial_i\Lambda(z)&=&-\epsilon_{ij}\partial_j\ln|z|
\ea
\ee
where the second equation follows from the Cauchy-Riemann
equations. This last condition on $\Lambda$ will restrict the physical
Hilbert space of the theory~\cite{AK_1,AK_2,AK_3,AK_4}. Let us
define a new local operator
\be
V(\vb{z_0})=\exp\left\{-i\int_\Sigma d^2z\left[\left(E^i+\frac{k}{4\pi}\epsilon^{ij}A_j\right)\epsilon^{ik}\partial_k\ln|\vb{z_0}-\vb{z}|-\Lambda(\vb{z_0}-\vb{z})\rho_0\right]\right\}
\lb{V}
\ee
The physical states of the theory must be gauge invariant (under $U$)
as well as eigenstates of this new local operator.
Using the identity $\partial_k\partial_k\ln|z|=2\pi\delta(z)$ and the
commutation relations~(\r{com_EB}) for $E$ and $B$ is obtained the relation 
\be
\left[B(\vb{z}),V^n(\vb{z_0})\right]=2\pi nV^n(\vb{z_0})\delta^{(2)}(\vb{z}-\vb{z'})
\ee
This means that the operator $V$ creates a pointlike magnetic vortex
at $\vb{z_0}$ with magnetic flux
\be
\int_\Sigma B=2\pi n\ \ \ \ \ n\in\mathbb{Z}
\lb{flux}
\ee

Note that this operator only exists in the gauge
theory where the boundary conditions on the fields are such that allow
large gauge conditions. If one wants to enlarge the gauge group to include a
non-compact gauge field sector (as will be discussed below) these quantum
operators and the induced quantum transitions induced by them will not
be present in that sector. Also, as stated before the $F^2$ term is
fundamental for the existence of these tunneling processes that hold
local charge non conservation, see~\cite{KL_1,AK_1} for further
details.  Instantons in three dimensions are the monopoles in four
dimensions. So in the rest of the manuscript they will be called
either instanton and monopole without any distinction.

Using the functional Schr\"{o}dinger representation
$\Pi^i=i\delta/\delta A_i$ and imposing the condition that phase
acquired by a physical state under a gauge transformation be single
valued the charge spectrum is obtained
\be
q=m+\frac{k}{8\pi}\int_\Sigma B=m+\frac{k}{4}\,n \ \ \ \ \ m,n\in\mathbb{Z}  
\lb{q}
\ee
As it can be seen from the above formula for a generic, non-integer,
value of $k$ the allowed charges are non-integers. In principle in the
Abelian theory one expects that the charges are quantized as
happens in the Maxwell theory. But the existence of the Chern-Simons
term changes this picture, the charges are quantized through $m$ and
$n$ dependence but here $k$ itself is considered not to be
quantized (this point of view differs from many other works in the
subject). As will be explained below this fact is only compatible with
(and demands) the existence of a new gauge sector which fields do not
allow large gauge transformations.

A charge $q$ propagating in the bulk can interact with one monopole
with flux~(\r{flux}) changing by an amount
\be
\Delta q =\frac{k}{2}\, n
\lb{dq}
\ee

The path of a charge in the bulk can be thought as a Wilson line.    
The phase induced by the linking of two Wilson lines carrying charges
$q_1$ and $q_2$ is
\be
<W_{q_0}W_{q_1}>=\exp\left\{2\pi i \frac{2}{k}\, q_0\, q_1\, l\right\}
\lb{df}
\ee
where $l$ is the linking number between the lines. The above computation
was done in the limit of vanishing Maxwell term and with the assumption that
there is no self-linking for the individual Wilson lines.
The connection between the boundary CFT's and the bulk theory can be
achieved by noting that the bulk gauge fields become pure gauge in the
boundary and the Wilson lines on the boundary are none other than the 
vertex operators on $\Sigma_0$ and $\Sigma_1$ with momenta $q$
\be
\ba{c}
V_{0,q_0}=\left.\exp\left\{-iq\int A_\nu dx^\nu\right\}\right|_{\Sigma_0}=\exp\left\{-iq_0\lambda\right\}\vspace{.2 cm}\\
V_{1,q_1}=\left.\exp\left\{-iq\int A_\nu dx^\nu\right\}\right|_{\Sigma_1}=\exp\left\{-iq_1\chi\right\}
\ea
\lb{vertex}
\ee
Note that generally, as will be considered in the next chapter, the
charge $q$ figuring in the Wilson line is not constant and may change 
over time due to tunneling effects, i.e. monopole processes.

The conformal dimensions of these vertices are
\be
\ba{c}
\Delta_0=\frac{q_0^2}{k}\vspace{.2 cm}\\
\Delta_1=\frac{q_1^2}{k}
\ea
\lb{confd}
\ee
$k$ appears because the induced action on the boundary is that of a
chiral string action multiplied by $k$ such that it plays the role of
the (inverse) Regge slope and has dimensions of (target)
space-time. To see the Wilson line and vertex operator correspondence
let us obtain the above result from the bulk theory.  Consider two
Wilson lines carrying charges $q$ and $-q$ propagating from one
boundary to the other corresponding to two vertex insertions on the
boundary with momenta $q$ and $-q$, see figure (\r{fig0}).

\fig{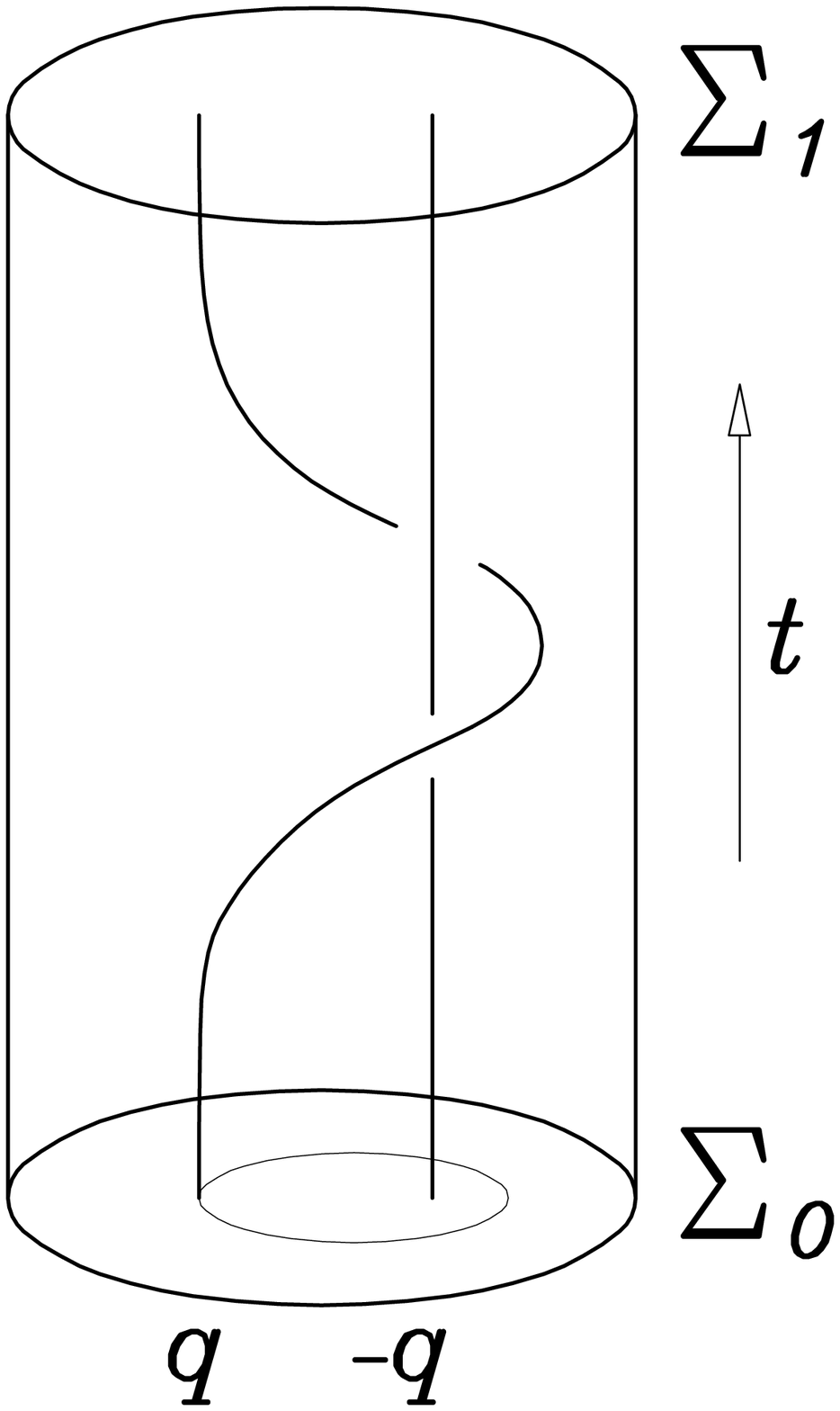}{Two charges propagating through the bulk are
represented as two Wilson lines, By a rotation of one charge around
the other one, a linking ($l=1$) is induced in the bulk.}{fig0}

Two point correlation functions of these two vertices follows as usual
\be
\ba{c}
\displaystyle<V_{0,q}(\vb{z_1})V_{0,-q}(\vb{z_2}))>=\frac{1}{{\vb{z_{12}}^{2\Delta}}}\vspace{.1 cm}\vspace{.2 cm}\\
\displaystyle <V_{1,q}(\vb{\bz_1})V_{1,-q}(\vb{\bz_2}))>=\frac{1}{{\vb{\bz_{12}}^{2\bar{\Delta}}}}\\
\ea
\lb{OPE2}
\ee
A rotation of one charge (vertex) around the other in one boundary
induces a phase of $-4\pi i \Delta$ in (\ref{OPE2}).  In the bulk this
rotation induces one linking of the Wilson lines ($l\rightarrow l+1$),
so (\r{df}) gives an Aharonov-Bohm phase change of $-4\pi i q^2/k$.
By identifying these two phases it is concluded that (\r{confd}) follows.
Let us recall that in terms of the three dimensional theory
Polyakov~\cite{P_1} pointed out that $\Delta$ is the transmuted spin
of a charged particle which exists because of the interaction with
the Chern-Simons term. So the conformal dimension of the boundary fields
is identified with the transmuted spin of the bulk charges.  For later
use let this result be generalized to three charges $q_1$, $q_2$ and
$q_3=-q_1-q_2$. The net charge will always be considered to be zero.
The correlation function of three vertices reads
\be
<V_{q_1}(\vb{z_1})V_{q_2}(\vb{z_2})V_{q_3}(\vb{z_3}))>=\frac{1}{{\vb{z_{12}}^{\Delta_1+\Delta_2-\Delta_3}{\vb{z_{13}}^{\Delta_1+\Delta_3-\Delta_2}}}{\vb{z_{23}}^{\Delta_2+\Delta_3-\Delta_1}}}
\lb{OPE3}
\ee
where the charge-conformal dimension identifications are  
\be
\ba{rcl}
\Delta_1+\Delta_2-\Delta_3&=&-\frac{2}{k}q_1q_2\vspace{.2 cm}\\
\Delta_1+\Delta_3-\Delta_2&=&\frac{2}{k}q_1(q_1+q_2)\vspace{.2 cm}\\
\Delta_2+\Delta_3-\Delta_1&=&\frac{2}{k}q_2(q_1+q_2)
\ea
\lb{confd3}
\ee

The antiholomorphic three point correlation function follows trivially.

It is clear that if any one of the vertices is rotated around one of
the other two; the phase factor induced in the three-point function
will be the same as~(\r{df}) which comes from the linking of Wilson
lines in the bulk.  From the point of view of CFT the sum of conformal
factors must be integer in order to have single valued OPE's. This
result is going to be derived from the bulk theory in the next section
independently of the boundary. By a simple argument let us show that
there can exist only integer conformal dimensions for each vertex.  At
the level of the two point correlation functions there can exist two
integer $\Delta$'s or two half integer $\Delta$'s.  For three point
functions it is necessary to have three integer $\Delta$'s or two
integer and one half integer $\Delta$'s. One can then consider the
four point function and decompose it to two pairs which are well
separated from each other. In this case the conformal dimensions can
be integers and half-integers. Suppose that at least two of the
vertices have half-integer conformal dimensions.  Then one vertex from
a pair can be adiabatically moved to the other pair and a three point
function with a total half-integer conformal dimension will be
formed. But this is not permitted as argued above.  For this reason
the existence of vertices with half integer conformal dimension is
excluded.  The same arguments follow for the charges from the point of
view of the $3D$~bulk theory.

Let us remember that there are two independent chiral conformal field
theories on two different surfaces up to this point. In order to get a
non-chiral CFT (as in string theory) is necessary to identify in some way
the two boundaries. The most obvious way to define non-chiral vertex
operators is
\be
V_{q_0,q_1}(\vb{z},\vb{\bz})=V_{0,q_0}(\vb{z})V_{1,q_1}(\vb{\bz})
\ee
with the conformal dimension
\be
\Delta_0+\Delta_1=\frac{1}{k}(q_0^2+q_1^2)
\lb{confdLR}
\ee

As long as $V_0$ is considered to be holomorphic and $V_1$
antiholomorphic or vice-versa, the full vertex is indeed  non-chiral.

Note that there is a little bit of difference in the nomenclature in
relation to the standard one. In most of the literature
$\Delta_0=\Delta$ and $\Delta_1=\bar{\Delta}$ are called conformal
weights, $\Delta-\bar{\Delta}$ is called the spin and the
sum~(\r{confdLR}) $\Delta+\bar{\Delta}$ is called the conformal
dimension. Here $\Delta$ and $\bar{\Delta}$ are called conformal
dimensions as well when referring to the chiral vertices operators
since they can exist in each of the boundaries \textit{independently}
of each other.  From now on the notation will be definitively changed
to $\Delta=\Delta_0$ and $\bar{\Delta}=\Delta_1$ and the same for the
charges, $q_0=q$ and $q_1=\bar{q}$.

\section{Canonical Formalism\lb{ch.quant:sec.can}}

In this section a review will be given on the canonical quantization
of Maxwell-CS theories. Again a $U(1)$ Topologically Massive
Gauge Theory defined on a $3D$ manifold $M=\Sigma\times[0,1]$ with two
boundaries $\Sigma_0$ and $\Sigma_1$ will be considered
\be
S_{TMGT,1}=\int_Mdtd^2z\left[-\frac{\sqrt{-g}}{\gamma}F_{\mu\nu}F^{\mu\nu}+\frac{k}{8\pi}\epsilon^{\mu\nu\lambda}A_\mu \partial_\nu A_\lambda\right]
\ee
where $A$ is a gauge connection field obeying $\mathcal{B}$ boundary
conditions and $F$ its curvature, $M=\Sigma\times[0,1]$ has
two boundaries $\Sigma_0$ and $\Sigma_1$. $\Sigma$ is taken to be a
compact manifold, $t$ is in the interval $[0,1]$ and $(z,\bz)$ stand
for complex coordinates on $\Sigma$ as stated before.

As has been widely studied, this theory induces new degrees of freedom on the
boundaries, which are fields belonging to $2D$ chiral CFTs
living on $\Sigma_1$ and $\Sigma_2$.

The previous action can be rewritten in a time-space (i.e. $A_0$ and
$A_i$) splitting as
\be
S=\displaystyle\int_M\left[-\frac{\sqrt{-g}}{2\gamma}F_{0i}F^{0i}-\frac{\sqrt{-g}}{4\gamma}F_{ij}F^{ij}+\frac{k}{4\pi}\epsilon^{ij}A_0F_{ij}+\frac{k}{2\pi}\epsilon^{ij}A_iF_{j0}\right]
\ee

The canonical momenta conjugate to $A_i$ are given by~(\r{mom_i})
\be
\pi^i=-\frac{\sqrt{-g}}{\gamma}F^{0i}+\frac{k}{8\pi}\epsilon^{ij}A_j
\ee

The canonical momentum conjugate to $A_0$ is identically zero and
imposes the Gauss Law
\be
0=\int_\Sigma
d^2x\left(-\frac{\sqrt{-g}}{\gamma}\partial_iF^{0i}+\frac{k}{4\pi}\epsilon^{ij}F_{ij}\right)-\oint_{\partial\Sigma}\left(-\frac{\sqrt{-g}}{\gamma}F^{0i}+\frac{k}{4\pi}\epsilon^{ij}A_{j}\right)n_i
\lb{can.gauss}
\ee
where $n$ is a normal vector to the boundary of $\Sigma$. Note that the
boundary term is only present when the $2D$ boundary $\Sigma$ of the
$3D$ manifold $M$ has a boundary. In principle this doesn't happen
since the boundary of a boundary is null, $\partial\partial M=\emptyset$. 
But once one orbifolds the theory a new boundary may emerge as extensively
studied in Chapter~\r{ch.opun}.

The Hamiltonian of the theory is computed to be
\be
\ba{rcl}
H&=&\Pi^i\partial_0A_i-L=\vspace{.1 cm}\\&=&\displaystyle-A_0\left[\partial_i\left(\pi^i-\frac{k}{4\pi}\epsilon^{ij}A_j\right)+\frac{k}{8\pi}\epsilon^{ij}F_{ij}\right]+\partial_i(A_0\pi^i)\vspace{.1 cm}\\
&+&\displaystyle\frac{\sqrt{-g}}{16\gamma}(\epsilon^{ij}F_{ij})^2+\frac{\gamma}{2}h_{ij}\left(\pi^i-\frac{k}{8\pi}\epsilon^{ik}A_k\right)\left(\pi^j-\frac{k}{8\pi}\epsilon^{jl}A_l\right)
\ea
\ee
As usual $A_0$ can be considered to be a Lagrange multiplier which
imposes the Gauss law.

The electric and magnetic fields are defined as in~(\r{EB})
\be
\ba{rcl}
E^i&=&\displaystyle\frac{1}{\gamma}F^{0i}\vspace{.1 cm}\\
B  &=&\partial_z A_\bz-\partial_\bz A_z
\ea
\ee
and the Gauss law (without boundary terms) reads simply
\be
\partial_iE^i+\frac{k}{4\pi}B=\rho_0
\ee
as already given in~(\r{path.gauss})

Upon quantization the charge spectrum is
\be
Q=m+\frac{k}{4}n
\lb{charge}
\ee
for some integers $m$ and $n$.  Furthermore, it has been
proved~\cite{TM_07,TM_10} that under the correct relative boundary
conditions, one insertion of $Q$ on one boundary (corresponding to a
vertex operator insertion on the boundary CFT) will necessarily demand
an insertion of the charge
\be
\bar{Q}=m-\frac{k}{4}n
\lb{ccharge}
\ee
on the other boundary. This statement is going to be derived in
chapter~\r{ch.tor} and will be assumed through out the rest of this section.
Some hints on how to rederive these results from a canonical perspective
are given in this section which is based on work in progress~\cite{II}.

\subsection{Wave Function as Boundary Conditions}

The functional approach of~\cite{AFC_1} is followed next in order to
derive the wave functions of the theory.

The functional Gauss law constraint takes the form
\be
\left[D_\bz\left(-i\frac{\delta}{\delta A_\bz}+\frac{k}{8\pi}\epsilon^{z\bz}A_z\right)+D_z\left(-i\frac{\delta}{\delta A_z}-\frac{k}{8\pi}\epsilon^{z\bz}A_\bz\right)+\frac{k}{4\pi}\epsilon^{z\bz}F_{z\bz}\right]\Psi[A_z,a_\bz]=0
\ee

In TMGT the wave functions are not gauge invariant, as they transform
with a one-cocycle. The way to make them gauge invariant is to
integrate the cocycle condition such that they decompose into three
factors
\be
\Psi[A_z,A_\bz]=\left\{-\frac{ik}{4\pi}\int d^2z\sqrt{h}\tilde{\epsilon}^{z\bz}A_zA_\bz\right\}\psi[A_z]\Phi[B]
\lb{Psi}
\ee
where $B=\partial_z A_\bz-\partial_\bz A_z$ is the magnetic field and
$ds^2_{(3)}=g_{ij}dx^idx^j=-dt^2+h_{z\bz}dzd\bz$.  From now on the
metric is fixed to be of this form with $h_{z\bz}=1$ (the TMG sector
is not considered).  Note that $\sqrt{-g}=-i$ and the $2D$
antisymmetric tensor $\tilde{\epsilon}$ is purely imaginary and
induced from the bulk by
$\tilde{\epsilon}^{ij}=\frac{\epsilon^{0ij}}{\sqrt{-g}}$ such that
$\epsilon^{z\bz}=i$.

The factor $\Phi[B]$ is the solution for the pure Maxwell theory such
that the Gauss law constraint is obeyed
\be
\left[D_\bz\frac{\delta}{\delta A_\bz}+D_z\frac{\delta}{\delta A_z}\right]\Phi=0
\ee
For further details in
the treatment of this factor see~\cite{AFC_1} and references therein.
If the fields have nontrivial magnetic charge $\int B \neq 0$
the wave function is $0$. Note that this result is the statement
of overall charge conservation on the closed surface one is
considering. Of course it is possible that non-zero magnetic field
distributions exist locally.

$\psi[A_z]$ is a \textit{topological} term (due to the Chern-Simons term
in the Lagrangian) and obeys the Schr\"odinger equation
\be
\left[D_z\frac{\delta}{\delta A_z}-\frac{i k}{4\pi}\tilde{\epsilon}^{z\bz}\partial_\bz A_z\right]\psi[A_z]=0
\ee
Note that this equation corresponds in the topological limit (the
ground state), with $\Phi=1$, to the equation in $\Psi$
\be
\left[D_\bz\frac{\delta}{\delta A_\bz}+\frac{k}{8 \pi}D_\bz A_z-\frac{k}{4\pi}F_{\bz z}\right]\left.\Psi[A]\right|_{CS}=0
\label{WW}
\ee
where $2D$ Euclidean space-time is considered with
$\tilde{\epsilon}^{z\bz}=i$.

As presented in~\cite{W_1} one solution of~(\r{WW}) for some fixed time $t$,
compatible with WZWN construction for an Abelian gauge group, is
\be
\Psi[A]=\int{\mathcal{D}}\tilde{\phi}\ \exp\left\{\frac{k}{8\pi}\int_{\Sigma(t)} \left[A_\bz A_z - 2 A_\bz \partial_z\tilde{\phi}+\partial_\bz\tilde{\phi}\partial_z\tilde{\phi}\right]\right\}
\label{WWS1}
\ee

Here it is enough to consider the lowest Landau level (the ground state)
of the theory since it is the most probable state. This means the
topological limit is being considered, in which case $\Phi=1$ and
the solution for the pure Chern-Simons case is retrieved. This solution
corresponds to configurations with weak magnetic field,
$\epsilon^{z\bz}F_{z\bz}\simeq 0$.
Nevertheless it is important to stress that here $A_z$ and $A_\bz$ are not
canonically conjugate variables, unlike the pure CS case.

Next it will be shown that these wave functions are the
\textit{building blocks} of the boundary theories; by inserting such
states on the boundaries they \textit{act} as boundary
conditions and are effectively selecting our boundary world. In this
way, through $\tilde{\phi}$, new degrees of freedom are inserted
on the boundaries. Furthermore, these wave functions are necessary for the
consistency of the full theory (in a bounded manifold) as a
well-defined gauge theory. Basically they play the same role as the
actions inserted on the boundary of the previous section.

Consider, for the time being, two states of the form~(\r{WWS1})
inserted at the boundaries $\Sigma(t=0)$ and $\Sigma(t=1)$. The
partition function is then the correlator
\be
Z=\left<\Psi_0|\Psi_1\right>=\int{\mathcal{D}}A_z{\mathcal{D}}A_\bz e^{iS}\bar{\Psi}_0\Psi_1
\lb{Z}
\ee
where $\Psi_1$ is given by~(\r{WWS1}) with $\Sigma(t)=\Sigma_1$ while
$\bar{\Psi}_0$ is
\be
\bar{\Psi}_0[A]=\int{\mathcal{D}}\tilde{\phi}\ \exp\left\{\frac{k}{8\pi}\int_{\Sigma_0} \left[-A_\bz A_z + 2 A_z \partial_z\tilde{\phi}-\partial_\bz\tilde{\phi}\partial_z\tilde{\phi}\right]\right\}
\label{WWS0}
\ee
The overall minus sign comes from the change $z\leftrightarrow\bz$
in the measure of the integral due to the change of relative
orientations from boundary to boundary.

The importance of the insertion of these boundary wave functions is
that they constrain the theory assuring
that the path integral~(\r{Z}) is both, gauge invariant and has a
classical extremum.

Performing a gauge transformation $A_i\to A_i+\partial_i\Lambda$, the
bulk exponential factor in the partition function (similarly to~(\r{gauge1}))
changes as
\be
S\to S- \left.\frac{k}{8\pi}\int_\Sigma d^2z \left[A_z\partial_\bz \Lambda-A_\bz\partial_z \Lambda\right]\right|_{\Sigma_0}^{\Sigma_1}
\ee
where, as in~(\r{WW}), a Euclidean $2D$ structure was taken.
Here is where the non gauge invariance due to boundaries resides. Taking the
transformation of the one on $\Sigma_1$ gives
\be
\ba{l}
\displaystyle\frac{k}{8\pi}\int_{\Sigma_1} \left[A_\bz A_z - 2 A_\bz \partial_z\tilde{\phi}+\partial_\bz\tilde{\phi}\partial_z\tilde{\phi}\right]\to\vspace{.2cm}\\
\displaystyle\frac{k}{8\pi}\int_{\Sigma_1} \left[A_\bz A_z + A_\bz \partial_z\Lambda +\partial_\bz\Lambda A_z +\partial_z\Lambda\partial_\bz\Lambda\right.\vspace{.1 cm}
\left.- 2 A_\bz \partial_z\tilde{\phi}- 2 \partial_\bz\Lambda \partial_z\tilde{\phi}
+\partial_\bz\tilde{\phi}\partial_z\tilde{\phi}\right]
\ea
\lb{Sgauge2}
\ee
Combining all the factors, the exponential factor corresponding to
$\Sigma_1$ is simply
\be
\frac{k}{8\pi}\int_{\Sigma_1} \left[A_\bz A_z - 2 A_\bz \partial_z(\tilde{\phi}-\Lambda)+\partial_z(\tilde{\phi}-\Lambda)\partial_\bz(\tilde{\phi}-\Lambda)\right]
\lb{Sgauge3}
\ee
The gauge parameter $\Lambda$ is now easily eliminated by redefining
the field corresponding to the degree of freedom at the boundary,
$\tilde{\phi}\to\tilde{\phi}+\Lambda$.
This redefinition does not change the measure $D\tilde{\phi}$.
In this elegant way, by inserting \textit{ad hoc} new
degrees of freedom one manages to ensure gauge invariance of the full
theory. Another way to explain things is that gauge transformations
will necessarily induce new degrees of freedom on the boundaries as
argued in the last section. Both ways of arguing are equivalent.

Further to this discussion, the boundary wave functions ensure that
the theory has a classical extremum. An infinitesimal variation of the
fields induce exponential boundary terms of the form
\be
i\delta S_{\partial M}=\frac{ik}{8\pi}\int_{\Sigma_1}\epsilon^{ij}A_i\delta A_j-\frac{ik}{8\pi}\int_{\Sigma_0}\epsilon^{ij}A_i\delta A_j
\ee

After a change of measure to $\partial_z\tilde{\phi}$ and
$\partial_\bz\tilde{\phi}$ the integrals on the boundary can be
performed resulting in terms of the form
\be
\int D(\partial_\bz\tilde{\phi}(1)\partial_z\tilde{\phi}(0))\delta_{\Sigma_1}(2A_\bz-\partial_\bz\tilde{\phi})\delta_{\Sigma_0}(2A_z-\partial_z\tilde{\phi})\exp\left\{\frac{k}{8\pi}\int_{\Sigma_1}A_\bz A_z-\frac{k}{8\pi}\int_{\Sigma_0}A_\bz A_z\right\}
\ee
In this way $A_\bz$ is fixed on one boundary while $A_z$ is fixed on
the other boundary such that $\left.\delta A_\bz\right|_{\Sigma_1}=0$
and $\left.\delta A_z\right|_{\Sigma_0}=0$.
The net boundary variation containing $\delta A_i$ vanishes
since the term which comes from the bulk cancels the one coming from
boundary exponential term in the wave functions. In this way the theory
has a well defined classical limit. Note as well that the
fields $\tilde{\phi}$ living on $\Sigma_0$ correspond to holomorphic
degrees of freedom while the ones living on $\Sigma_1$ are
antiholomorphic. It is with this construction that the two chiral CFT's
on the boundary are obtained.

\subsection{Conformal Blocks and the CFT Partition Function}

Turning to specific geometries, consider a Hodge decomposition
of the gauge fields $A_i$ at each time slice $A=a+d\phi_R+*d\phi_I$
such that
\be
\ba{rcl}
A_z&=&a_z + \partial_z\bar{\phi}\vspace{.1 cm}\\
A_\bz&=&a_\bz + \partial_\bz\phi
\ea
\lb{HD}
\ee
where $\phi=\phi_R+i\phi_I$ is a complex scalar field and $a$ is a
harmonic form, $da=*da=0$ such that $\partial_za_\bz\pm\partial_\bz a_z=0$.

For a torus with modular parameter $\tau=\tau_1+i\tau_2$
the parameterization for the harmonic form is
\be
\ba{lcr}
a_z&=&\pi ia\tau_2^{-1}\bar{\omega}(z)\vspace{.1 cm}\\
a_\bz&=&-\pi i\bar{a}\tau_2^{-1}\omega(z)
\ea
\ee
$\tau_2$ is the imaginary part of $\tau$.  $\omega$ is an holomorphic
one-form such that $\int_\alpha\omega=1$, $\int_\beta\omega=\tau$ and
$\int d^2z\omega\bar{\omega}=-2i\tau_2$. $\alpha$ and $\beta$ are
closed non contractible contours in the torus which generate its first
homology.  Considering this parameterization and the rational $k=2p/q$
(with even $p$), one can build an orthonormal basis with $pq$ elements
for the wave functions~\cite{BN_1}
\be
\Psi_{0,\lambda}=C\exp\left\{\frac{k\pi}{4}a\tau_2^{-1} (a-\bar{a})\right\}\Theta\left[\ba{c}\frac{\lambda}{q}\\0\ea\right]\left(\sqrt{2}a|\frac{2\tau}{k}\right)
\lb{WT2}
\ee
with $\lambda=0,1,\ldots,pq-1$ and $\Theta$ are modified Jacobi
theta-functions
\be
\Theta\left[\ba{c}\frac{\lambda}{q}\\0\ea\right](\sqrt{2}a|\frac{2\tau}{k})=
\sum_{s\in{\mathbb{Z}}}\exp\left\{2\pi i
\tau\frac{1}{k}\left(sp+\frac{\lambda}{q}\right)^2+2\sqrt{2}\pi i a\left(sp+\frac{\lambda}{q}\right)\right\}
\lb{theta}
\ee
where the sum is considered to run only over multiples of $p$ (it is
in this sense it was called \textit{modified}).  Note the different
normalization compared to reference~\cite{BN_1}. There Bos and Nair
considered the Chern-Simons coefficient to be $\tilde{k}=2p'q$
($p=2p'$). Taking the normalization used here the $U(1)$ charges
carried by Wilson lines have to belong to the spectrum of the theory
as given by~(\r{charge}). In this way large gauge transformations,
$a\to a+s+\tau r$, are restricted to the ones which have $s,r=0\mod
p$. Note that $\lambda/q$ are the primary charges and that for
rational values of $k$ there is one charge independent monopole
process corresponding to $\Delta Q=p$. In terms of the CFT these shifts of
charge are inside the same family as will be shown in
section~\r{ch.tor:sec.cb}. Then the above restriction on large gauge
transformations is nothing else then the allowed monopole processes
which shifts the charges inside the same family or block. The
remaining numerical factors in the $\Theta$ functions come from the
factor $k/8\pi$ instead of $\tilde{k}/4\pi$.

As has been widely studied~\cite{BN_1,LR_1} there is a one-to-one
correspondence between the $3D$ QFT wave functions and the blocks of
the $2D$ CFT. In this work the wave functions are \textit{interpreted} 
as being labeled by charges in such a way that, again, there is a
one-to-one correspondence between the primary charges of the theory
(or each family of charges) and the wave functions (and necessarily
the conformal blocks).

Consider the pair $(m,n)$ (obeying the Bezout lemma) such that for a
given primary charge $\lambda/q=m+kn/4$ on one boundary, as will be
derived in detail in section~\r{ch.tor:sec.bulk}, there will be a
corresponding charge $\bar{\lambda}/q=m-kn/4$ on the other boundary.
Then the wave function on $\Sigma_1$ is
\be
\bar{\Psi}_{1,\bar{\lambda}}=C\exp\left\{\frac{k\pi}{4}\bar{a}
\tau_2^{-1}(\bar{a}-a)\right\}\Theta\left[\ba{c}\frac{\bar{\lambda}}{q}\\0\ea\right]\left(\sqrt{2}\bar{a}|\frac{2\bar{\tau}}{k}\right)
\lb{WT3}
\ee

This raises a problem since so far the literature has not considered
how to introduce the monopole processes in the path integral using
this formalism. These monopole processes can be thought of as
insertions of a local \textit{gauge} operator $V$
(see~\cite{AK_3,AK_4}). The main point in this discussion is that the
physical wave functions are the ones which are invariant under the
action of any possible combination of operators $V$\footnote{The
author acknowledges Alex Kovner for this useful remark.}. Moreover
the total amplitude of combinations of wave function (on the
boundaries) which are not invariant under those operators average to
zero as will be shown.

In this way the effective partition function (path integral) has to
take into account this phenomenon and be of the form
\be
Z=\left<\Psi_0\left|\otimes V\right|\Psi_1\right>
\ee

This issue is not going to be developed in detail here, the proper and 
detailed treatment using a different formalism will be postponed until
Chapter~\r{ch.tor}.  It is enough to consider the shift of the charge
in $\Sigma_1$ by the amount $\Delta Q=-kn/2$ or $\Delta Q=0$ depending
on the boundary conditions.  The effective wave function
on that boundary will then be of the form
\be
\Psi_{\bar{\lambda}=m-kn/4}=\exp\left\{i\frac{k}{2}n\oint_\beta A.dx\right\}\Psi_{\lambda=m+kn/4}
\lb{Psi-W}
\ee
or simply the same present in $\psi_0$ for $\Delta Q=0$.  The
non-perturbative processes, Wilson line braiding and monopole induced
processes, will take account of the charge difference between the two
boundaries.

In this way the free boundary partition function will be of the form
\be
Z_{\mathrm{free}}=\sum\left<\psi_{m+nk/4}\psi_{m+nk/4}\right>+\sum\left<\psi_{m+kn/4}\psi_{m-kn/4}\right>
\ee
This is a modular invariant of the boundary CFT. Depending on the
boundary conditions imposed on the membrane one can set it to be
the first or the second sum only. Of course for free boundary
conditions both are present.

Identifying the fields on both boundaries and writing
$a=\rho+\tau\gamma$, with $\rho, \gamma \in [0,1]$ 
the partition function can be computed explicitly.
Combining the first exponential factors in the wave
functions the factor $-k\pi\gamma^2/{\rm Im}\tau$ is obtained.
The $\rho$ integration imposes the constraint
through a Dirac delta-function for each $(s,s')$ pair
\be
\delta\left(p(s-s')\right)
\lb{delta2}
\ee
and the $\gamma$
integration can be performed recombining the remaining factors
into a Gaussian integral under a shift
$\gamma\to\gamma-2\sqrt{2}(sp+\lambda/q)$.
Finally, considering the sum over the $pq$ wave functions
the partition function is obtained
\be
\ba{rl}
Z=&\displaystyle\frac{1}{|\eta(\tau)|^2}\sum_{s,\lambda=0}^{\lambda=pq-1}
\exp\left\{2\pi i \tau\frac{1}{k}\left(ps+\frac{\lambda}{q}\right)^2\right\}
\exp\left\{-2\pi
i\bar{\tau}\frac{1}{k}\left(ps+\frac{\bar{\lambda}}{q}\right)^2\right\}\vspace{.1cm}\\
=&\displaystyle\sum_{\lambda=0}^{pq-1}|\chi_\lambda|^2
\ea
\ee
$\chi_\lambda=\sum_s\exp 2\pi i\tau(ps+\lambda/q)$ are the
characters of the conformal algebra.
To ensure that the wave functions are normalized to $1$
the constant in~(\r{WT2}) is set to be
\be
C=\frac{1}{|\eta(\tau)|}\left(k{\rm Im}\tau\right)^\frac{1}{4}
\ee

So the partition function of the $2D$ boundary CFT is obtained
as a sum of several possible transition amplitudes from boundary to
boundary.

\chapter{Toroidal Compactification Spectrum and the Heterotic String}
\lb{ch.tor}

This Chapter is based on the original published work by the author,
Ian Kogan and Bayram Tekin~\cite{TM_15}.

In~\cite{TM_07} the charge non conservation induced by monopoles and
linking was discussed and those ideas (in the framework of TM) were
applied to T-duality and Mirror Symmetry in~\cite{TM_10}. Nevertheless
the problem of building the correct left-right spectrum has never been
properly solved. The main goal of this chapter is to determine the
charge lattice structure allowed by TMGT. Essentially it is shown
that, upon imposing suitable boundary conditions, the theory demands a
charge lattice which is exactly of the form of the string theory momentum
lattice. From the non-perturbative dynamics of three dimensional gauge
theory the Narain lattice spectrum is derived.

In section~\r{ch.tor:sec.string} a short review of the CFT and string
theory aspects necessary to the development of the ideas
presented here is given.

Next, in section~\r{ch.tor:sec.bulk}, a
model which describes the dynamics of charges propagating in the $3D$
bulk theory is built. Some well know results in string theory are
derived purely from the dynamics of the bulk $3D$ theory. Namely the
mass spectrum of toroidally compactified closed string theory emerges.

In section~\r{ch.tor:sec.bc} the relevant issue of gluing both
$2D$ boundaries of the $3D$ manifold in order to get a single
non chiral conformal field theory on the boundary is studied.

In section~\r{ch.tor:sec.cb} the underlying conformal
block structure of the $c=1$ compactified bosonic RCFT and the
corresponding fusion rules are found as a result of monopole-instanton
induced interactions in the bulk.

Finally in section~\r{ch.tor:sec.lat} the spectrum of the heterotic string and
possible backgrounds are rederived in the light of these new
results.

\section{\lb{ch.tor:sec.string}Introduction}

\subsection{Several $U(1)$'s and Mass shell condition}

For generic $k$ and $q$'s, $\Delta$ in~(\r{confd}) and the sums
in~(\r{confd3}) are not integers. This is not a problem, actually it
is very welcome since it is simply the statement that it is necessary
to add something else in the theory, a new gauge group sector which
fields don't allow large gauge transformations, hence obeying
$\tilde{\mathcal{B}}$ boundary conditions, or in terms of string
theory, non-compactified dimensions. So in addition to the
action~(\ref{STMGT1}) describing TMGT with one single $U(1)$
as gauge group with the fields obeying $\mathcal{B}$ boundary
conditions, consider the following $U(1)^D$ action describing TMGT
with gauge fields obeying $\tilde{\mathcal{B}}$ boundary conditions
\be
S_{D}=\int_M d^2z\,dt\left[-\frac{1}{4\gamma'}f^{\mu\nu}_Mf_{\mu\nu}^M+\frac{k'\delta_{MN}}{8\pi}\epsilon^{\mu\nu\lambda}a^M_\mu\partial_\nu a^N_\lambda\right]
\lb{SD}
\ee
where $N$ and $M$ run from $1$ to $D$. For the purposes of this work
it is enough to consider the couplings given by
$K'_{MN}=k'\delta_{MN}$.  The actions~(\ref{STMGT1}) and~(\ref{SD})
will be considered together such that the full gauge group is
$U(1)\times U(1)^D$ with the fields corresponding to the first $U(1)$
factor obeying $\mathcal{B}$ boundary conditions and the remaining $D$
fields obeying $\tilde{\mathcal{B}}$ boundary conditions. In general
the $\mathcal{B}$ sector could contain a product of several $U(1)$'s with
action given by~(\r{STMGTN}). For the discussion that follows it is
enough to consider a single field.

A generic (i.e. not necessarily tachyon) non-chiral vertex of the
boundary CFT is of the form
\be
\partial_z^s\zeta\partial_\bz^{s'}\zeta\left(\prod_i\partial_z^{r_i}\Omega^M\right)\left(\prod_i\partial_\bz^{r'_i}\Omega^M\right)\exp\left\{-i\left[(q+\bar{q})\zeta+p_M\Omega^M\right]\right\}
\lb{VD}
\ee
The fields $\zeta$ and $\Omega^M$ correspond to the gauge parameters
of $A$ and $a^M$ respectively. The levels of the vertex operators are
integers defined as
\be
\ba{rcl}
L      &=&s+\sum_i r_i\vspace{.2 cm}\\
\bar{L}&=&s'+\sum_i r'_i
\ea
\ee

The exponential part may be represented as the bulk Wilson
line propagating from boundary to boundary
\be
W_{D}=\exp\left\{-i\int\left[qA_\nu+p_Ma^M_\nu\right]dx^\nu\right\}
\lb{WD}
\ee
while the remaining factors have to be considered as products
of $A$ and $a$ fields in the bulk. They will not be discussed here.

The mass shell condition for the boundary vertex is simply
\be
p_M p^M=- \mbox{mass}^2
\lb{mass}
\ee
where the boundary CFT momenta corresponds to the bulk charges as
shown before.

The mass spectrum in CFT's is built out of the allowed values for the
conformal dimension of the fields or the operators. In particular the
vertex operators (\r{VD}) have conformal dimensions
\be
\ba{rcl}
\Delta&=&\displaystyle \frac{q^2}{k}+\frac{p^2}{k'}+L\vspace{.2 cm}\\
\bar{\Delta}&=&\displaystyle\frac{\bar{q}^2}{k}+\frac{p^2}{k'}+\bar{L}
\ea
\lb{weights}
\ee
Due to conformal invariance one has $\Delta=\bar{\Delta}=1$.  To get
the usual String theory normalization it is necessary to replace
$k'=4/\alpha'=k/R^2$ and take the sum of both equations
(\r{weights})
\be
\mbox{mass}^2=-p^2=\frac{(q^2+\bar{q}^2)k'}{2k}+\frac{k'}{2}(L+\bar{L}-2)=\frac{m^2}{R^2}+\frac{R^2n^2}{\alpha'^2}+\frac{2}{\alpha'}(L+\bar{L}-2)
\lb{mshell}
\ee
Subtracting one equation from the other in~(\r{weights}) gives
\be
\frac{q^2-\bar{q}^2}{k}+L-\bar{L}=nm+L-\bar{L}=0
\lb{spin}
\ee
Let us call~(\r{mshell}) the mass shell condition and~(\r{spin}) the
spin (or level-matching) condition. For further details
see~\cite{POL_1}.  In the normalization used, the explicit form of the
charges (string momenta) is
\be
\ba{c}
\displaystyle q=m+\frac{k}{4}n\vspace{.2 cm}\\
\displaystyle \bar{q}=m-\frac{k}{4}n
\ea
\ee
But remember that the allowed charges in the theory are of the
$q$-form. In principle $\bar{q}$ is not related to $q$ and it should
be of the form $\bar{q}=m'+n'k/4$. As it will be explained in detail
in this chapter the reader should have the following picture in
mind: A charge $q$ is inserted in one of the boundaries and it goes
through the bulk interacting with the gauge fields until it reaches
the other boundary. During its journey through the bulk its
interactions with (in)finitely many monopole-instantons induce a change
of its charge by $\Delta q=Nk/2$. The charge emerges in the other
boundary as $\bar{q}=q+Nk/2=m+(n+2N)k/4$. In the particular case that $N=-n$,
$\bar{q}=m-nk/4$. This is the aim if one wants to describe the bosonic 
string spectrum!  But why N must be equal to
$-n$? How do the monopoles know a priori that they have to interact by this
amount with the initial charge $q$?  In principle there could exist any
other process holding $N\neq-n$ that would generate some charge
$\bar{q}$ leading us to disaster. The pair $(q,\bar{q})$ would be obtained
with no correspondence with any physically sensible momentum pair
$(p,\bar{p})$ of string theory. So it is necessary to show that the full
$3D$-amplitudes $q\rightarrow\bar{q}$ in the theory corresponding to
unwanted processes $\bar{q}\neq m-nk/4$ are vanishing!  This crucial
property shows that seemingly independent chiral field theories on
different boundaries are actually related when the non-perturbative
excitations in the bulk are taken into account. This fact is closely
related to considering the Maxwell-CS theory instead of the
pure Chern-Simons and the presence in the bulk of monopole
processes.

\subsection{RCFT's and Fusion Rules}

A CFT is rational when its infinite set of primary fields (vertex
operators) can be organized into a finite number $N$ of families usually
called \textit{primary blocks}. In each of those blocks one minimal
field can be chosen such that it is the generator of that
family. There is an algebra between these families, or in other words,
between the $N$ minimal fields called the fusion algebra. The fusion
rules define this second algebra of fields.

In what follows we will discuss only the holomorphic part of the c=1
RCFT of a bosonic field $\phi$ living in a circle of radius
$R=\sqrt{2p'/q}$, where $p'$ and $q$ are integers. The study of the
antiholomorphic sector of the theory follows in pretty much the same
way.

The vertex operators are
\be
V_{Q_c}=\exp(2\pi i Q_c \phi)\ \ \ \ \ \ Q_c=\frac{r}{R}+s\frac{R}{2}\ \ \ \ \ \ r,s\in \mathbb{Z}  
\ee
where $Q_c$ are the charges (or momenta) of the theory. The conformal
dimensions of these vertex operators are $\Delta=Q_c^2/2$. There are
$N=2p'q$ primary blocks (or families). The generators $V_{Q_\lambda}$
of such families are chosen such that their conformal dimensions are
the lowest allowed by the theory. In terms of their charges these are
\be
Q_\lambda=\frac{\lambda}{\sqrt{N}}\ \ \ \ \lambda=0,1,2,\ldots,N-2,N-1
\ee
From now on they will be called primary charges. In this way
$Q_\lambda$ runs from $0$ to $\sqrt{N}-1/\sqrt{N}$. The remaining
fields of the theory are obtained by successive products of the
generators. The charges in a family $\lambda$ are
\be
Q_{\lambda,L}=Q_\lambda+L\sqrt{N} \ \ \ \ \ \ L\in \mathbb{Z}
\ee

The generators form the fusion algebra given by the fusion rules
\be
Q_{\lambda}+Q_{\lambda'}=Q_{(\lambda+\lambda')\mod N}
\ee
Here the fusion rules are expressed in terms of
the charges (momenta in CFT). Formally they are expressed in terms of
fields or vertex operators $V_{\lambda}\times V_{\lambda'}=V_{(\lambda+\lambda')\mod N}$.
This simply means that
two primary charges are picked out of two families and added together.
A charge from a different family will be obtained, but not necessarily the primary one.
Basically the fusion rules pick the primary charge of the family
corresponding to the new charge. For
further details on these subjects see~\cite{FMS}.

The relation between the standard normalization of CFT and the
normalization used in this work is computed to be
\be
k=2R^2\ \ \ \ Q=R Q_c
\ee
where $R$ is the radius of the boson as stated before.

\section{Propagation in the bulk \lb{ch.tor:sec.bulk}}

The aim in this section is to investigate the changes that a charge
undergoes when it travels from one boundary to the other. As it will be
shown, the bulk theory imposes, independently of what the boundary
conditions are, restrictions on the charges. Irrational values of $k$
are assumed in this section. Only in section~\r{ch.tor:sec.cb} will
the case of rational $k$ be addressed.

As stated before, $M=\Sigma\times[0,1]$ is a compact manifold such
that at each fixed time slice $\Sigma(t)$ the total charge is zero.
The monopole processes (instantons) can induce only a local charge
non-conservation. So let us insert a pair of charges
$Q_1=-Q_2$ in one boundary, say $\Sigma_0$. They will travel through
the bulk and emerge in the opposite boundary, $\Sigma_1$, as two new
charges $\bar{Q}_1=-\bar{Q}_2$. In their paths the charges can
interact with the bulk fields. Their paths can link but it is assumed
there is no self-linking and the charges do not interact with each
other.

Translating this to a more formal language, there are two open Wilson
lines, each of them have their ends attached to different
boundaries. The Wilson line represents the path of a charged
particle. Leaving aside the perturbative interactions in the bulk,
there are charge non-conserving interactions with the
monopole-instantons. These interactions will be represented as
insertions of instantons along the Wilson lines. So the physical
picture is that there is a Wilson line on which the charge varies from
point to point. But this variation is not random and it is induced by
instantons. One can in principle consider chains of Wilson lines which
is essentially the same picture. In between two insertions, the two
lines (minimal Wilson line segments with constant charges on them) can
link to each other changing the correlator of the lines according
to~(\r{df}). So it is assumed that the insertions of instantons and
the linkings of the lines can be separated on space-time. Another
way to put it is that insertions of instantons are done at the ends of
the tiny Wilson line segments.

\fig{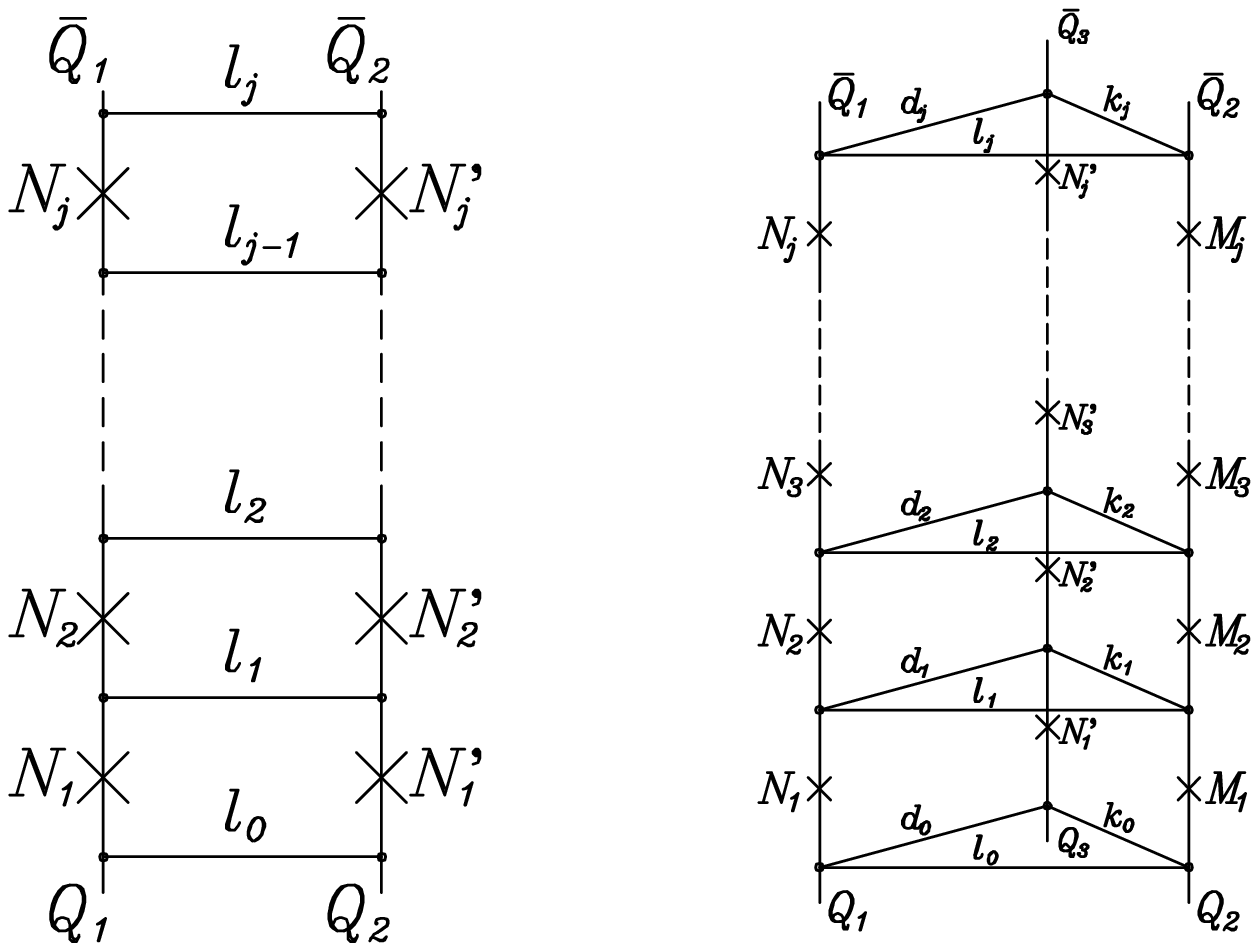}{Two and three charge propagation through the bulk. The $N$'s, $M$'s,
and $N'$'s represent the flux of instantons and the $l$'s, $d$'s and
$k$'s the linking number of the Wilson lines between
insertions.}{fig1}

A generic propagation of two such charges with $j$ instanton
interactions and $j+1$ sets of linkings in the bulk is represented as
the planar diagram in figure~\r{fig1}. The insertions are represented by
crosses and the linkings by horizontal lines. The full physical
picture is obtained by taking the limit $j\rightarrow\infty$ which
allows infinitely many instanton insertions on a line. Since
global charge conservation is assumed, $\bar{Q}_1-Q_1=Q_2-\bar{Q}_2$ and
the upper charges are the result of all the monopole contributions
$\bar{Q}_1=Q_1+\sum k/2 N_i$. So the condition
\be
\sum_{i=1}^{j}N_i=N =-\sum_{i=1}^{j}N'_i
\lb{N}
\ee
has to be imposed. Each pair of charges can be formally interpreted
as a quantum state $(Q_1,-Q_1)$ and its propagation from one boundary
to the other boundary can be considered as a time evolution to the
state $(\bar{Q}_1,-\bar{Q}_1)$.  Then for each initial $Q_1$ and final
$\bar{Q}_1$ a transition coefficient $C[Q_1,N]$ can be built. It is
to be interpreted as the square root of a probability exactly as in a
quantum system.
\be
(Q_1,-Q_1)\rightarrow C[Q_1,N]\,(Q_1+\frac{k}{2} N,-Q_1-\frac{k}{2} N)
\ee
where condition~(\r{N}) was already used.

$C$ is a function of $Q_1$ and $N$ only but, in order to find it, it is
necessary to take into account all the possible processes in the bulk,
monopole interactions and linkings.  To be able to extract some
information about the boundary let us insert a new variable $l$, which
denotes the total linking number. Up to a normalization factor, $C$ can
be written as
\be
C[Q_1,N]=\sum_{\tilde{l}=-\infty}^\infty\int dl\, \delta(\tilde{l}-l)\,c[Q_1,N,l]
\lb{C}
\ee
with the definition
\be
l=l_0+\sum_{i=1}^jl_i
\lb{l}
\ee

It is now time to determine how the infinitely many bulk processes
contribute to this coefficient. Due to the interaction with the
monopoles labeled by $i$, the pair of charges will go through the
transition $(q,q')\rightarrow(q+k/2 N_i,q'+k/2 N'_i)$ according to
(\r{dq}). Between each monopole insertion the linking will induce a
phase change of $4\pi i qq'l_i/k$ according to~(\r{df}). The total
change of the charges is already taken into account through the $N$
dependence of $C$. But the processes that are being summed over must be
selected in order to fulfill the requirement that the total monopole
contribution is $N$ and that the total linking number is $l$. The
coefficient $c$ is then the sum of the phases
\be
c[Q_1,N,l]=\sum_{N_a,N'_b}\sum_{l_i,l_0}\exp\{2\pi i \Phi[N_a,N'_b,l_i,l_0]\}
\lb{c}
\ee
where the sums over $N_a$ and $N'_b$ stand for all the allowed
configurations obeying (\r{N}) and the sum over $l_i$ and $l_0$ for
the ones obeying (\r{l}).  The phase, $\Phi$, for each process is
simply the sum of the several phases induced at each step $i$ along
the propagation in the bulk (see figure~\r{fig1})
\be
\Phi=\frac{2}{k}\left\{-Q_1^2l_0+\sum_{i=1}^{j}\left(Q_1+\frac{k}{2}\sum_{b=1}^iN_b\right)\left(-Q_1+\frac{k}{2}\sum_{a=1}^iN'_a\right)l_i \right\}
\ee 
Writing the last term of the sum in $i$ by using (\r{N}), i.e. for $i=j$,
and expanding the products, are obtain
\be
\ba{rcl}
\Phi&=&\displaystyle
    \frac{2}{k}\left\{-Q_1^2\,l_0-Q_1^2\sum_{i=1}^{j}\,l_i+\left(Q^2_1-\left(Q_1+\frac{k}{2}N\right)^2\right)l_j\right. \vspace{.2 cm}\\
&+&\displaystyle
   \left.\frac{k}{2}\sum_{i=1}^{j-1}\left(Q_1\sum_{a=1}^{i}(N'_a-N_a)+\frac{k}{2}\sum_{a=1}^iN'_a\sum_{b=1}^iN_b\right)l_i\right\}
\ea
\lb{f}
\ee
Note that the phase is no longer $N_j$ dependent, as it was replaced by
the $N$ dependence.  Furthermore it is necessary to get the $l$ dependence of
$c$. In the following discussion it is investigated in which cases
this coefficient is non vanishing.

Bearing in mind that these coefficients are actually formal series
(which are divergent in general), different ways of factorizing the
sums is safer. So three different factorizations will be considered
below (it is argued that these are the only possible
factorizations). $c$ will factorize into an $l$ dependent phase and one
independent of $l$. Unless otherwise stated, from now on, the
index $i$ will run from $1$ to $j-1$.

The most obvious way to factorize the sum is to eliminate $l_0=l-\sum
l_i-l_j$. Then (\r{c}) factorizes as
\be
\ba{rcl}
c[Q_1,N,l]&=&\displaystyle\exp\left\{-2\pi i \frac{2}{k} Q_1^2\,l\right\}\times\vspace{.2 cm}\\
&\times&\displaystyle\sum_{N_{a}=-\infty}^{\infty}\sum_{N'_{b}=-\infty}^{\infty}\prod_{i=1}^{j-1}\sum_{l_{i}=-\infty}^\infty\sum_{l_j=-\infty}^\infty\exp\{2\pi i \Phi'_0[N_a,N_b,l_i,l_j]\}
\ea
\lb{c0}
\ee
The indices $a$ and $b$ run from $1$ to $j-1$.
Since the phase $\Phi'_0$ is the sum of several phase changes $\phi$,
the second factor can be rewritten as
\be
\sum_{N_{a},N'_{b}}\left(\prod_{i=1}^{j-1}\sum_{l_{i}=-\infty}^\infty\exp\{2\pi i\phi_i\,l_i\}\right)\left(\sum_{l_j=-\infty}^\infty\exp\{2\pi i \phi_j\,l_j\}\right)
\ee
where
\be
\ba{rcl}
\phi_i&=&\displaystyle
      Q_1\sum_{a=1}^{i}(N'_a-N_a)+\frac{k}{2}\sum_{a=1}^iN'_a\sum_{b=1}^iN_b\vspace{.2 cm}\\
\phi_j&=&\displaystyle
      \frac{2}{k}\left\{Q_1^2-\left(Q_1+\frac{k}{2}N\right)^2\right\}
\ea
\lb{f0}
\ee

Now each of the sums over $l_j$ and the $l_i$'s can be considered
independently. If one of them is zero $c$ will be vanishing. Note that
although normalization factors have not been discussed here, each of
these phase sums must be normalized to a number between $0$ and $1$
such that $C^2$ is interpreted as a probability, because $C$ is the
coefficient of a quantum state. Further, if one takes the limit of
$j\rightarrow\infty$ these sums will become integrals. It can be
investigated if they are zero or not by using the identity
\be
\sum_{q=-\infty}^\infty\exp\{2 \pi i\, \phi\, q\}=\sum_{p=-\infty}^\infty\delta(\phi-p)
\lb{fourier}
\ee
The sum over delta functions is zero if
$\phi$ is not an integer. The conclusion is then that, for every $i$,
$\phi_i$ must be an integer.
Summing over $l_j$ the restriction
\be
\frac{2}{k}\left\{Q_1^2-\left(Q_1+\frac{k}{2}N\right)^2\right\}\in \mathbb{Z}
\lb{int0}
\ee
is obtained. Replacing $Q_1$ by its form~(\r{q}), expanding and
getting rid of the even integer term $2mN$ which does not change
anything concerning the condition being imposed, the former condition
reads
\be
\frac{k}{2}N\left(n+N\right)\in \mathbb{Z}
\lb{int01}
\ee
For irrational $k$ the only two solutions are
\bea
N&=&0\lb{nN00}\vspace{.2 cm}\\
N&=&-n\lb{nN0n}
\eea
The physical meaning of these results will be given a little later.
The remaining conditions on the $\phi_i$'s will allow us to build the
intermediate processes, which will be addressed at the end of this section.

An other way to carry out the sums is to
eliminate $l_j=l-l_0-\sum l_i$. Then~(\r{c}) factorizes as
\be
\ba{rcl}
c[Q_1,N,l]&=&\displaystyle\exp\left\{-2 \pi i \frac{2}{k}\left(Q_1+\frac{k}{2}N\right)^2l\right\}\times\vspace{.2 cm}\\
&\times&\displaystyle\sum_{N_{a}=-\infty}^{\infty}\sum_{N'_{b}=-\infty}^{\infty}\prod_{i=1}^{j-1}\sum_{l_{i}=-\infty}^\infty\sum_{l_0=-\infty}^\infty\exp\{2\pi i \Phi'_j[N_a,N_b,l_i,l_0]\}
\ea
\lb{cj}
\ee
where the phase is now
\be
\ba{rcl}
\Phi'_j&=&\displaystyle
    \frac{2}{k}\left\{-\left(Q_1^2-\left(Q_1+\frac{k}{2}N\right)^2\right)l_0\right. \vspace{.2 cm}\\
&+&\displaystyle
   \left.\frac{k}{2}\sum_{i=1}^{j-1}\left[Q_1\left(\sum_{a=1}^{i}(N'_a-N_a)+2N\right)+\frac{k}{2}\left(\sum_{a=1}^iN'_a\sum_{b=1}^iN_b+N^2\right)\right]l_i\right\}
\ea
\lb{fj}
\ee
By performing the sum over $l_0$ the same conditions~(\r{int0})
and~(\r{int01}) on $N$ are obtained.  There are extra $2N$ and $N^2$
factors but they do not change the coefficient $c$. This can be checked
explicitly by replacing $Q_1$ by its form~(\r{q}). Apart from an
irrelevant factor of $4\pi i m N$ one gets the same result as
for~${\Phi'}_0$.

A third factorization is studied next. Rewriting~(\r{f}) in terms
of $l^\pm$ defined by
\be
l_0^+ = l_0+l_j  \hskip 1.5cm l_0^- = l_0-l_j
\lb{lpm}
\ee
and taking into account that $l_0^+=l-\sum l_i$,
(\r{c})~factorizes as
\be
\ba{rcl}
c[Q_1,N,l]&=&\displaystyle\exp\left\{-2 \pi i \frac{2}{k}\left(Q_1^2+\left(Q_1+\frac{k}{2}N\right)^2\right)\frac{l}{2}\right\}\times\vspace{.2 cm}\\
&\times&\displaystyle\sum_{N_{a}=-\infty}^{\infty}\sum_{N'_{b}=-\infty}^{\infty}\sum_{l_{i}=-\infty}^\infty\sum_{l_0^-=-\infty}^\infty\exp\{2\pi i \Phi'_+[N_a,N_b,l_i,l_0^-]\}
\ea
\lb{c+}
\ee
The new phase is
\be
\ba{rcl}
\Phi'_+&=&\displaystyle
    \frac{2}{k}\left\{-\left(Q_1^2-\left(Q_1+\frac{k}{2}N\right)^2\right)\frac{l_0^-}{2}\right. \vspace{.2 cm}\\
&+&\displaystyle
   \left.\frac{k}{2}\sum_{i=1}^{j-1}\left[Q_1\left(\sum_{a=1}^{i}(N'_a-N_a)+N\right)+\frac{k}{2}\left(\sum_{a=1}^iN'_a\sum_{b=1}^iN_b+\frac{1}{2}N^2\right)\right]l_i\right\}
\ea
\lb{f+}
\ee
By performing the sum over $l_0^-$ a new different condition is obtained
\be
\frac{2}{k}\left\{Q_1^2-\left(Q_1+\frac{k}{2}N\right)^2\right\}\in 2\mathbb{Z}
\lb{even}
\ee
By expanding $Q_1$ this is simply
\be
\frac{k}{2}N\left(n+N\right)\in 2\mathbb{Z}
\lb{int0+}
\ee

For generic $k\in\mathbb{R}$ the same solutions as in~(\r{int0}) are
obtained and apparently this condition turns out to hold the same
results of~(\r{int01}). It will become clear that this last result is
different from the previous two cases. As before the extra $N$ and
$N^2/2$ factors do not change the coefficient.

It is now time to explain the physical implications of the three
previous results. The phase factors in front of $l$ are none other
than the conformal dimensions~(\r{confd}) of the vertex operators
inserted on the boundary CFT's. When the sum over $l_0$ was eliminated
it gave~(\r{c0}). Extracting the $l$ phase factor gives
\be
\frac{2}{k}Q_1^2=2\Delta=\Delta_1+\Delta_2
\lb{confd0}
\ee
Upon elimination of the $l_j$ sum the factorization~(\r{cj}) was
obtained such that the phase factor reads
\be
\frac{2}{k}\bar{Q}_1^2=2\bar{\Delta}=\bar{\Delta}_1+\bar{\Delta}_2
\lb{confdj}
\ee
while for the factorization~(\r{c+}), where the sum over $l_0^+$ was
eliminated, we obtained
\be
\frac{1}{k}(Q_1^2+\bar{Q}_1^2)=\Delta+\bar{\Delta}
\lb{confd+}
\ee

Note that no other factors can be extracted. Trying to eliminate any
other combination of $l_0$ and $l_j$ would yield the result $c=0$.
Eliminating some combination of $l_i$'s will always end up
effectively in extracting one of the three above factors or, again,
give the result $c=0$. The solutions for the monopole
interactions~(\r{ab}) and~(\r{bb}) computed below assure this result.

In order for $C$ not to be zero it is necessary to get integer conformal
dimensions as stated in section~\r{ch.tor:sec.string}. Consider then
adding extra type $\tilde{\mathcal{B}}$ gauge fields. One Wilson line~(\r{WD})
carries now charges from all the $U(1)$'s. Summing over $l$ for the
three previous cases the conditions
\be
\ba{cccc}
r&=&\displaystyle 2\left(\Delta+\frac{p^2}{k'}\right)&\in\mathbb{Z}\vspace{.2 cm}\\
s&=&\displaystyle 2\left(\bar{\Delta}+\frac{p^2}{k'}\right)&\in\mathbb{Z}\vspace{.2 cm}\\
t&=&\displaystyle \Delta+\bar{\Delta}+\frac{2p^2}{k'}&\in \mathbb{Z}
\ea
\lb{rst}
\ee
are obtained. Subtracting the second equation from the first gives
the equation
\be
r-s=2(\Delta-\bar{\Delta})=\frac{2}{k}(Q_1^2-\bar{Q}_1^2)=-2mN
\lb{r-s}
\ee
where $N=0,-n$ according to the allowed solutions~(\r{nN00})
and~(\r{nN0n}). By averaging the first two conditions, the third one is
obtained as long as the identification
\be
t=(r+s)/2
\ee
is considered. According to the discussion in
section~\r{ch.tor:sec.string} $r$ and $s$ have to be even in order for
the full conformal dimension to be integer. Making the identifications
\be
\ba{ccc}
\displaystyle \frac{r}{2}=1-L \vspace{.2 cm}\\
\displaystyle \frac{s}{2}=1-\bar{L}\vspace{.2 cm}
\ea
\ee
the mass shell condition~(\r{mshell}) is retrieved. Furthermore,
(\r{r-s})~becomes the spin condition~(\r{spin}).

Below the possible intermediate processes are investigated by
performing the sums over each of the $l_i$'s. Remembering that the
total charge at each level is null, a chain of conditions $\phi_i \in
\mathbb{Z}$ is obtained. These conditions allow the building of the full set of
possible diagrams contributing to the transitions.  Starting with
$i=1$ a simple condition is found
\be
n(N'_1-N_1)+2N_1N'_1=0
\ee
It has two possible solutions
\be
\ba{lrcccr}
a\ \ \ \ \ &N_1&=&0&=&N'_1\vspace{.2 cm}\\
b          &N_1&=&-n&=&-N'_1
\ea
\lb{ab}
\ee

For $i=2$ the condition reads
\be
n(N'_1-N_1)+n(N'_2-N_2)+2N_1N'_1+2N_2N'_2+2N_1N'_2+2N_2N'_1=0
\ee
Choosing $a$ for $N_1$ and $N'_1$ the same two previous
solutions for $N_2$ and $N'_2$ are obtained. Choosing the $b$
solution for $N_1$ and $N'_1$, again two solutions are valid for $N_2$
and $N'_2$: $a$ or a new solution
\be
\tilde{b}\ \ \ \ \ N_2=n=-N'_2\\
\lb{bb}
\ee
If this solution is chosen for $i=2$, an $a$ or
$b$ solution will be allowed for $i=3$. So no new solutions emerge and
the solution for generic $i$ can be built using the same arguments.

Note that a solution $a$ will reduce the diagram to one of type $i-1$,
as the linkings before and after that monopole may be joined since it
doesn't change the charge (or in other words the monopole actually
doesn't exist!). Without loss of generality the $a$ solution will be
disregarded in the following discussion.

By induction the $i$th condition reads then
\be
n(1-2\#b+2\#\tilde{b})(N'_i-N_i)+2N_iN'_i=0
\ee
with the restriction that $\#b-\#\tilde{b}$ takes only the values $0$
and $1$.  Then a chain of alternating solutions $b\tilde{b}\ldots
b\tilde{b}b\tilde{b}\ldots$ will be obtained as the only possible
solution. Some diagrams are presented in figure~\r{fig2}.

The above results are consistent with global charge conservation, and
moreover they show that local charge violation is quite
restricted. Given some charge $q$ on the first boundary it can change
to $\bar{q}$ and return to its previous form $q$ again as many times
as it wants but it can become nothing else.

\fig{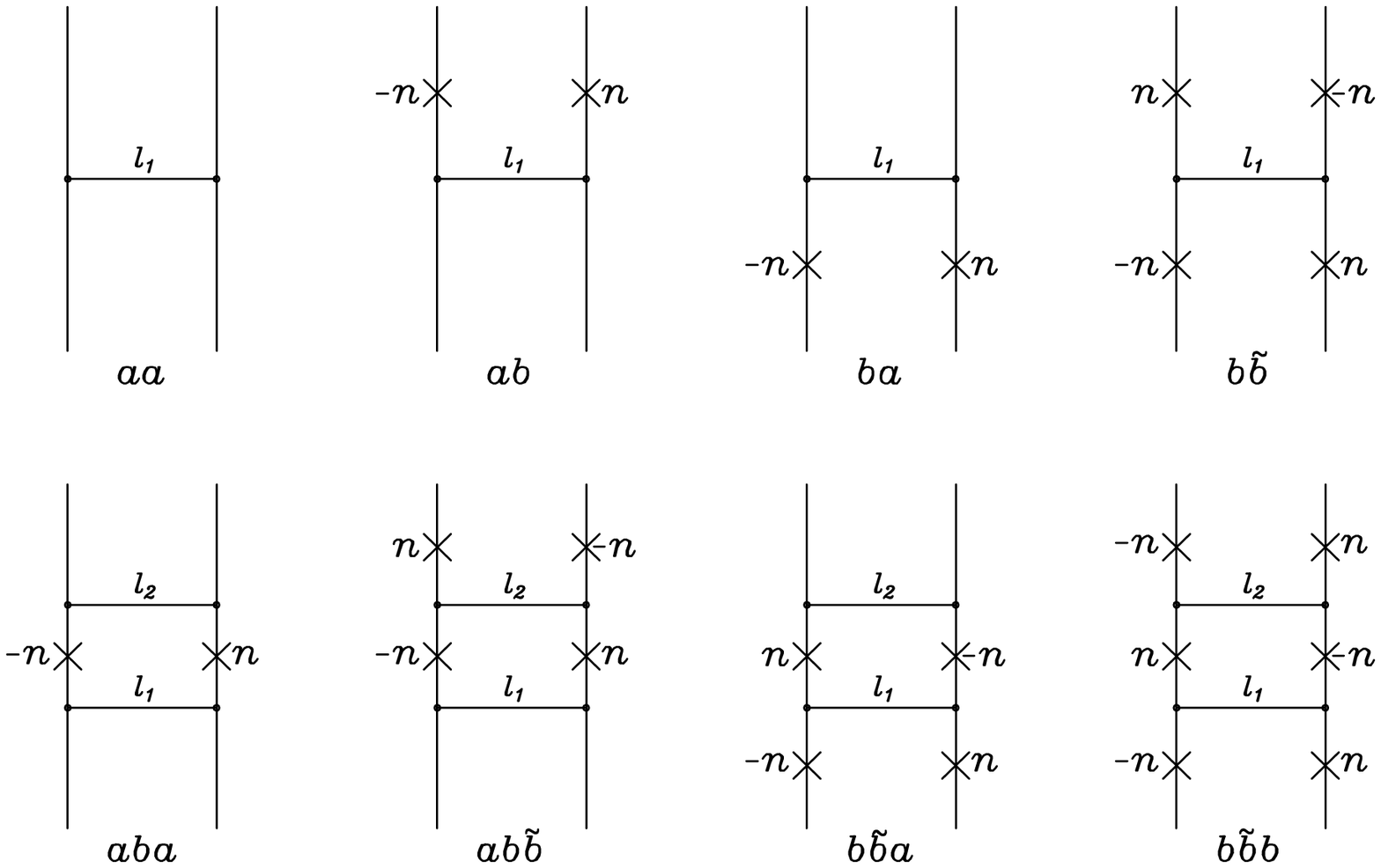}{Some of the possible diagrams for two charge
propagation, $i=2$ and $i=3$.}{fig2}

Lets go on and consider now 3 charges propagating from one boundary to
the other (or equivalently evolving in time)
\be
\ba{ccc}
Q_1&\rightarrow&\bar{Q}_1\vspace{.2 cm}\\
Q_2&\rightarrow&\bar{Q}_2\vspace{.2 cm}\\
Q_3=-Q_1-Q_2&\rightarrow&\bar{Q}_3=-\bar{Q}_1-\bar{Q}_2
\ea
\ee
as pictured in figure~\r{fig1}. The phase is now 
\be
\ba{rcl}
\Phi_3=&\displaystyle
      \left.\frac{2}{k}\,\right\{&Q_1Q_2 l_0 -Q_2(Q_1+Q_2)d_0-Q_1(Q_1+Q_2)k_0+ \vspace{.2 cm}\\
      &+&\displaystyle
      \sum_{i=1}^{j-1}\left[\bar{Q}_{1(i)}^2l_i-\bar{Q}_{1(i)}\bar{Q}_{3(i)} d_i-\bar{Q}_{2(i)}\bar{Q}_{3(i)}k_i\right]\vspace{.2 cm}\\
      &+&\displaystyle\left.\bar{Q}_1\bar{Q}_2 l_j -\bar{Q}_2(\bar{Q}_1+\bar{Q}_2)d_j-\bar{Q}_1(\bar{Q}_1+\bar{Q}_2)k_j\ \right\}
\ea
\ee
where the following definitions
\be
\ba{rclcrcl}
\bar{Q}_{1(i)}&=&\displaystyle Q_1+\frac{k}{2}\sum_{a=1}^iN_i&\ \ \ \ \ \ &\bar{Q}_{1}&=&\displaystyle Q_1+\frac{k}{2}N \vspace{.2 cm}\\
\bar{Q}_{2(i)}&=&\displaystyle Q_2+\frac{k}{2}\sum_{a=1}^iM_i& &\bar{Q}_{2}&=&\displaystyle Q_2+\frac{k}{2}M \vspace{.2 cm}\\
\bar{Q}_{3(i)}&=&\displaystyle Q_1+Q_2-\frac{k}{2}\sum_{a=1}^iN'_i& & \bar{Q}_{3}&=&\displaystyle Q_1+Q_2-\frac{k}{2}N'
\ea
\ee
were used. Now it can be chosen to factor out some combination of $l_0$,
$d_0$, $k_0$, $l_j$, $d_j$ and $k_j$. But it is now clear, from the
previous discussion of two charge propagation, what factorizations one
should look for. So the sensitive choices are to eliminate all the $0$'s, all
the $j$'s or some particular combination of $0$'s and $j$'s for the three
linkings.

The dependence on $l_0$, $d_0$ and $k_0$ in the sums can be replaced
by sums over the total linking numbers defined as $l=l_0+\sum
l_i+l_j$, $d=d_0+\sum d_i+d_j$ and $k=k_0+\sum k_i+k_j$.  Then the
phases (conformal dimensions) appearing in the three field
correlation functions given in~(\r{OPE3}) for $z_{12}$, $z_{23}$ and $z_{13}$
correspond to the phase factors containing $l$, $d$ and $k$
respectively.

By considering extra type $\tilde{\mathcal{B}}$ gauge fields with
charges $p_1$, $p_2$ and $p_3$ and summing over the previous variables
gives the conditions that the combinations of conformal dimensions
in~(\r{OPE3}) must be integer. The conformal dimensions are to be read
off by considering these extra gauge fields and charges
$\Delta=Q^2/k+p^2/k'$ and $\bar{\Delta}=\bar{Q}^2/k+p^2/k'$. From now
on let us use this definition of conformal dimensions. As discussed in
section~\r{ch.tor:sec.string}, from the point of view of the boundary
CFT's, these factors must indeed be integer to ensure that the three
point OPE's are single valued.

In terms of the individual vertices it is not so clear what these
conditions mean.  Without loss of generality and in order to clarify
them let us replace the sums over $l$, $d$ and $k$ by the sums over
three new variables $l_+$, $d_+$ and $k_+$ such that $l=l_++k_+$,
$d=d_++l_+$ and $k=k_++d_+$. By summing over these new variables one
obtains the following conditions
\be
2\Delta_1,\ 2\Delta_2,\ 2\Delta_3\ \in\mathbb{Z}
\lb{l03}
\ee
In a similar way the dependence over $l_j$, $d_j$ and $k_j$ can be
replaced by a dependence over $l_+$, $d_+$ and $k_+$ and, upon summation,
the respective conditions are obtained
\be
2\bar{\Delta}_1,\ 2\bar{\Delta}_2,\ 2\bar{\Delta}_3\ \in\mathbb{Z}
\lb{lj3}
\ee
Or using the definition~(\r{lpm}) for $l^\pm_0$, $d^\pm_0$ and
$k^\pm_0$ and replacing the $l^+_0$, $d^+_0$ and $k^+_0$ dependences by
$l_+$, $d_+$ and $k_+$ in the same way as before (performing the
sums in these last variables) we get the conditions
\be
\Delta_1+\bar{\Delta}_1,\ \Delta_2+\bar{\Delta}_2,\ \Delta_3+\bar{\Delta}_3\ \in\mathbb{Z}
\lb{l+3}
\ee
Similarly to the previous discussion for two charge propagation these
conditions turn out to be equivalent to the mass shell and spin conditions of
string theory.

The fundamental differences come from the remaining conditions.
By taking the first two cases~(\r{l03}) and~(\r{lj3}), where the $0$'s
and $j$'s linking number sum dependences were replaced, and summing
over $(l_{+j},d_{+j},k_{+j})$ and $(l_{+0},d_{+0},k_{+0})$ gives,
in both cases, three conditions.  The first two are redundant, as they
correspond to (\r{int0}) which has been obtained from the two charge
propagation, while the third one is new
\be
\frac{4}{k}\left(Q_1Q_2-\bar{Q}_1\bar{Q}_2\right) \in \mathbb{Z}
\lb{I3a}
\ee

Nevertheless the solutions for
irrational $k$ end up being of the same kind as before.
There are two solutions
\be
\ba{ccc}
\left\{\ba{ccc}N&=&0\\M&=&0\ea\right.&\ \ \ &\left\{\ba{ccc}N&=&-n_1\\M&=&-n_2\ea\right.
\ea
\lb{N3}
\ee
For the case of~(\r{l+3}), where the sum over $+$'s linking number was
replaced, one finds, upon summation over
$(l_{+0}^-,d_{+0}^-,k_{+0}^-)$, the two conditions~(\r{even}) which
were obtained in the two charge case.  The third one is again new
\be
\frac{2}{k}\left(Q_1Q_2-\bar{Q}_1\bar{Q}_2\right) \in \mathbb{Z}
\lb{I3}
\ee

The solutions for the full diagrams are very similar to the ones of
two charges,  with the charges oscillating simultaneously between $Q_1$,
$Q_2$ and $\bar{Q}_1=Q_1-n_1k/2$, $\bar{Q}_2=Q_2-n_2k/2$. For four
charges there are no new conditions.

After all this algebra let us summarize what was obtained so far. In the
start there was a theory which, at the perturbative level, had the
allowed charges of the form $Q = m + (k/4)n$. It was shown that, under
certain assumptions, if the non-perturbative processes are taken
into account, the charge spectrum is modified (restricted) quite
drastically. The allowed left/right pair of charges are in one-to-one
correspondence with the processes which have non vanishing quantum
amplitudes. All these processes can be organized in a lattice $\Gamma$
with elements $l=(Q,\bar{Q}=Q+(k/2)N)$ with $N$ either $0$ or $-n$.
Moreover a Lorentzian inner product $\circ$ of signature ($+$,$-$) emerges
naturally from~(\r{int0}),~(\r{even}) and~(\r{I3}) defined by
\be
l\circ l'=\frac{2}{k}(QQ'-\bar{Q}\bar{Q}')
\lb{lorprod}
\ee

There is more to add to it, going even further and extracting very
important properties of the lattice, it is {\bf even} due
to~(\r{even})
\be
l\circ l=\frac{2}{k}(Q^2-\bar{Q}^2)\in2\mathbb{Z}
\lb{EVEN}
\ee
and {\bf integer} due to~(\r{I3})
\be
l\circ l'=\frac{2}{k}(QQ'-\bar{Q}\bar{Q}')\in\mathbb{Z}
\lb{INT}
\ee
Note that~(\r{I3a}) is redundant since it is necessarily obeyed if~(\r{I3}) is.

By inspection, for $N=-n$ the lattice is {\bf self-dual} as well but
not for $N=0$. Note that the dual lattice $\Gamma^*$ is defined as the
set of all points in ${\mathbb{R}}^{1,1}$ (or ${\mathbb{R}}^{d,d}$
generally) which have integer inner product with all the points in the
original lattice $\Gamma$. In this way the condition of the lattice
being integer is only the statement that $\Gamma\subset\Gamma^*$.

In the next section it will be explained how to exclude the $N=0$ case and
obtain what is sought. The lattice is then exactly the one for
bosonic string theory with one compact dimension. It is the Narain
lattice for compactification on $S^1$.  For toroidal compactification
of several dimensions it is enough to consider several compact gauge
fields. This situation is described in
section~\r{ch.tor:sec.lat}.

\section{Boundary conditions \lb{ch.tor:sec.bc}}

So it remains to understand what mechanism selects between the $N=0$ and
$N=-n$ cases. The key is to consider different combinations of the
boundary conditions introduced in~(\r{bc}). Furthermore, to obtain a CFT
on the boundary (whether it is chiral or not) as an effective $2D$
theory of the $3D$ TMGT, it is necessary to identify the two boundaries
$\Sigma_0$ and $\Sigma_1$ in some way.  These are the mechanisms that
allow us to build several string theories out of the same $3D$ theory.
Note that these constructions are only possible due to the Maxwell term
in the action and the existence of monopoles.

In the last section it was shown that the bulk theory only allows the
charge $q= m+kn/4$ to become $\bar{q}=m-kn/4$ or remain unchanged.
What is actually $n$? Going back to (\r{q}),
$2\pi n$ is the flux of the magnetic field. So upon some kind of
boundary identification the theory only admits the ones that support one
of the conditions
\be
\ba{lcl}
N=0& &\displaystyle\int_{\Sigma_0}d^2z_0B=\int_{\Sigma_1}d^2z_1B\vspace{.2 cm}\\
N=-n&\ \ \ &\displaystyle\int_{\Sigma_0}d^2z_0B=-\int_{\Sigma_1}d^2z_1B
\ea
\lb{id0}
\ee
Consider a map (coordinate transformation) that transforms a vector in
$\Sigma_1$ into another vector in $\Sigma_0$, defining in this way the
identification \textit{rule}. There will be two kinds of maps. One
which maintains the relative orientation of both $2D$ boundaries, let
us call it parallel (${\small//}$), and one that reverses
the relative orientation, let us call it perpendicular ($\perp$).  The
names are chosen by the relative identification of the axes from
boundary to boundary as pictured in figure~\r{fig_id01}.

Note that the induced $\tilde{\epsilon}$ does not change from boundary
to boundary. From the $3D$ point of view time is being reversed on one
boundary and the complex coordinates are swapped (in $3D$ this is a
rotation). From the point of view of the bulk nothing special is
happening.

\fig{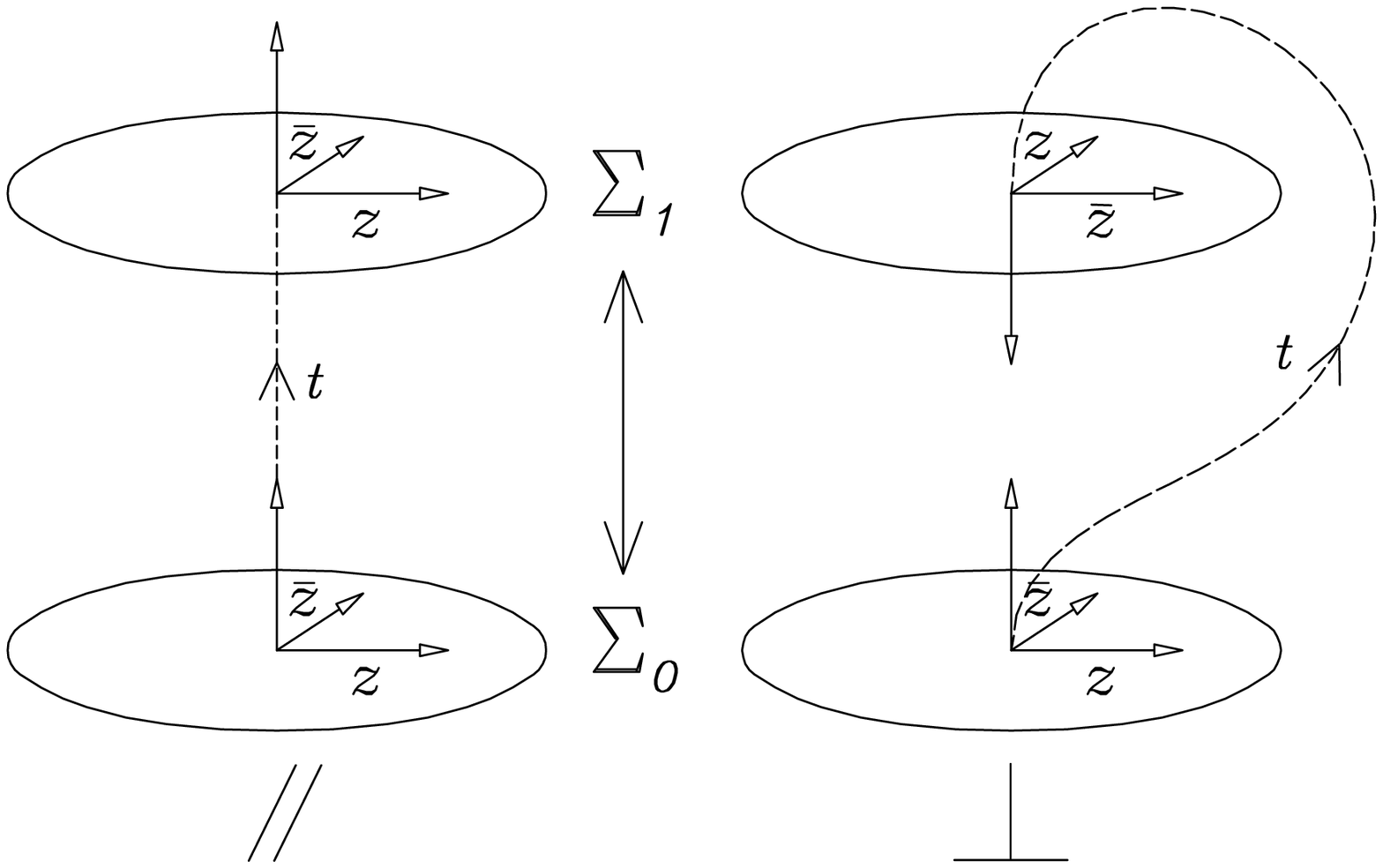}{Boundary identification with the same orientation (parallel; ${\small//}$)
and reversed orientation (perpendicular; $\perp$)}{fig_id01}

Let the difference of fluxes be defined as
\be
\delta\phi=\int_{\Sigma_1}d^2z_1 B-\int_{\Sigma_0}d^2z_0 B=
\int_{\Sigma_1}d^2z\epsilon^{ij}\partial_iA_j-\int_{\Sigma_0}d^2z\epsilon^{ij}\partial_iA_j
\ee
Given our boundary conditions these differences can be evaluated
explicitly by writing out what the magnetic field is for each of the
allowed boundary conditions
\be
\ba{lrl}
C.&\phi_C=&-\int\partial_\bz A_z\vspace{.2 cm}\\
\bar{C}.&\phi_{\bar{C}}=&\int\partial_z A_\bz
\ea
\ee
In these equations the magnetic field definition is
$B=\partial_zA_\bz-\partial_\bz A_z$ and the respective boundary
conditions are imposed.

Taking into account the boundary identifications and labeling the
different combinations of boundary conditions accordingly
the flux difference is computed to be
\be
\ba{lc}
CC_{\small//}.&\delta\phi=0\vspace{.2 cm}\\
CC_\perp.&\delta\phi=-\int(\partial_\bz A_z+\partial_z A_\bz)=-2\pi(2n)\vspace{.2 cm}\\
C\bar{C}_{\small//}.&\delta\phi=-\int(\partial_\bz A_z+\partial_z A_\bz)=-2\pi(2n)\vspace{.2 cm}\\
C\bar{C}_\perp.&\delta\phi=0
\ea
\ee
Note that for the parallel type of identifications the fields and
integrals are summed without any relative change. For the
perpendicular type the space indices of the fields have to be changed
to their conjugates and the measure in the integral changes sign (in
one boundary, say the right one).  The results are summarized in
figure~\r{fig_id02}.

\fig{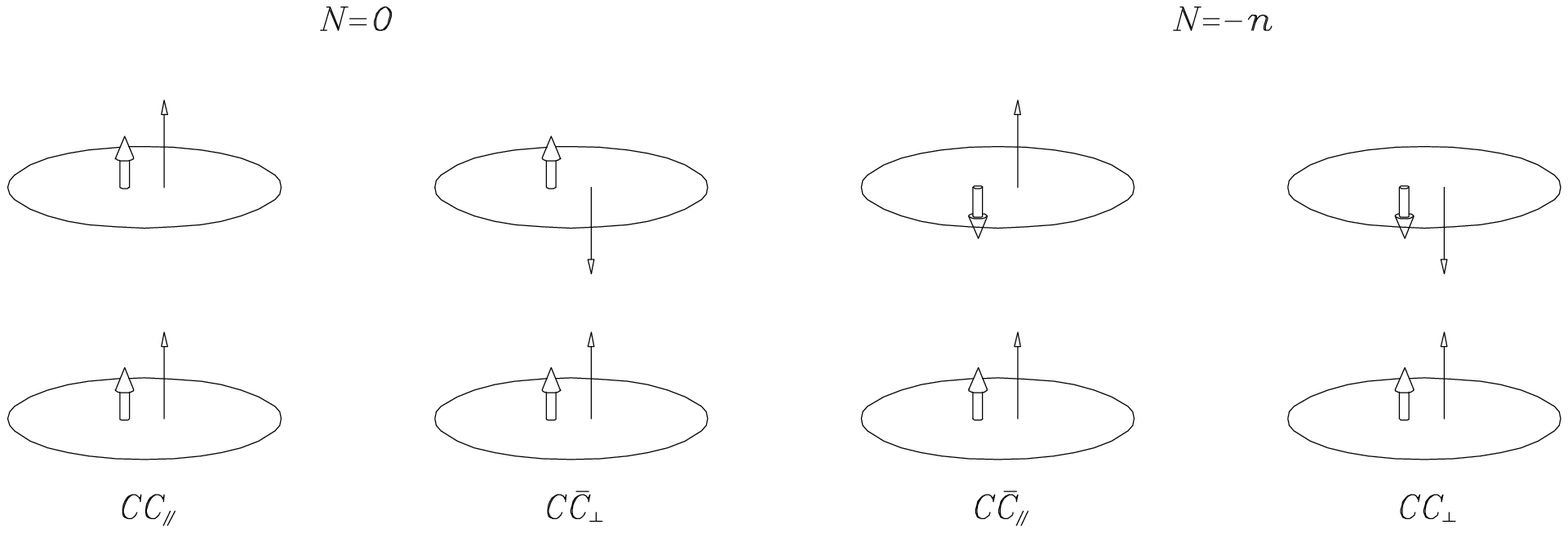}{Boundary identifications for several combinations of
boundary conditions. The fat arrow represents the magnetic flux and the
thin one the boundary orientation.}{fig_id02}

So the conclusion is that $Q=\bar{Q}$ corresponds to $CC_{\small//}$,
$\bar{C}\bar{C}_{\small//}$, $C\bar{C}_\perp$ or $\bar{C}C_\perp$ and
$Q-\bar{Q}=(k/2) n$ to $CC_\perp$, $\bar{C}\bar{C}_\perp$,
$C\bar{C}_{\small//}$ or $\bar{C}C_{\small//}$ boundary conditions.
Note that in principle there may exist other choices of the maps which
would give different boundary theories.

Note that the spectrum for $NC$ or $N\bar{C}$ boundary conditions must
be obtained by truncating the spectrum of $CC_\perp$ or
$C\bar{C}_{\small//}$. Simply consider the charge on the $N$ boundary
to be restricted to $\bar{Q}=0$. This means
\be
m=\frac{k}{4}n
\lb{QN}
\ee
Then the charges on the other boundary become
\be
Q=\frac{k}{2}n
\lb{QQN}
\ee

Of course for irrational $k$ the only solution for the
condition~(\r{QN}) is $m=n=0$ and a very poor and empty theory is
obtained in the boundary. In section~\r{ch.tor:sec.lat} it will be
clarified in which cases the condition~(\r{QN}) allows some dynamics
on the boundary.  Trying to truncate the spectrum of $CC_{\small//}$
or $C\bar{C}_\perp$ will set straight away $Q=\bar{Q}=0$ since the
charges are equal on both boundaries killing the hope of finding any
dynamics on the boundaries.

From now on it is chosen to work with $CC_\perp$ or
$C\bar{C}_{\small//}$ type of boundary conditions by default since
they are the ones which give us the desired spectrum on the boundary
CFT's. Furthermore, as just explained, $N$ boundary conditions can
easily be obtained from them.

\section{RCFT's and Fusion Rules\lb{ch.tor:sec.cb}}

In section~\r{ch.tor:sec.string} a short overview on RCFT and fusion
rules was given. In what follows it will be explained how the fusion
rules and the RCFT block structure emerge, with some naturalness, from
the bulk theory. Take $k=2p/q$ to be a rational number. Note that in
the previous discussion $2p'=p$, so take even $p$.  By inspection it
can be checked that the $(m,n)$ space is divided in diagonal bricks
containing $p\times q$ charges. They are distributed in diagonal
layers of identically valued bricks as symbolically pictured in
figure~\r{figbricks}.  To build this \textit{diagonal structure}
explicitly take a generic charge labeled by the pair $(m,n)$. It can
be represented by any other pair $(m-pn'/2 ,n+q n')$
\be
Q=m+\frac{p}{2q}n=m-\frac{p}{2}\ n'+\frac{p}{2q}(n+q\ n')
\lb{Qp}
\ee
This simply represents a diagonal translation of $n'$ bricks
in the figure.

\fig{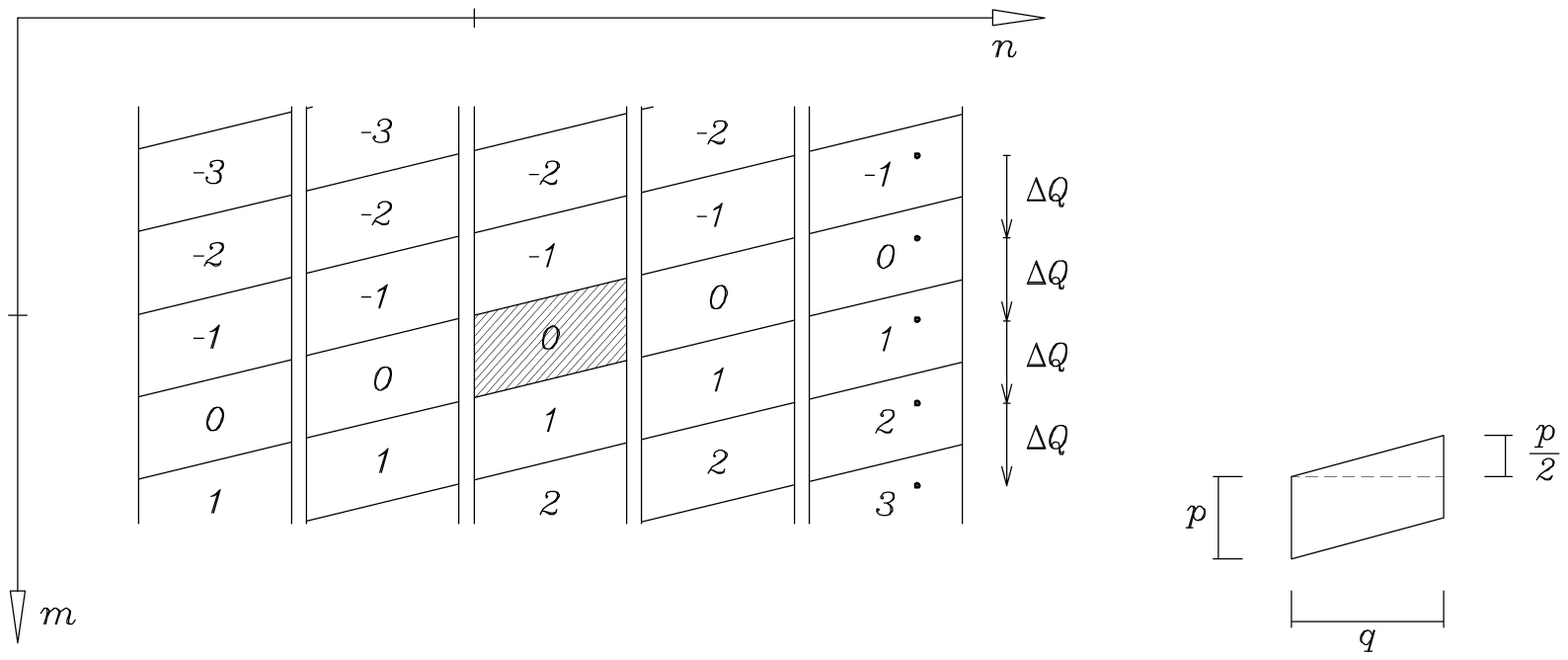}{$(m,n)$ structure for even $p$. The dashed region
is the primary brick $(r,s)$, some examples are presented in
figure~\r{figbrick}.
The numbers inside the blocks represent the $L \Delta Q$ shift of the
charge values as given by (\r{DQ}). Blocks with
the same $L$ have the same charge entries. In each brick there is
exactly one element of level $L$ belonging to each family.}{figbricks}

Note that this choice of the brick shape is not unique. It could be taken
to be some other choice as long as it has dimension $pq$ (e.g. a parallelogram
of sides $p$ and $q$) and the following results would hold
nevertheless. The reason for this particular choice is in order to get
a direct parallel with the usual results of RCFT's.  Furthermore
one of these bricks is considered to be the primary brick. It corresponds to
the minimal charges (chosen to be positive) allowed by the theory
built out of the lowest pairs of integers $(r,s)$
\be
Q_{rs}=r+\frac{p}{2q}s=\frac{\lambda}{q}\ \ \ \ \ \ \lambda=rq+\frac{p}{2}s=0,1,2,\ldots,pq-2,pq-1
\ee

Examples for some values of $k$ are given in figure~\r{figbrick}. The
geometric rule to organize the charges inside the brick is to order
them in ascending order by their distance to the upper diagonal line
of slope $k/4$.

\fig{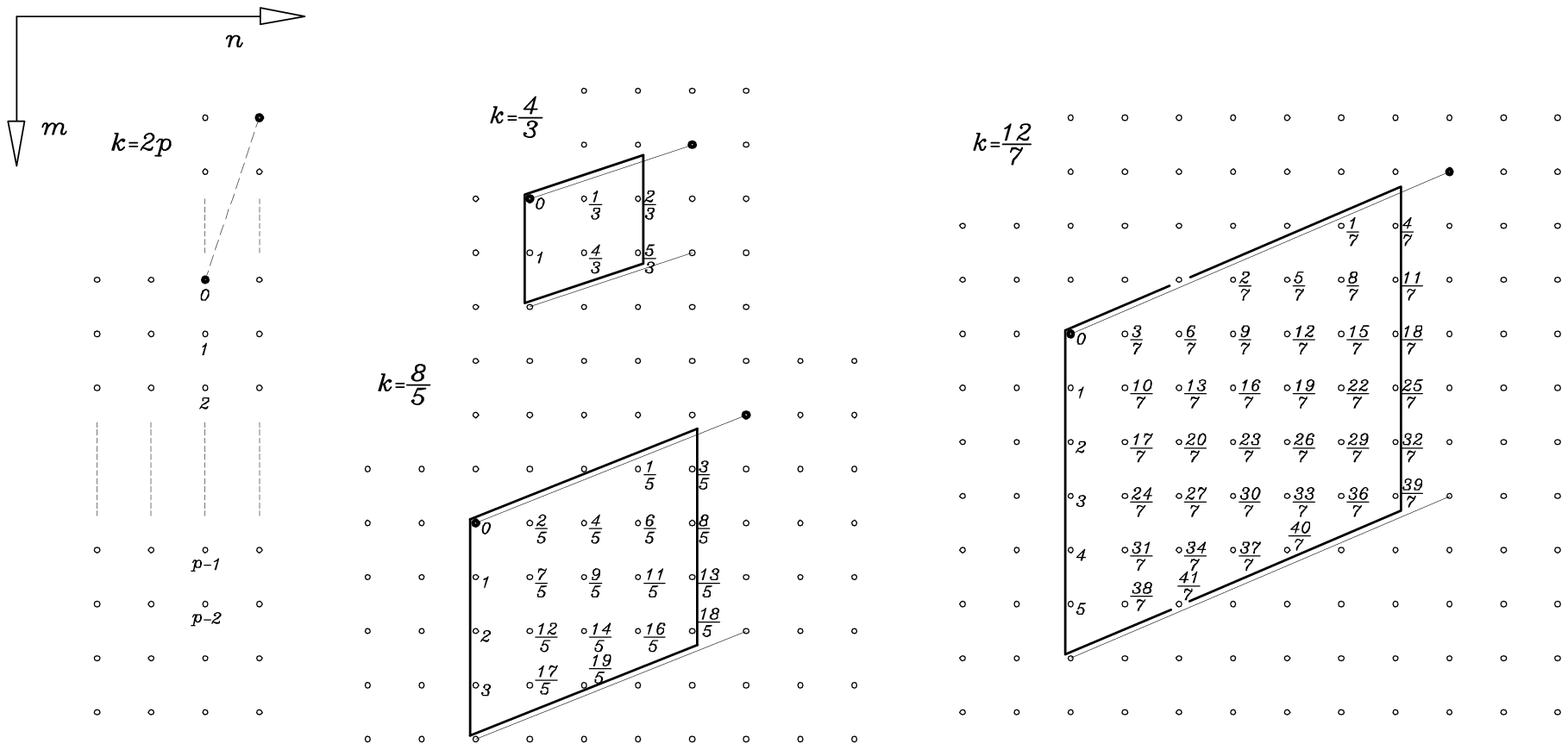}{Primary charge distribution in the $(m,n)$ plane
constituting the primary brick $(r,s)$. The rule to order the charges is
by their distance to the (thin) line connecting the charges $Q=0$
(filled dots) of slope $k/4=p/2q$.}{figbrick}

So far the well know charge structure of RCFT's is reproduced. It is
now time to return to the bulk theory and justify it.

Take again condition (\r{even}). Setting $k=2p/q$, it reads then
\be
\frac{p}{2q}\left(nN-N^2\right) \in \mathbb{Z}
\ee
which can be reexpressed as
\be
N(n-N)=0\mod q
\lb{evenrac}
\ee
The solutions for (\r{evenrac}) can be easily computed to be
\be
\ba{ccc}
N&=& 0\mod q\vspace{.2 cm}\\
N&=&-n\mod q
\ea
\ee
which are equivalent to (\r{int0}) and (\r{int01}).
There is one important lesson to take from this result.
Besides the previously allowed monopole-instanton
process $n\rightarrow-n$ which is charge dependent, there is a new
charge independent one:
\be
\Delta Q=\pm p
\lb{DQ}
\ee
This is actually the physical process that spans each of the families!
Or in other words, the process that builds up each of the conformal
blocks of the theory.  To obtain the charges of some family in terms
of $m$ and $n$ from the primary one, either $n$ or $m$ can be shifted
as $m\rightarrow m+pL$ or $n\rightarrow n+2qL$.  This is due to an
infinite degeneracy in the $(m,n)$ plane of the charge values
expressed by~(\r{Qp}).  Figure~\r{figcharges} presents the
distribution of charges for $k=2p$. In this simplified case the
families are simply organized along the $m$ axis for even values of
$2n$. This structure is repeated by shifts on the $m$ axis of magnitude
$p$.

\fig{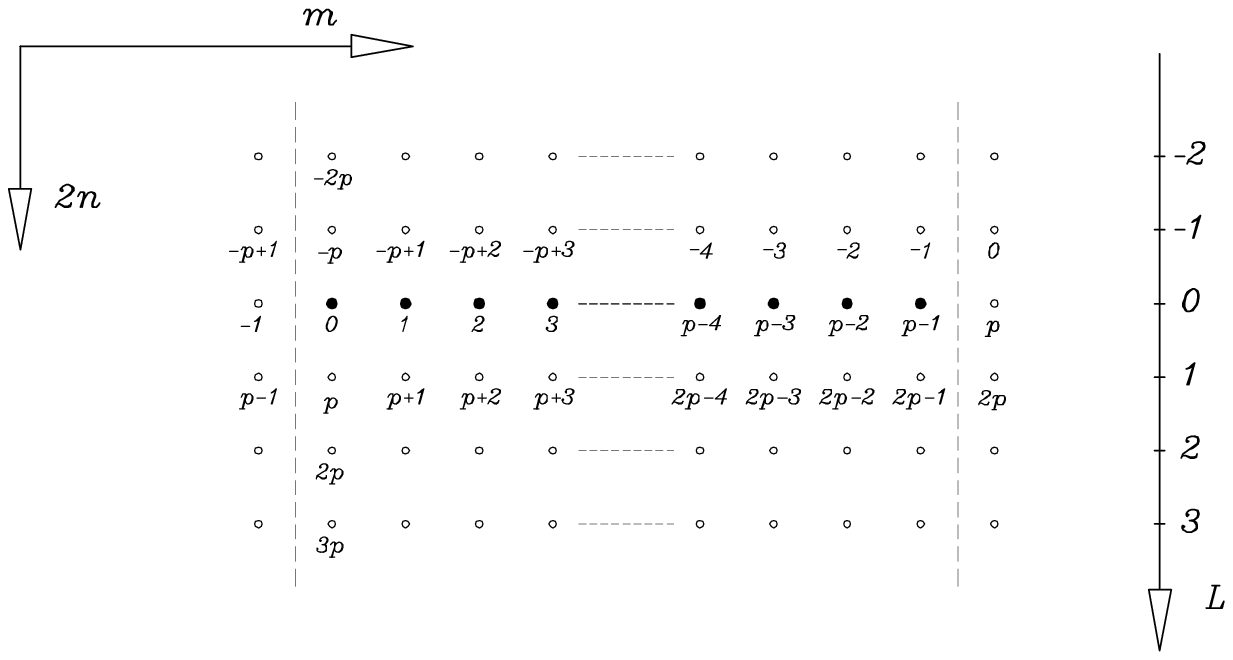}{Charge structure for $k=2p$. Only even values are considered
on the $n$ axis}{figcharges}

For odd $p$ all the structure is similar but with bricks of dimension
$p\times 2q$. Take again a generic charge Q
\be
Q=\frac{p\ n}{2q}+m=\frac{p\ (n+2q\ n')}{2q}+m-p\ n'
\lb{Qpp}
\ee
Because $p$ is odd, equal-valued charges correspond to pairs related by
$(m,n)\rightarrow(m-pn',n+2qn')$. The slope of the brick is the same
$k/4$ but they are twice as large in the $n$ direction. Everything else
works in the same way.

\section{Lattices and The Heterotic String \lb{ch.tor:sec.lat}}

Consider now an $U(1)^D$ action of the type~(\r{STMGTN}) with gauge
fields obeying $\mathcal{B}$ boundary conditions (replacing $k$ by a
generic matrix $K_{IJ}$) with action
\be
S=\int_M d^2z\,dt\left[-\frac{\sqrt{-g_{(3)}}}{4\gamma}F_I^{\mu\nu}F^I_{\mu\nu}+
\frac{G_{IJ}+B_{IJ}}{8\pi}\epsilon^{\mu\nu\lambda}A^I_\mu\partial_\nu A^J_\lambda\right]
\lb{SK}
\ee
$G_{IJ}$ and $B_{IJ}$ stand for the symmetric and antisymmetric parts
of $K_{IJ}$.

Using the same procedure outlined in section~\r{ch.quant:sec.path}
the effect of monopole-instantons is computed to induce a change of
the charge
\be
\Delta Q_I=\frac{G_{IJ}}{2}N^J
\lb{DDQQ}
\ee
Using the Schr\"{o}dinger picture the charge spectrum is computed to be
\be
Q_I=m_I+\frac{K_{IJ}}{4}n^J
\ee
Note that the antisymmetric part $B_{IJ}$ is present in the charge but
not in the monopole effects. This is due to the fact that it is
manifested only at the level of the boundary. Also in the action
the term corresponding to $B_{IJ}$ is a total derivative and can be
completely integrated out to the boundary
\be
\int_{M}B_{IJ}\epsilon^{\mu\nu\lambda}A^I_\mu\partial_\nu A^J_\lambda=\frac{1}{2}\left[-\int_{\Sigma_0}B_{IJ}\tilde{\epsilon}^{ij}A_i^IA_j^J+\int_{\Sigma_1}B_{IJ}\tilde{\epsilon}^{ij}A_i^IA_j^J\right]
\ee
Let us analyze first the case for $B_{IJ}=0$. The previous discussion
of section~\r{ch.tor:sec.bc} follows in the same fashion. Choosing
$CC_\perp$ or $CC_{\small//}$ boundary conditions, the desired
relative spectrum in each boundary is obtained, that is,
$\vb{Q}=\vb{m}+K\vb{n}/4$ and $\bar{\vb{Q}}=\vb{m}-K\vb{n}/4$. This
means that every monopole contribution of the form~(\r{DDQQ}) has
exactly $N^J=-n^J$.

What about if $B_{IJ}\neq 0$?  Returning to a more careful analysis of
the boundary identifications, for $CC_\perp$ the measures of the
integrals in opposite boundaries change their relative sign, say
$\int_1d^2z\rightarrow-\int_1d^2z$, due to the $2D$ measure changing
sign. The fields in one boundary (say $\Sigma_1$) are swaped,
$A_i\leftrightarrow A_j$. Note that neither $B_{IJ}$ nor
$\tilde{\epsilon}^{ij}$ in the right integral are being changed, they
are induced from the bulk and do not change by these kinds of boundary
identifications.  Thus this transformation has no effect in the
$B_{IJ}$ term
\be
\int_{\Sigma_0}B_{IJ}\epsilon^{ij}A_i^IA_j^J\rightarrow \int_{\Sigma_0}B_{IJ}\epsilon^{ij}A_i^IA_j^J
\ee
For $CC_{\small//}$ nothing changes either.  This means that the
$B_{IJ}$ term does not change sign under any of our boundary
identifications.

So the left/right spectrum obtained is
\be
\ba{c}
\displaystyle Q_I=\frac{G_{IJ}+B_{IJ}}{4}n^J+m_I\vspace{.2cm}\\
\displaystyle \bar{Q}_I=\frac{-G_{IJ}+B_{IJ}}{4}n^J+m_I
\ea
\ee
such that the charge difference $Q_I-\bar{Q}_I=G_{IJ}\ n^J/2$
is indeed the monopole contribution.

Similarly the lattice obtained is of the form $l=(\vb{Q},\vb{\bar{Q}})$
with the Lorentzian product of signature $(\vb{+},\vb{-})$ defined as
\be
l\circ l'=2G^{(-1)IJ}(Q_I{Q'}_J-\bar{Q}_I{\bar{Q}'}_J)
\lb{lorprod2}
\ee
where $G^{-1}$ stands for the inverse of $G_{IJ}$.
The signature of the product has $D$ plus and $D$ minus signs.
The properties of this lattice are the same as the ones for the
previous $D=1$ case and follow in a similar way from the bulk theory
as presented in section~\r{ch.tor:sec.bulk}. The lattice is integer, even and 
self-dual.

It is now time to analyze which lattices do exist for $CN$ boundary
conditions. As explained before in section~\r{ch.tor:sec.bc} they can
be obtained by truncating the lattices where $N^J=-n^J$.  Imposing
$CN$ is then equivalent to truncating the lattice by choosing
$\vb{\bar{Q}}=0$. Similarly to~(\r{QN}), this means
elements of the lattice are selected such that
\be
m_I=\frac{G_{IJ}-B_{IJ}}{4}n^J
\lb{QQN2}
\ee

This sublattice has elements $l=(\vb{Q},0)$ with $\vb{Q}$ built out of
$\vb{n}$ and $\vb{m}$ obeying~(\r{QQN2}).  Once $\bar{\vb{Q}}=0$ the
Lorentzian product~(\r{lorprod2}) becomes, for this particular
sublattice, simply $2G^{IJ} Q_I Q_J$. So it becomes Euclidean for this
particular choice of elements. But the properties of being even,
integer and self dual are inherited from the full lattice.  It is a
known fact that the only even integer self dual Euclidean lattices are
of dimension $0\mod 8$. Of dimension $8$ there is only one,
$\Gamma^8$, the root lattice of $E_8$.  From~(\r{QQN2})
the spectrum of allowed $\vb{Q}$'s is computed to be
\be
Q_I=\frac{G_{IJ}}{2}n^J
\lb{QN2}
\ee
Note that the resulting spectrum is independent of $B_{IJ}$.
Due to the spectrum~(\r{QN2}) and by closure of the lattice under the
product~(\r{lorprod2}) $\vb{n}$ is any vector with integer entries,
therefore it cannot be constraint.

Anyhow, as stated before, the charge lattice must be the root lattice
of $E_8$, and it is clear that the only choice that leads to this result is
$G=2C$. $C$ is the standard Cartan matrix of this group (the diagonal
elements have the value $2$ and the off diagonal $-1$)
\be
C=\left[\ba{cccccccc}
2&-1&0&0&0&0&0&0\\
-1&2&-1&0&0&0&0&0\\
0&-1&2&-1&0&0&0&0\\
0&0&-1&2&-1&0&0&0\\
0&0&0&-1&2&-1&0&-1\\
0&0&0&0&-1&2&-1&0\\
0&0&0&0&0&-1&2&0\\
0&0&0&0&-1&0&0&2\\
\ea\right]
\ee

The
corresponding Dynkin diagram is pictured in figure~\r{fige8}.
Returning to~(\r{QQN2}), it must be a realizable condition for every
$\vb{n}$, thus there is no choice but to impose $G_{IJ}-B_{IJ}=0\mod
4$ for $J>I$.

The only possibilities left are
\be
\ba{rclc}
G_{IJ}&=&2C_{IJ}& \vspace{.2 cm}\\
B_{IJ}&=&(2+4 r)C_{IJ}&J>I,r\in\mathbb{Z}
\ea
\ee

\fig{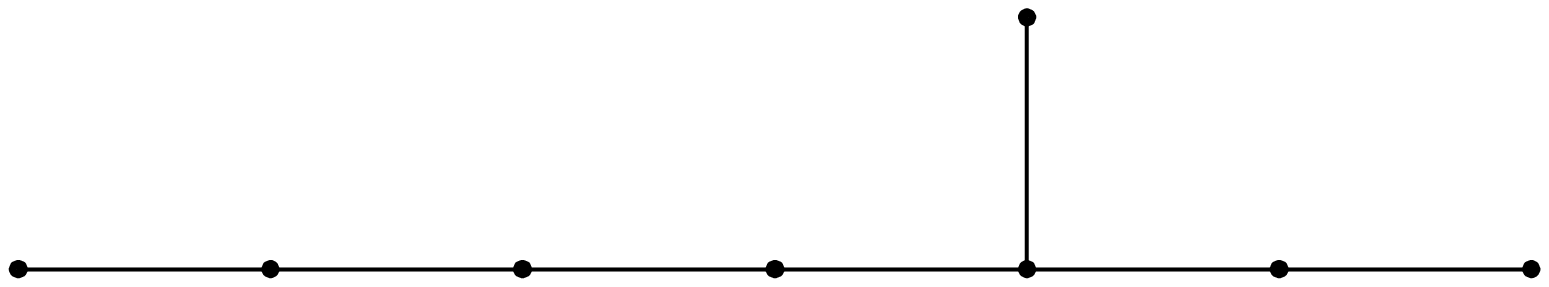}{Dynkin Diagram of $E_8$}{fige8}

There are two lattices of dimension $16$ obeying these properties:
$\Gamma^8\times\Gamma^8$ with elements in the root lattice of
$E_8\times E_8$ and $\Gamma^{16}$. The well known results of string
theory are retrieved. Note that in the first works of the non-interacting
Heterotic string~\cite{GHMR_1} there is no mention of the tensor
$B_{IJ}$ (the free spectrum depends solely on $G_{IJ}$). Only after
the introduction of interactions and an effective action for the
theory~\cite{GHMR_2} does it becomes clear that there is a need for
$B_{IJ}$.  For $r=0$ the result presented in~\cite{R_1} and references
therein is obtained up to the normalization of $1/4$ on $K_{IJ}$.

\chapter{Open and Unoriented Strings}
\lb{ch.opun}

This Chapter is based on the original work of the author and
Ian Kogan~\cite{TM_16}.

Although originally (and historically) open string theories were
considered as theories by themselves, it soon become evident that,
whenever they are present, they come along with closed (non-chiral)
strings. Moreover open string theories are obtained from closed string
theories by gauging certain discrete symmetries of the closed theory
(see~\cite{SBH_00} and references therein for a discussion of this
topic). The way to get open strings from closed strings is by gauging
the world-sheet parity~\cite{SBH_00,SBH_01,SBH_02},
$\Omega:z\rightarrow-\bz$. That is, the identification
$\sigma_1\cong-\sigma_1$ is imposed, where $z=\sigma_1+i\sigma_2$ and
$\bz=\sigma_1-i\sigma_2$ is the complex structure of the world-sheet
manifold.  The spaces obtained in this way can be of two types: closed
unoriented and open oriented (and unoriented as well). These last ones
are generally called orbifolds and the singular points of the
construction become boundaries.  The states (operators and fields of
the theory in general) of the open/unoriented theory are obtained from
the closed oriented theory by projecting out the ones which have
negative eigenvalues of the parity operator. This is obtained by
building a suitable projection operator $(1+\Omega)/2$ such that only
the states of positive eigenvalue are kept in the theory. Namely the
identification $X^{I}(z,\bz)\cong X^{I}(\bz,z)$ or $X^{I}_{L}(z)\cong
X^{I}_{R}(\bz)$ (in terms of the holomorphic and antiholomorphic parts
of $X=X_L+X_R$) holds.

Another construction in string theory is orbifolding the target space
of the theory under an involution of some symmetry of that space.  In
this work only a $Z_2$ involution is going to be considered, imposing
the identification $X^{I}\cong-X^{I}$, where $X^{I}$ are the target
space coordinates.  When combining both constructions, world-sheet and
target space orbifolding, open/unoriented theories on
orbifolds~\cite{SBH_03,SBH_04,SBH_05,SBH_05a} or orientifolds
($X^{I}(z,\bz)=-X^{I}(\bz,z)$) are obtained, implying the existence of twisted
sectors in the open/unoriented theories.

Further to the previous discussion
both sectors (twisted and untwisted) need to be
present for each surface in order to ensure modular invariance of the
full partition function~\cite{SBH_00,SBH_06,SBH_07}.
One point that must be stressed is that twisting in open strings can, for
the case of a $Z_2$ target space orbifold, be simply interpreted as
the choice of boundary conditions: Neumann or Dirichlet.

Toroidal compactification is an important construction in string
theories and in the web of target space dualities.  Early works
considered also open string constructions in these toroidal
backgrounds~\cite{SBH_07,SBH_08}.  In these cases some target space
coordinates are compactified, say
$X^{J}(z+2\pi i,\bz-2\pi i)\cong X^{J}(z,\bz)+2\pi R$
($R$ is the radius of compactification of
$X^{J}$), and the twisted states in the theory are the ones corresponding
to the points identified under
$X^{J}(z+2\pi i,\bz-2\pi i)\cong-X^{J}(z,\bz)+2\pi R$,
or in terms of the holomorphic and
antiholomorphic parts of $X$,
$X_L^{I}(z)\cong-X_R^{I}(\bz)$.

An important result coming from these constructions is that the gauge
group of the open theory, the Chan-Paton degrees of freedom carried by 
the target space photon Wilson lines (only present in open theories), is
constrained, due both to dualities of open string theory~\cite{SBH_07}
and to modular invariance of open and unoriented
theories~\cite{SBH_07,SBH_08,SBH_09,SBH_10}. This will result in the
choice of the correct gauge group that cancels the anomalies in the theory.

One fundamental ingredient of string theory is modular invariance.
Although for bosonic string theory the constraints coming from genus 1
amplitudes are enough to ensure modular invariance at generic genus
$g$, it becomes clear that once the fermionic sector of superstring
theory is considered it is necessary to consider genus $2$ amplitude
constraints.  For closed strings (types II and 0) the modular group at
genus $g$ is $PSL(2g,{\mathbb{Z}})$ and the constraints imposed by
modular invariance at $g=2$ induce several possible projections in the
state space of the theory~\cite{MI_01,MI_02,MI_03,MI_04,MI_05} such
that the resulting string theories are consistent. Among them are the
well known GSO projections~\cite{GSO} that insure the correct
spin-statistics connection, project out the tachyon and ensure a
supersymmetric effective theory in the $10D$ target space.

Once an open superstring theory (type I) is
\textit{created} by orbifolding the world-sheet parities,
for each open (and/or unoriented) surface
a Relative Modular Group still survives the orbifold
at each genus $g$~\cite{SBH_06}.
Again in a similar way to the closed theory the modular invariance under
these groups will result in generalized
GSO projections~\cite{SBH_06,SBH_06a,SBH_06b,SBH_06c}.

For a more recent overview of the previous topics
see~\cite{SBH_11,SBH_12} (see also~\cite{POL_1} for an extensive
explanation of them).

The purpose of this Chapter is to build open, open unoriented, and closed
unoriented string theories (with and without orbifolding of the target
space) from the Topological Membrane.

Closed string theories are obtained as the effective boundary theory,
their world-sheet is the closed boundary $\partial M$.  Obtaining open
string theory raises a problem because it is necessary to have an open
world-sheet to define them. But the boundary of a boundary is null,
$\partial\partial M=\emptyset$. So naively it seems that TM cannot
describe open strings since world-sheets are already a boundary of a
$3D$ manifold. The way out is to consider orbifolding of the bulk
theory. In this way the fixed points of the orbifold play the role of
the boundary of the $2D$ boundary of the $3D$ membrane. This proposal
was first introduced by Horava~\cite{H_1} in the context of pure
Chern-Simons theories. His results are going to be extended to TMGT
and the orbifold group reinterpreted as the discrete symmetries of
the full gauge theory.

Other works have developed Horava's idea. For a recent study on
WZWN orbifold constructions see~\cite{FS_1} (and references therein)
For an extensive study, although in a more formal way than our
work, of generic Rational Conformal Field Theories (RCFT)
with boundaries from pure $3D$ Chern-Simons theory
see~\cite{FS_2} (and references therein).
Nevertheless in these previous works the monopole
processes were not studied. These are crucial for describing the
winding modes and T-duality in compact RCFT from the TM point of view
and, therefore, in compactified string theories.

An orbifold of the Topological Membrane (and Topologically Massive
Gauge Theory) [TM(GT)] is considered such that one new boundary is
created at the orbifold fixed point. To do this the discrete
symmetries of the $3D$ theory are gauged, namely $PT$ and
$PCT$. Several $P$'s are going to be defined as generalized parity
operations. $C$ and $T$ are the usual $3D$ QFT charge conjugation and
time inversion operations (see~\cite{GD_1} for a review). The
orbifolding of the string target space corresponds in pure
Chern-Simons membrane theory to the quotient of the gauge group by a
$Z_2$ symmetry~\cite{MS_1}.  As will be shown, in the full TM(GT), the
discrete symmetry which will be crucial in this construction is charge
conjugation $C$. Besides selecting between twisted and untwisted
sectors in closed unoriented string theory it will also be responsible
for setting Neumann and Dirichlet boundary conditions in open string
theory.  More generic orbifold groups will not be considered.

There are two main new ideas introduced here. Firstly the use
of all possible realizations of $P$, $C$ and $T$ combinations, which
constitute discrete symmetries of the theory, as the orbifold
group. Although the mechanism is similar to the one previously studied
by Horava for pure Chern-Simons theory, the presence of the Maxwell
term constrains the possible symmetries to $PT$ and $PCT$ type
only. Also the interpretation of the orbifold group as the discrete
symmetries in the quantum theory is new, as is the interpretation of
charge conjugation $C$ which selects between Neumann and Dirichlet
boundary conditions. This symmetry explains the T-duality of open
strings in the TM framework. It is a symmetry of the $3D$ bulk which
exchanges trivial topological configurations (without monopoles) with
nontrivial topological configurations (with monopoles). In terms of
the effective boundary CFT (string theory) this means exchanging
Kaluza-Klein modes (no monopole effects in the bulk) with winding
number (monopole effects in the bulk).

In section~\r{ch.opun:sec.riemman} we introduce the Riemann surfaces
of genus 0 (the sphere), and genus 1 (the torus), and their possible
orbifolds under discrete symmetries which are identified with generalized
parities $P$.

Section~\r{ch.opun:sec.cft.bc} gives an account of Neumann and Dirichlet
boundary conditions in usual CFT using the Cardy method~\cite{CD_1}
of relating $n$ point full correlation functions in boundary
Conformal Field Theory with $2n$ chiral correlation functions in the
theory without boundaries.

Then, in section~\r{ch.opun:sec.tmgt} a brief overview of the discrete
symmetries of $3D$ QFT is given. They are used to orbifold TM(GT) and
the $3D$ configurations compatible with several orbifolds are
considered, both at the level of the field configurations and of the
particular charge spectra corresponding to the resulting theories.  It
naturally emerges from the $3D$ membrane that the configurations
compatible with $PCT$ correspond to Neumann boundary conditions (for
open strings) and to untwisted sectors (for closed unoriented).  The
configurations compatible with $PT$ correspond to Dirichlet boundary
conditions (for open strings) and twisted sectors (for closed
unoriented). The genus 2 constraints are discussed here although a
more detailed treatment is postponed for future work.  Further it is
shown that Neumann (untwisted) corresponds to the absence of monopole
induced processes while for Dirichlet (twisted) these processes play a
fundamental role. A short discussion on T-duality shows that it has
the same bulk meaning as modular invariance, in that both of them
exchange $PT\lra PCT$.

\newpage
\section{\lb{ch.opun:sec.riemman}Riemann Surfaces:\\ from Closed Oriented to
Open and Unoriented}
Any open or unoriented manifold $\Sigma_u$ can, in general, be obtained
from some closed orientable manifolds  $\Sigma$ under identification
of a $Z_2$ (or at most two $Z_2$) involution(s)
\be
\ba{cccl}
\pi:&\Sigma&\rightarrow&\Sigma_u=\Sigma/Z_2\\
    &(x,-x)&\rightarrow&x
\ea
\lb{proj}
\ee
such that each point in $\Sigma_u$ has exactly two corresponding
points in $\Sigma$ conjugate in relation to the $Z_2$ involution(s).
The pair $(x,-x)$ in the last equation is symbolic, where the second
element stands for the action of the group $Z_2$, $z_2(x)=-x$,
on the manifold. Usually this operation is closely related to
parity as will be explained bellow.
Although in this work the perspective adopted is that the starting
point is a full closed oriented theory and the open/unoriented theory
is obtained by orbifolding it, there is the 
reverse way of explaining things. This means that any theory
defined on an open/unoriented manifold is equivalently defined
on the closed/oriented manifold which \textit{doubles} (consists of
two copies of) the original open/unoriented.

Let us summarize how to obtain the disk $D_2$ (open orientable) and
projective plane $RP_2$ (closed unorientable) from the sphere $S^2$ and
the annulus $C_2$ (open orientable), and the M\"{o}bius Strip (open
unorientable) and Klein bottle $K_2$ (closed unorientable)
from the torus $T^2$.

\subsection{The Projective Plane and the Disk obtained from the Sphere}

For simplicity we work in complex stereographic coordinates
\mbox{$(z=x_1+ix_2$}, $\bz=x_1-ix_2)$ such that the sphere is identified with
the full complex plane.  The sphere has no moduli and the Conformal
Killing Group (CKG) is $PSL(2,\mathbb{C})$.  A generic element of this
group is $(a,b,c,d)$ with the restriction $ad-bc=1$. It acts on a
point $z$ as
\be
z'=\frac{a z+b}{c z+d}
\lb{PSLC}
\ee

It has then six real parameters, that is, six generators. Thus
the sphere has six Conformal Killing Vectors (CKV's).
It is necessary to use two coordinate charts to cover the full sphere, 
one including the north pole and the other one including the south
pole. Usually it is enough to analyze the theory defined on the sphere
only for one of the patches but it is necessary to check that the
transformation between the two charts is well defined. In
stereographic complex coordinates the map between the two charts
(with coordinates $z,\bz$ and $u,\bar{u}$) is given by $z\to1/u$ and
$\bz\to 1/\bar{u}$.

The {\bf disk $D_2$} can be obtained from the sphere under the identification
\be
z\cong\bz
\lb{D2_id}
\ee
This result is graphically pictured in figure~\r{fig_D2} and consists
in the involution of the manifold $S^2$ by the group $Z^{P_1}$,
$D_2=S^2/Z^{P_1}$.  There is one boundary corresponding to the real
line in the complex plane and the disk is identified with the upper
half complex plane.
\fig{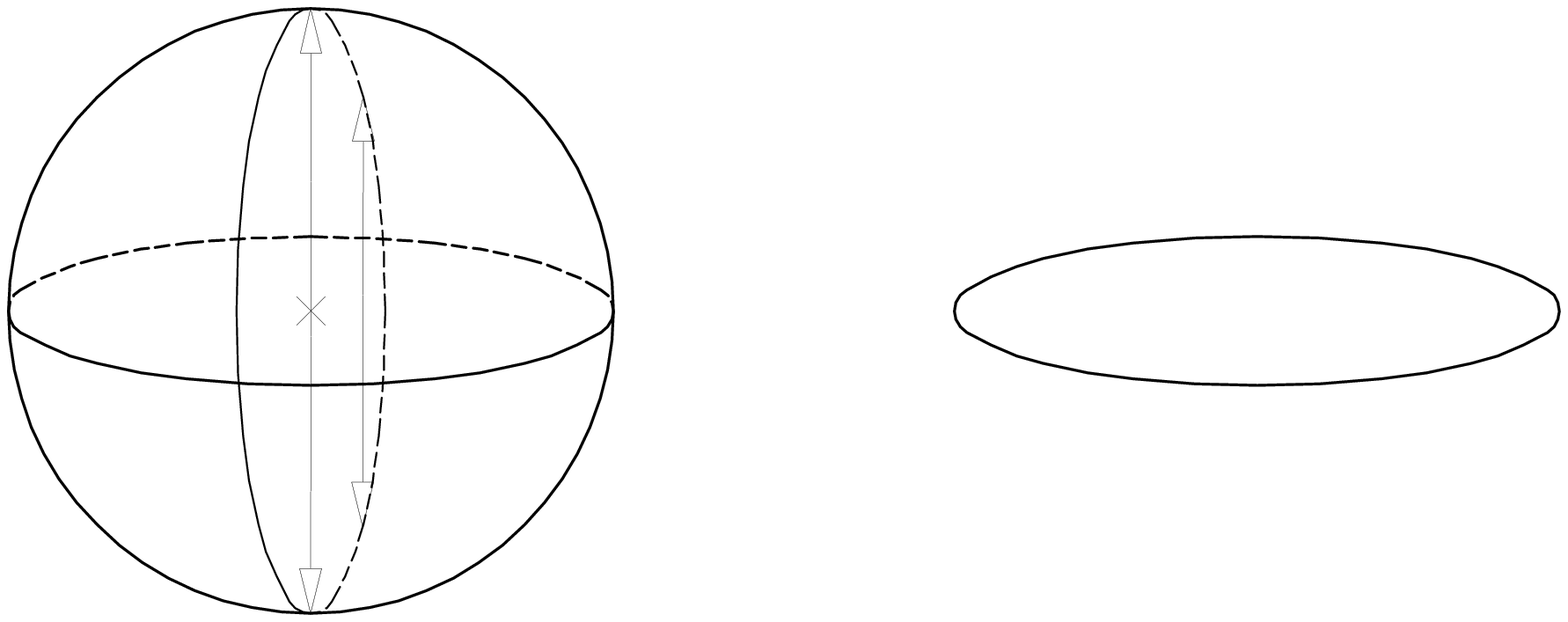}{The disk $D_2=S^2/P_1$ generated from the sphere under
the involution generated by the parity operation $P_1$.}{fig_D2}

It is straightforward to see that the non trivial element of
$Z^{P_1}$ is nothing else but the usual $2D$ parity transformation
\be
\ba{cccc}
P_1:&z      &\rightarrow&\bz\vspace{.1 cm}\\
       &\bz    &\rightarrow&z
\ea
\lb{P}
\ee
The CKG of the disk is the subgroup of $PSL(2,\mathbb{C})$ which
respects the constraint (\r{D2_id}), that is $PSL(2,\mathbb{R})$.

From the point of view of the fields defined on the sphere this
corresponds to the usual $2D$ parity transformation. In order that the
theory be well defined in the orbifolded sphere it is necessary to
demand that the fields of the theory be compatible with the
construction
\be
\ba{rcl}
f(z)&=&f(P_1(z))\vspace{.1 cm}\\
\phi_i[x_j]&=&P_1\phi_i[P_1(x_j)]
\ea
\lb{Of}
\ee
where the first equation applies to scalar fields and the second to
vectorial ones. For tensors of generic dimensions $d$ (e.g. the metric
or the antisymmetric tensor) the transformation is easily generalized
to be $T(x)=P_1^dT(P_1(x))$.

In order to orbifold the theory defined on the sphere it is enough
to introduce the projection operator
\be
P_{1,\mathrm{proj}}=\frac{1}{2}(1+P_1)
\ee
which projects out every operator with odd parity eigenvalue and keeps
in the theory only field configurations which are compatible with the
$Z_2$ involution.

To obtain the {\bf projective plane $RP_2$} it is necessary to make the
identification
\be
z\cong -\frac{1}{\bz}
\lb{sphere_id}
\ee
This result is graphically pictured in figure~\r{fig_RP2} and again is
an involution of the sphere $RP_2=S^2/Z_2^{P_2}$.  The resulting space
has no boundary and no singular points. But it is now an unoriented
manifold.
\fig{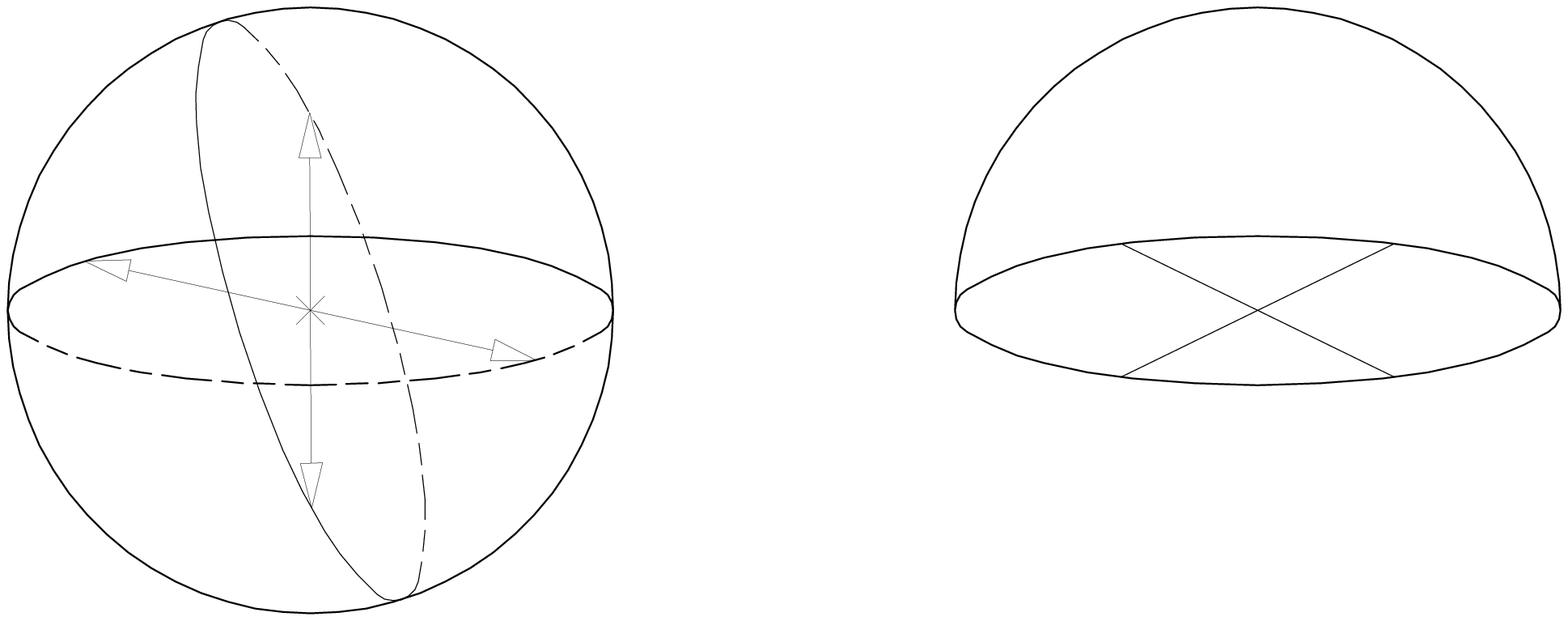}{The projective plane $RP_2=S^2/P_2$ obtained from the sphere
under the involution generated by the parity operation $P_2$.}{fig_RP2}

This identification can be thought of as two operations. The action of
the element $\alpha=(0,-1,1,0)\in Z_2^\alpha\subset SL(2,{\mathbb{C}})$
followed by the operation of parity as given by~(\r{P}). Note that
$\alpha(z)=-1/z$ but $P_1\alpha(z)=-1/\bz$ as
desired.  In this case one can define a new parity operation
$P_2\in Z_2^{P_2}=Z_2^{P_1}\times Z_2^\alpha$ as
\be
\ba{rrcl}
P_2:&z  &\rightarrow&\displaystyle-\frac{1}{\bz}\vspace{.1 cm}\\
  &\bz&\rightarrow&\displaystyle-\frac{1}{z}
\ea
\lb{PP}
\ee
From the point of view of the fields defined on the sphere one can
use the usual parity transformation since any theory defined on the
sphere should be already invariant under transformation~(\r{PSLC})
such that $PSL(2,\mathbb{C})$ is a symmetry of the theory.
But in order to have a more transparent picture
let us use the definition~(\r{PP}) of $P_2$ and demand that
\be
\ba{rcl}
f(z)&=&f(P_2(z))\\
\phi_i[x_j]&=&P_2\left(\,\phi_i[P_2(x_j)]\,\right)
\ea
\lb{Pf}
\ee
where the first equation concerns scalar fields and the second
vectorial ones. For tensors of generic dimensions $d$ (such as the metric
or the antisymmetric tensor) the transformation is again easily
generalized to be $T(x)=P_2^dT(P_2(x))$.

The CKG is now $SO(3)$, the usual rotation group. It is the subgroup of
$PSL(2,{\mathbb{C}})/Z_2^\alpha$ that respects the constraint~(\r{D2_id}).

\subsection{\lb{sec:torus_inv}The annulus, M\"{o}bius strip and Klein bottle from the Torus}

We proceed to the genus one closed orientable manifold, the torus.
It is obtained from the complex plane under the identifications
\be
z\cong z+2\pi\cong z+2\pi (\tau_1+i\tau_2)
\ee
There are two modular parameters $\tau=\tau_1+i\tau_2$ and two CKV's.
The action of the CKG, the translation group in the
complex plane, is
\be
z'=z+a+ib
\ee
with $a$ and $b$ real.
The metric is simply $ds^2=\left|dx^1+\tau dx^2\right|^2$ and the
identifications on the complex plane are invariant under the two
operations
\be
{\mathcal{T}}:\tau'=\tau+1\ \ \ \ \ \ \ \ \ \ \ {\mathcal{S}}:\tau'=-\frac{1}{\tau}
\ee
These operations generate the \textbf{modular group}
$PSL(2,{\mathbb{Z}})$. That is
\be
\tau'=\frac{a\tau+b}{c\tau+d}
\ee
with $a,b,c,d\in \mathbb{Z}$ and $ad-bc=1$.

The {\bf annulus $C_2$} (or its topological equivalent, the cylinder) is
obtained from the torus with $\tau=i\tau_2$ under the identification
\be
z\cong-\bz
\lb{C2_id}
\ee
This result is symbolically pictured in figure~\r{fig_C2}.
\fig{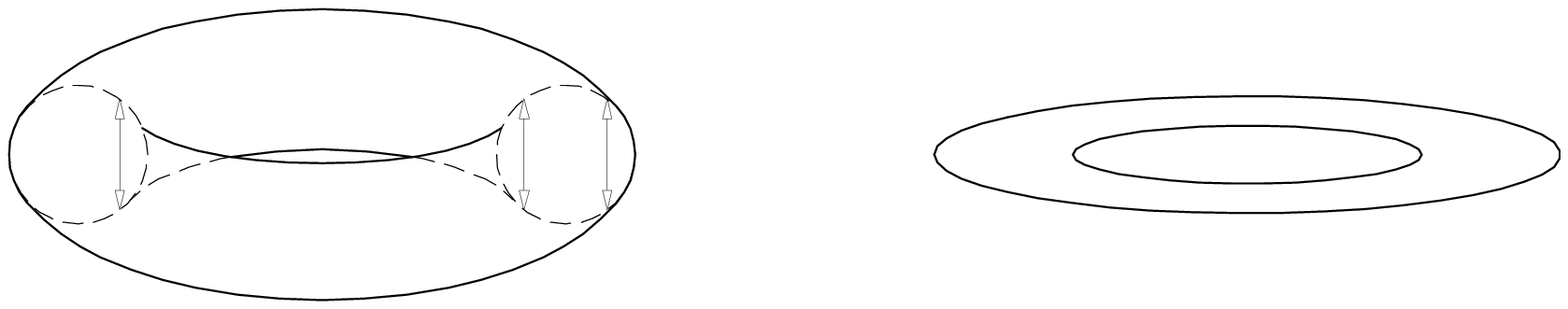}{The annulus (or cylinder) $C_2=T^2/\Omega$ obtained
from the torus under the involution generated by the parity
operation $\Omega$.}{fig_C2}

There is now one modular parameter $\tau_2$ and no modular
group. There is only one CKV being the CKG action given by $z'=z+ib$,
translation in the imaginary direction.
In terms of the fields defined on the torus this correspond to the
projection under the parity operation
\be
\ba{rrcl}
\Omega:&z      &\rightarrow&-\bz\vspace{.1 cm}\\
       &\bz    &\rightarrow&-z
\ea
\lb{omega}
\ee

The {\bf M\"{o}bius strip $M_2$} can be obtained from the annulus
(obtained from the torus with $\tau=2i\tau_2$) by the
identification under the element $\tilde{a}$~\cite{POL_1} of
the translation group
\be
\tilde{a}:z\to z+2\pi\left(\frac{1}{2}+i \tau_2\right)
\lb{M2_id}
\ee
Note that $\tilde{a}$ belongs to the translation group of the torus,
not of the disk, and that $\tilde{a}^2=1$. This construction
corresponds to two involutions, so the orbifold group is
generated by two $Z_2$'s,
$M_2=T^2/(Z_2^{\Omega}\subset\hspace{-12.9pt}{\times}Z_{2}^{\tilde{a}})$,
where $\subset\hspace{-12.9pt}{\times}$ stands
for the semidirect product of groups. Thus the ratio of areas between
the M\"{o}bius strip and the original torus is $1/4$ compared to the
$1/2$ of the remaining open/unoriented surfaces obtained from the
torus, due to the extra projection operator $(1+\tilde{a})/2$ taking
the annulus to the strip.

In terms of the fields living on the torus this identification can be thought 
of as the projection under a new discrete symmetry which here is
also called parity
\be
\tilde{\Omega}\equiv\tilde{a}\circ\Omega
\lb{omegat}
\ee
Although this operation does not seem to be a conventional parity
operation note that, applying it twice to some point the same point is
retrieved, $\tilde{\Omega}^2=1$. It is in this sense a generalized
parity operation.

\figh{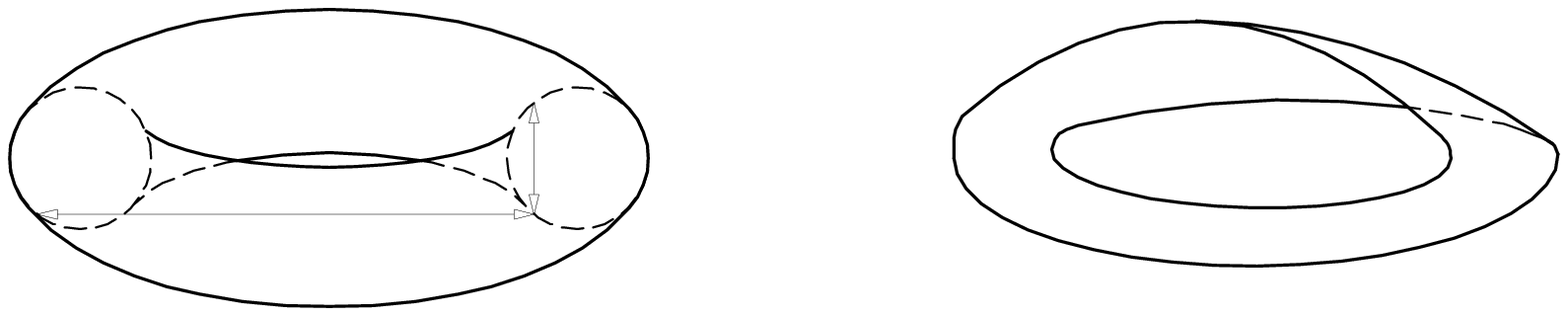}{The M\"{o}bius strip $M_2=T^2/\tilde{\Omega}$ obtained
from the torus under the involution generated by the parity
operation $\tilde{\Omega}$.}{fig.M2}

The previous construction is presented, for example, in Polchinski's
book~\cite{POL_1}. Note however that
one can build the M\"{o}bius strip directly from a torus~\cite{SBH_00}
with modulus $\tau=1/2+i\tau_2$ under the involution $\Omega$ given
in~(\r{omega}).
In this case the ratio of areas between the original torus and the
involuted surface is $1/2$. As will be shown later
both constructions correspond to the same region on the
complex plane. The first one results from two involutions
of a torus ($\tau=2i\tau$) with twice the area
of the second construction ($\tau=i\tau$). In this
sense both constructions are equivalent.
The M\"{o}bius strip orbifolding is pictured in figure~\r{fig.M2}.

Again there is one modular parameter $\tau_2$ and no modular
group. The only CKV is again the translation in the imaginary direction.

The {\bf Klein bottle $K_2$} is obtained from the torus with $\tau=2i \tau_2$ under the identification
\be
z\cong-\bz+2\pi i \tau_2
\lb{torusid}
\ee
This result is pictured in figure~\r{fig.K2}.
\fig{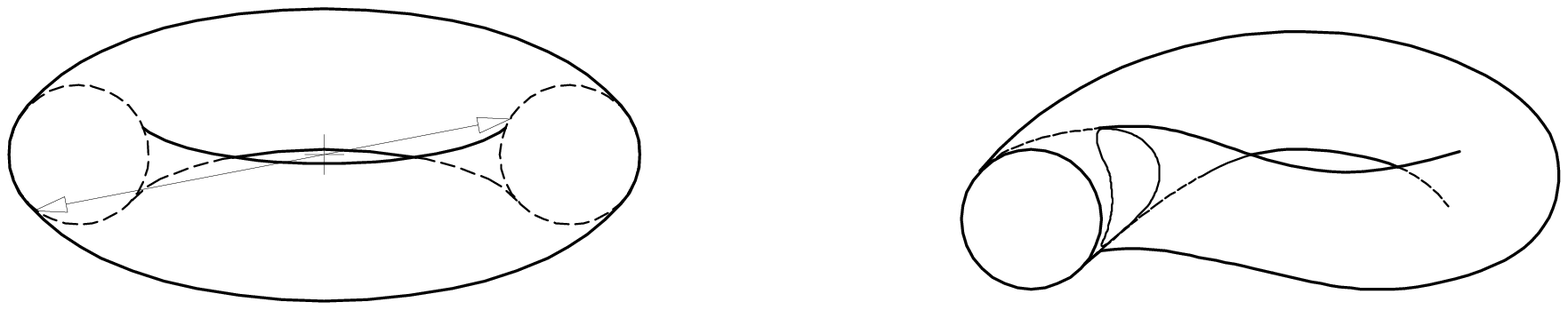}{The Klein bottle $K_2=T^2/\Omega'$ obtained from the
torus under the involution generated by the parity operation
$\Omega'$.}{fig.K2} The bottle is the involution of the torus
$K_2=T^2/Z^{\Omega'}_2$, which has a one parameter CKG with one CKV,
translations in the imaginary direction. There is one modulus $\tau_2$
and no modular group.  The resulting manifold has no boundary and no
singular points but is unoriented.  Again a new parity transformation
$\Omega'$ can be defined
\be
\ba{cccc}
\Omega':&z  &\rightarrow&\displaystyle-\bz+2\pi i\tau_2\vspace{.1 cm}\\
          &\bz&\rightarrow&\displaystyle -z+2\pi i\tau_2
\ea
\lb{omega'}
\ee
\bt{|cccc|cccc|}
\hline$\vb{S_2}$&$P_1$  &$P_2$          &\ \ \ \ &$\vb{T_2}$ &$\Omega$  &$\tilde{\Omega}$                       &$\Omega'$\\\hline
&&&&&&&\\
        &$z\lra\bz$&$z\lra-1/\bz$&        &        &$z\lra-\bz$&$\tilde{a}\circ\Omega$&$z\to-\bz+2\pi i\tau_2$\\
        &          &             &        &        &           & &$\bz\to-z+2\pi i\tau_2$\\
&&&&&&&\\
$S^2/P=$&$D_2$     &$RP_2$       &        &$T^2/P=$&$C_2$      &$M_2$                      &$K_2$\\
C/O&O/O     &C/U          &        &C/O     &O/O        &O/U&C/U\\
&&&&&&&\\\hline
\et{\lb{tabpar}Parity operations for the topology $T^2\times[0,1]$.
The torus geometry considered is $\tau=i\tau_2$ for $\Omega$ and
$\tau=2 i\tau_2$ for $\tilde{\Omega}$ and $\Omega'$. Note that $M_2$
can also be obtained from the torus with $\tau=1/2+i\tau_2$
by considering the parity $\Omega$.
In the labels of the last line the first letter stands for {\bf O}pen
or {\bf C}losed surface while the second letter stands for
\textbf{O}riented or \textbf{U}noriented.}

The various parity operations just studied are presented in
table~\r{tabpar} together with the resulting involutions (or orbifolds).

\section{\lb{ch.opun:sec.cft.bc}Conformal Field Theory -\\ Correlation
Functions and Boundary Conditions on the Disk}

To study string theory it is necessary to know what the world-sheet
CFT is. In a closed string theory it is given by CFT on a closed
Riemann surface, the simplest of which is the sphere, or equivalently
the complex plane. To study open strings is necessary to study CFT on
open surfaces. As was shown by Cardy~\cite{CD_1}, n-point correlation
functions on a surface with a boundary are in one-to-one
correspondence with chiral $2n$ point correlation functions on the
double covering surface\footnote{One of the constructions presented to
obtain the M\"{o}bius strip uses the double involution under
$\tilde{\Omega}$.  In that case $n$ insertions on it correspond to
$4n$ on the original torus.} (for more details and references
see~\cite{FMS}).

The disk and the annulus will be studied, so that the number of charges
(vertex operators) are doubled by inserting charges $\pm q$ (vertex
operators with $\Delta=2q^2/k$) at the Parity conjugate points. Note
that the sign of the charges inserted depends on the type of boundary
conditions that one wants to impose but the conformal dimension of the
corresponding vertex operator is the same.

The 2, 3 and 4-point holomorphic correlation functions of
vertex operators for the free boson on a closed surface of genus $0$ are
\be
\ba{rcl}
<\phi(z_1)\phi(z_2)>&=&z_{12}^{-2\Delta}\vspace{.1 cm}\vspace{.1 cm}\\
<\phi(z_1)\phi(z_2)\phi(z_3)>&=&z_{12}^{-\Delta_1-\Delta_2+\Delta_3}z_{13}^{-\Delta_1+\Delta_2-\Delta_3}z_{23}^{\Delta_1-\Delta_2-\Delta_3}\vspace{.1 cm}\vspace{.1 cm}\\
<\phi(z_1)\phi(z_2)\phi(z_3)\phi(z_4)>&=&\prod_{i<j}z_{ij}^{2q_iq_j/k}
\ea
\lb{correl}
\ee
where in all the cases $\sum q_i=0$, otherwise they vanish, and $z_{ij}=|z_i-z_j|$.

The disk is taken to be the upper half complex plane. As explained before
it is obtained from the sphere (the full complex plane) by identifying
each point in the lower half complex plane with its complex conjugate in the
upper half complex plane. In terms of correlation functions
\be
\left<\phi_q(x,y)\right>_{D_2}=\left<\phi_q(z)\phi_{-q}(\bz)\right>_{S^2}
\ee
where $z=x+iy$ was used in the the first equation of~(\r{correl}),
$y$ is the distance to the real axis while $x$ is taken to be the
horizontal distance (parallel to the real axis) between vertex
insertions.

\subsection{Dirichlet Boundary Conditions}

As is going to be shown, when the mirror charges have opposite
sign the boundary conditions are Dirichlet.

The 2-point correlation function restricted to the upper half plane is
simply the expectation value
\be
\left<\phi_q(x,y)\right>=\frac{1}{(2y)^{2\Delta}}
\lb{1pD}
\ee

Insertion of vertex operators (except for the identity) on the boundary
is not compatible with the boundary conditions since the only charge
that can exist there is $q=0$ (since $q=-q=0$ on the boundary).
Taking the limit $y\to 0$ the expectation value~(\r{1pD}) blows up but
this should not worry us, since near the boundary the two charges
\textit{annihilate} each other. This phenomena is nothing else than
the physical counterpart of the operator fusion rules
$\phi_q(y)\phi_{-q}(-y)\to(2y)^{-2\Delta}\phi_0(y)$. That is,
$\left<\phi_0\right>_{\partial D_2}=\left<1\right>_{\partial D_2}$ on
the boundary of the disk.

3-point correlation functions cannot be used for the same reason, as one
of the insertions would need to lie on the boundary, but that would
mean $q_3=0$. The other two charges would have to be inserted symmetrically
with respect to the real axis and this would imply $q_1=-q_2$.
This reduces the 3-point correlator to a 2-point one in the full plane.

For 4-point vertex insertions consider $q_1$ and $q_3$ in the upper
half plane, $q_2$ (inserted symmetrically with respect to $q_1$) and
$q_4$ (inserted symmetrically with respect to $q_3$) in the lower half
plane. As pictured in figure~\r{fig.confd2}, the most generic
configuration is $q_1=-q_2=q$ and $q_3=-q_4=q'$. Taking
$z_2=\bz_1=-iy$ and $z_4=\bz_3=x-iy'$ the corresponding 2-point
correlators in the upper half plane are obtained
\be
\left<\phi_q(0,y)\phi_{q'}(x',y')\right>=\frac{1}{(2y)^{2\Delta}(2y')^{2\Delta'}}
\left(1-\frac{4yy'}{x^2+(y+y')^2}\right)^{\frac{2qq'}{k}}
\ee

Again note that boundary operators cannot be inserted without changing
the boundary conditions.  In the limit $x\to\infty$ both correlators
behave like
\be
\lim_{x\to\infty}\left<\phi(y_1)\phi(y_2)\right>= \frac{1}{(4y_1y_2)^{2\Delta}}
\ee

When the insertions approach the boundary the correlators go to infinity
independently of the value of $x$. This fact can be explained by the
kind of boundary conditions considered, since they are such that
when the fields approach the boundary they become infinitely correlated
independently of how far they are from each other.
Therefore these must correspond to Dirichlet boundary conditions,
as the fields are fixed along the boundary, furthermore as stated before
their expectation value is $\left<1\right>$. It doesn't matter
how far apart they are, they are always correlated on the boundary.
The tangential derivative to the boundary of the expectation value
$\partial_x\left.\left<\phi\right>\right|_{\partial D_2}=0$
also implies Dirichlet boundary conditions.

\subsection{Neumann Boundary Conditions}

For the case when the mirror charge has the same sign as the original
one, the boundary conditions will be Neumann. The expectation value for
the fields in the bulk vanishes since the 2-point function
$\left<\phi_q(z_1)\phi_q(z_2)\right>=0$ in the full plane.
Nevertheless, the non-zero 2-point correlation function on the boundary
can be directly evaluated
\be
\left<\phi_q(0)\phi_{-q}(x)\right>=\frac{1}{x^{2\Delta}}
\lb{2pN}
\ee
Note that contrary to the previous
discussion, concerning Dirichlet boundary conditions, in this case
$q\neq 0$ on the boundary since the mirror charges have the same sign
and the correlation function vanishes in the limit $x\to\infty$ indicating that
the boundary fields become uncorrelated.

The 3-point correlation function in the full plane must be considered
with one charge $-2q$ on the boundary and two other charges $q$
inserted symmetrically with respect to the real axis (see
figure~\r{fig.confd2}).
In the upper half plane this corresponds to one charge insertion on
the boundary and one in the bulk
\be
\left<\phi_{-2q}(0,0)\phi_q(x,y)\right>=\left(\frac{2y}{x^2+y^2}\right)^{2\Delta}
\lb{3pN}
\ee
Note that in the limit $y\to 0$ the fusion rules apply and~(\r{2pN})
is obtained with $\Delta$ replaced by $4\Delta$.

For the 2-point function in the upper half-plane it is necessary to
consider the 4-point correlation function in the full plane with
$q_1=q_2=-q_3=-q_4=q$, where $q_2$ is inserted symmetrically to $q_1$
with respect to the real axis, and $q_4$ to $q_3$. The bulk correlator
is obtained as
\be
\left<\phi_{q}(0,y)\phi_{-q}(x,y')\right>=\left(\frac{4yy'}{x^2(x^2+(y+y')^2)}\right)^{2\Delta}
\lb{4pN}
\ee
\fig{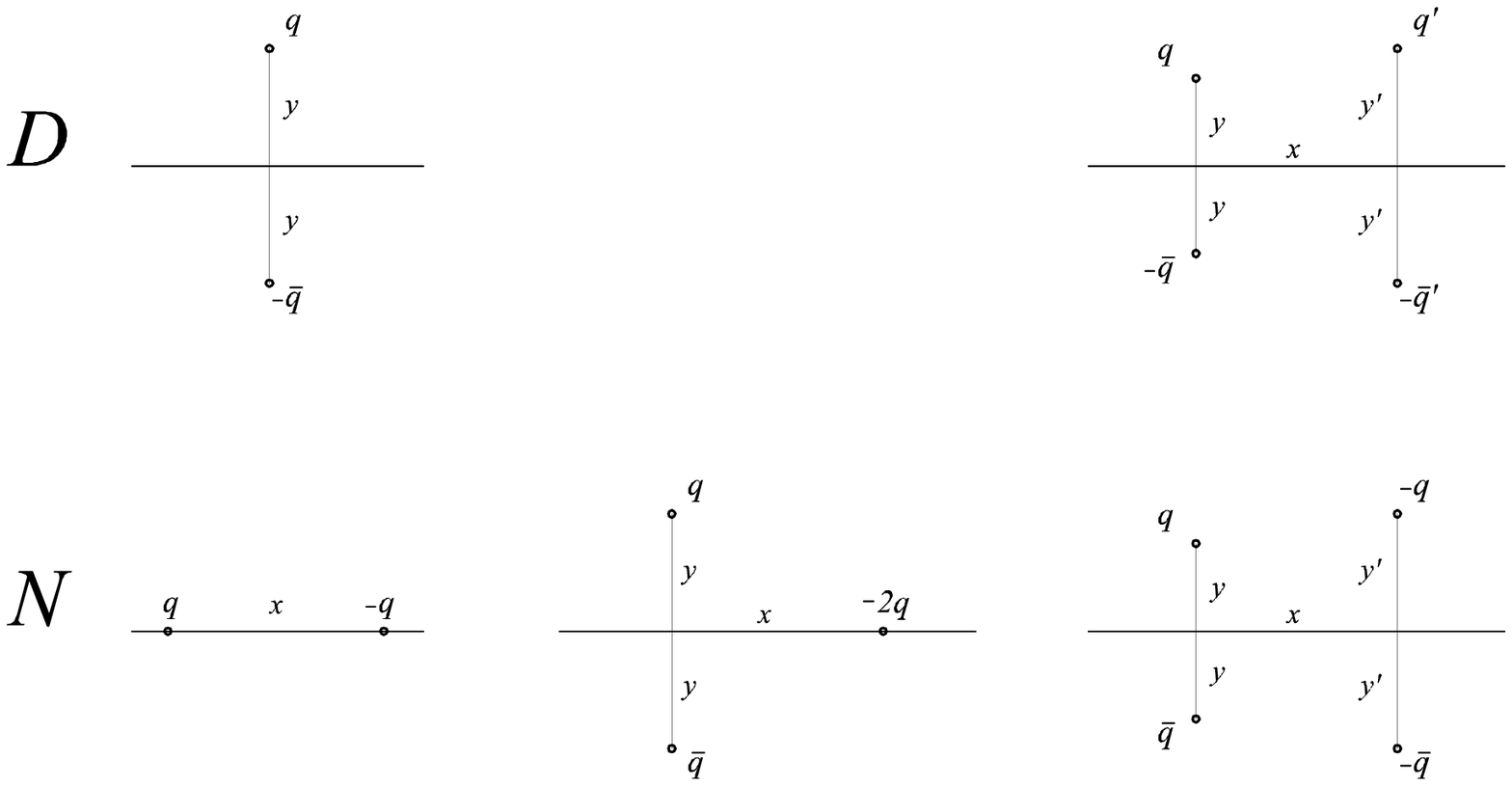}{The various possible analytical continuations of the disk (the
upper half plane) to the sphere (full plane) holding 2, 3 and 4-point
correlation functions for Neumann (N) and Dirichlet (D) boundary
conditions.}{fig.confd2}

Again in the limit $x\to\infty$ this correlator vanishes. This corresponds to
Neumann boundary conditions. The normal derivative to the boundary
of~(\r{4pN}) vanishes on the boundary
$\partial_y\left.\left<\phi(0)\phi(x)\right>\right|_{\partial D_2}=0$.

For the case of one compactified free boson the process follows in
quite a similar way. The main difference resides in the fact that the
right and left spectrum charges are different. Taking a charge
$q=m+kn/4$ its image charge is now $\pm\bar{q}$,
where $\bar{q}=m-kn/4$. In this way it is necessary to truncate the
spectrum holding $q=-\bar{q}=kn/4$ for Dirichlet boundary conditions, 
and $q=\bar{q}=m$ for Neumann boundary conditions, in a pretty much
similar way as happens in the Topological Membrane. The results
derived in this section are summarized in figure~\r{fig.confd2}.

\section{\lb{ch.opun:sec.tmgt}TM(GT)}

Is now time to turn to the $3D$ TM(GT). In this section the
results are derived directly from the bulk theory and its properties. The
derivations of the results presented here are in agreement with the CFT
arguments of the last section.

Initially only a single $U(1)$ of type $\mathcal{B}$ TMGT corresponding
to $c=1$ CFT with action given by~(\r{STMGT1}) will be considered
\be
S=\int_{M}dtd^2z\left[-\frac{\sqrt{-g}}{\gamma}F_{\mu\nu}F^{\mu\nu}+\frac{k}{8\pi}\epsilon^{\mu\nu\lambda}A_\mu F_{\nu\lambda}\right]
\ee
where, again, $M=\Sigma\times[0,1]$ has two
boundaries $\Sigma_0$ and $\Sigma_1$. $\Sigma$ is taken
to be a compact manifold, $t$ is in the interval $[0,1]$
and $(z,\bz)$ stand for complex coordinates on $\Sigma$.

As derived before upon quantization the charge spectrum is
\be
Q=m+\frac{k}{4}n
\ee
for some integers $m$ and $n$ as given by~(\r{charge}).  Furthermore
it has already been proved~\cite{TM_07,TM_15} (see the previous
chapter) that, for type $\mathcal{B}$ gauge groups and under the
correct relative boundary conditions, one insertion of $Q$ on one
boundary (corresponding to a vertex operator insertion on the boundary
CFT) will, necessarily, demand an insertion of the
charge~(\r{ccharge})
\be
\bar{Q}=m-\frac{k}{4}n
\ee
on the other boundary.
This fact will be assumed through out the rest of this chapter.

Our aim is to orbifold TM theory in a similar way to
Horava~\cite{H_1}, who obtained open boundary world-sheets through
this construction. A path integral approach will be taken and the
orbifold reinterpreted in terms of discrete $PT$ and $PCT$ symmetries
of the bulk $3D$ TM(GT).

\subsection{Horava Approach to Open World-Sheets\lb{ch.opun:sec.tmgt.horava}}

Obtaining open string theories out of $3D$ (topological)
gauge theories means building a theory in a manifold which has
boundaries (the $2D$ open string world-sheet) that is already a boundary
(of the $3D$ manifold). This construction raises a problem since
the boundary of a boundary is necessarily a null space.
One interesting way out of this dilemma is to orbifold the $3D$ theory,
such that its singular points work as the \textit{boundary} of the
$2D$ boundary.
Horava~\cite{H_1} introduced an orbifold group $G$ that combines the
world-sheet parity symmetry group $Z_2^{WS}$ ($2D$) with two elements
$\{1,\Omega\}$, together with a target symmetry
$\tilde{G}$ of the $3D$ theory fields
\be
G\subset \tilde{G}\times Z_2^{WS}
\ee
With this setup one can get three different kinds of
constructions. Elements of the kind $h=\tilde{h}\times 1_{Z_2^{WS}}$
induce twists on the target space (not acting in the world-sheet at
all), for elements $\omega=1_{\tilde{G}}\times\Omega$ the world-sheet
manifold is orbifolded (giving an open world-sheet) without affecting
the target space and for elements $g_1=\tilde{g}_1\times \Omega$
an \textit{exotic} world-sheet orbifold is obtained.  In this last case it
is further necessary to have an element corresponding to the twist in
the opposite direction $g_2=\tilde{g_2}\times\Omega$. To specify these
twists on some world-sheet it is necessary to define the monodromies
of fields on it. Taking the open string $C_o=C/Z_2$ as the orbifold of
the closed string $C=S^1$
\be
\pi_1(C_o)=D\equiv Z_2\ast Z_2\equiv Z_2\subset\hspace{-11.5pt}{\times}Z
\ee
with $\ast$ being the free product and $\subset\hspace{-11.5pt}{\times}$
the semidirect product of groups. $D$ is the infinite dihedral group,
the open string first homotopy group. So the monodromies of fields in
$C_o$ correspond to a representation of this group in the orbifold
group, $Z_2\ast Z_2\to G$, such that the following triangle is commutative
\be
\ba{lcr}
Z_2\ast Z_2&\longrightarrow&\ \ G\\
\hfill\searrow   &   &\swarrow\hfill\\
           &Z_2^{WS}&
\ea
\ee
The partition function contains the sum over all possible 
monodromies
\be
Z_C(\tau)=\frac{1}{|G|}\sum_{g_1,g_2,h}Z_C(g_1,g_2,h;\tau)
\ee
where $\tau$ is the modulus of the manifold. The monodromies
$g_1$, $g_2$ and $h$ are elements of $G$ as previously defined
satisfying $g_i^2=1$ and $[g_i,h]=1$.

It will be shown that $PCT$ plays the role of one of such symmetry
with $g_1=g_2$. It is in this sense one of the most simple cases of
\textit{exotic} world-sheet orbifolds.

The string amplitudes can be computed in two different pictures. The
loop-channel corresponds to loops of length $\tau$ of closed and open
strings and the amplitudes are computed as traces over the Hilbert
space. The tree-channel corresponds to a cylinder of length
$\tilde{\tau}$ created from and annihilated to the vacua through
boundary ($\left|B\right>$) and/or crosscaps ($\left|C\right>$) states.
Comparing both ways for the same amplitudes the following relations
are obtained
\be
\ba{lrcl}
{\rm Annulus\ } (C_2):&\Tr_{\rm open}\left(e^{-H_o \tau}\right)&=&\left<B\right|e^{-H_c\tilde{\tau}}\left|B\right>\vspace{.1 cm}\\
{\rm M\ddot{o}bius\ Strip\ } (M_2):&\Tr_{\rm open}\left(\Omega e^{-H_o \tau}\right)&=&\frac{1}{2}\left<B\right|e^{-H_c\tilde{\tau}}\left|C\right>+\frac{1}{2}\left<C\right|e^{-H_c\tilde{\tau}}\left|B\right>\vspace{.1 cm}\\
{\rm Klein\ Bottle\ } (K_2):\ &\Tr_{\rm open}\left(\Omega e^{-H_o \tau}\right)&=&\left<C\right|e^{-H_c\tilde{\tau}}\left|C\right>
\ea
\lb{tr_BC}
\ee
These equations constitute constraints similar to the modular
invariance constraints of closed string theories. The relation between 
the moduli are, for $K_2$ and $M_2$, $\tau=1/(2\tilde{\tau})$, and
for $C_2$, $\tau=2/(\tilde{\tau})$.

In terms of manifolds it is intended to obtain some open boundary
$\Sigma_{o}=\Sigma/I$ (where boundary refers to
$M=\Sigma\times[0,1]$) which is the involution under the symmetry $I$
of its double cover, $\Sigma$.
The resulting orbifolded manifold is
\be
M_{o}=(\Sigma\times[0,1])/I
\ee
where $I$ acts in $t$ as Time Inversion $t\to 1-t$.
This construction is pictured in figure~\r{fighorava}.
\fig{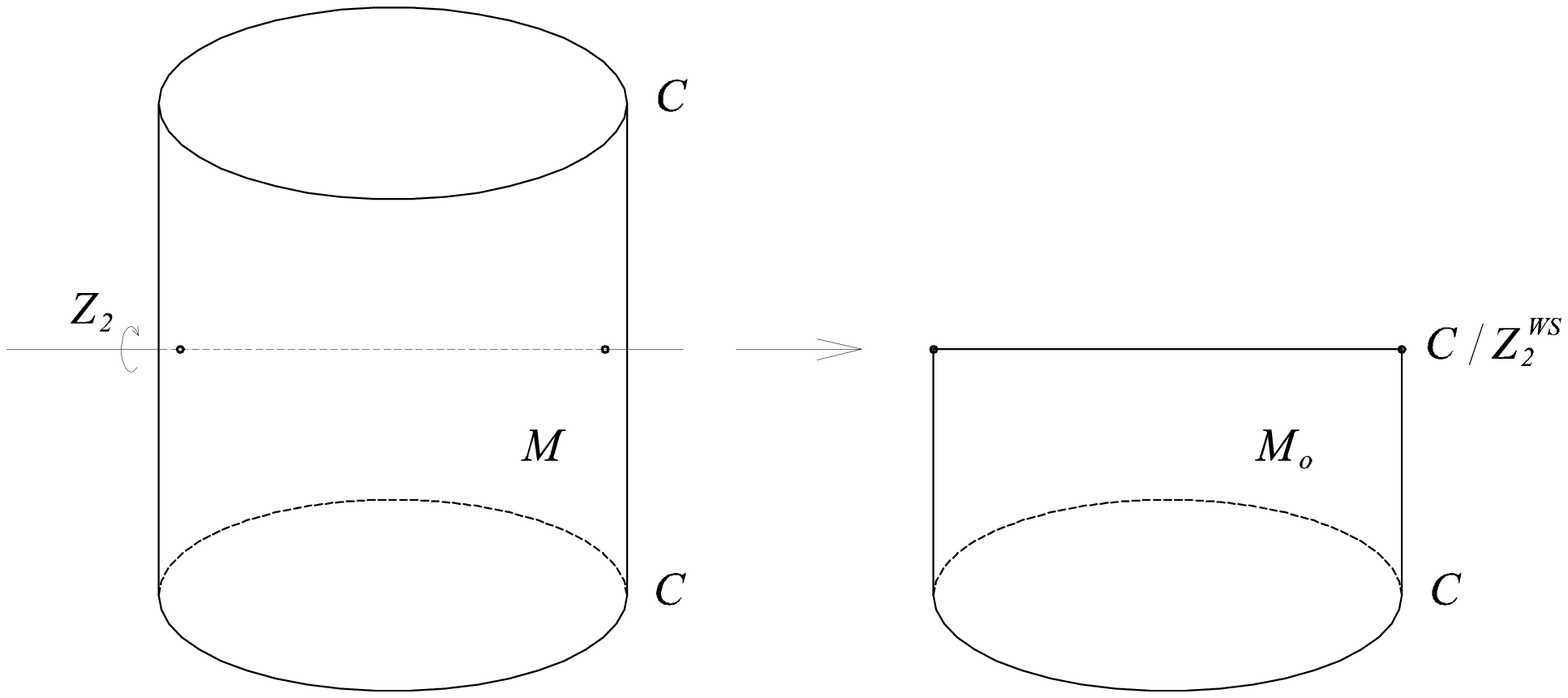}{The thickened open string $C_o$ as an orbifold of the
thickened closed string $C$ under an $I=Z_2$ symmetry. The boundaries
of $C_o=C/Z_2^{WS}$ are the singular points of the
orbifold.}{fighorava}

In terms of the action and fields in the theory Horava used the same
approach of extending them to the doubled cover manifold
\be
2S_o(A_o)=S(A)
\ee
In simple terms, $A$ stands for the extension of $A_o$ from
$M_o$ to its double cover $M$.

Since there is a one-to-one correspondence between the quantum states
of the gauge theory on M and the conformal blocks of the WZWN model, the
partition function may be written as
\be
Z_\Sigma=\sum h_{ij}\Psi_i\otimes\bar{\Psi}_j\in {\mathcal H}_\Sigma\otimes\bar{\mathcal H}_\Sigma
\ee
where $\Psi_i$ stands for a basis of the Hilbert space ${\mathcal H}_\Sigma$.
The open string counterpart in the orbifolded theory is
\be
Z_{\Sigma_o}=\sum a_i\Psi_i\in {\mathcal H}_\Sigma
\ee
which also agrees with the fact that in open CFT's the partition
function is the sum of characters (instead of the sum of squares) due
to the fact that the holomorphic and antiholomorphic sectors are not
independent.

\subsection{Discrete Symmetries and Orbifolds of TM(GT)}

Following the discussion of section~\r{ch.opun:sec.riemman} and
subsection~\r{ch.opun:sec.tmgt.horava}, it becomes obvious
that the parity operation plays a fundamental role in obtaining
open and/or non orientable manifolds out of closed orientable ones, and
hence open/unorientable theories out of closed
orientable theories.

Generally there are several ways of defining parity.  The ones of
interest have already been presented here.  For the usual ones, $P_1$
and $\Omega$ defined in~(\r{P}) and~(\r{omega}), the fields of our
$3D$ theory transform as
\be
\ba{ccc}
\ba{llcr}
P_1:     &z      &\leftrightarrow&\bz\vspace{.1 cm}\\
         &\Lambda&\rightarrow&\Lambda\vspace{.1 cm}\\
         &A_0    &\rightarrow&A_0\vspace{.1 cm}\\
         &A_z &\leftrightarrow&A_\bz\vspace{.1 cm}\\
         &E^z &\leftrightarrow&E^\bz\vspace{.1 cm}\\
         &B      &\rightarrow&-B\vspace{.1 cm}\\
         &Q      &\rightarrow&Q
\ea&\ \ \ \ &
\ba{llcr}
\Omega:  &z      &\lra       &-\bz\vspace{.1 cm}\\
         &\Lambda&\rightarrow&\Lambda\vspace{.1 cm}\\
         &A_0    &\rightarrow&A_0\vspace{.1 cm}\\
         &A_z    &\lra       &-A_\bz\vspace{.1 cm}\\
         &E^z    &\lra       &-E^\bz\vspace{.1 cm}\\
         &B      &\rightarrow&-B\vspace{.1 cm}\\
         &Q      &\rightarrow&Q
\ea
\ea
\lb{parity}
\ee
where $\Lambda$ is the gauge parameter entering into $U(1)$
gauge transformations. Under these two transformations the action
transforms as
\be
\int(F^2+kA\wedge F)\rightarrow\int(F^2-kA\wedge F)
\lb{PTS}
\ee

The theory is clearly not parity invariant. Let us then look for further
discrete symmetries which may be combined with parity in order to make
the action (theory) invariant. Introduce time-inversion, $T: t \to 1-t$,
implemented in this non standard way due to the compactness of time.
Note that $t=1/2$ is a fixed point of this operation.
Upon identification of the boundaries as described in~\cite{TM_15}
the boundary becomes a fixed point as well.

It remains to define how the fields of the theory change under these
symmetries.  There are two possible transformations compatible with
gauge transformations,
$A_\Lambda(t,z,\bz)=A(t,z,\bz)+\partial\Lambda(t,z,\bz)$. They are:
\be
\ba{cccr}
CT:&t      &\rightarrow&1-t\vspace{.1 cm}\\
    &\Lambda&\rightarrow&\Lambda\vspace{.1 cm}\\
    &A_0    &\rightarrow&-A_0\vspace{.1 cm}\\
    &\vb{A} &\rightarrow&\vb{A}\vspace{.1 cm}\\
    &\vb{E} &\rightarrow&-\vb{E}\vspace{.1 cm}\\
    &B      &\rightarrow&B\vspace{.1 cm}\\
    &Q      &\rightarrow&Q
\ea
\lb{T1}
\ee
and
\be
\ba{cccr}
T:&t      &\rightarrow&1-t\vspace{.1 cm}\\
    &\Lambda&\rightarrow&-\Lambda\vspace{.1 cm}\\
    &A_0    &\rightarrow&A_0\vspace{.1 cm}\\
    &\vb{A} &\rightarrow&-\vb{A}\vspace{.1 cm}\\
    &\vb{E} &\rightarrow&\vb{E}\vspace{.1 cm}\\
    &B      &\rightarrow&-B\vspace{.1 cm}\\
    &Q      &\rightarrow&-Q
\ea
\lb{T2}
\ee
where $C$, charge conjugation, is defined as $A_\mu\to-A_\mu$.
This symmetry inverts the sign of the charge, $Q\to-Q$, as usual.
These discrete symmetries together with parity $P$ or $\Omega$ are the
common ones used in $3D$ Quantum Field Theory.
When referring to parity in generic terms the letter $P$
will be used.

Under any of the $T$ and $CT$ symmetries the action changes in the
same fashion as it does for parity $P$ given by~(\r{PTS}).  In this
way any of the combinations $PT$ and $PCT$ are symmetries of the
action, $S\rightarrow S$.  Gauging them is a promising approach to
define the TM(GT) orbifolding.  It is now clear why extra
symmetries, besides parity, are necessary in order to have combinations of them
under which the theory (action) is invariant.  In general, whatever
parity definition is used, these results imply that $PT$ and $PCT$ are
indeed symmetries of the theory.

It can be concluded straight away that any of the two previous symmetries
exchange \textit{physically} two boundaries working as a mirror
transformation with fixed \textit{point} \mbox{$(t=1/2,z=\bz=x)$}
(corresponding actually to a line) as pictured in figure~\r{fig.mirror}.
It is considered that, whenever there is a charge insertion on
one boundary of $\mbox{q=m+kn/4}$, there exists an
insertion of $\bar{q}=m-kn/4$ on the other
boundary~\cite{TM_05,TM_15}.

Under the symmetries $PT$ and $PCT$ as given by~(\r{parity}), (\r{T1})
and~(\r{T2}) the boundaries will be exchanged as shown in
figure~\r{fig.mirror}. In the case of $PCT$ the charges will simply be 
swapped but in the case of $PT$ their sign will change $q\to-q$.
Note that $\Sigma_{\frac{1}{2}}=\Sigma(t=1/2)$ only feels $P$ or $CP$.

\fig{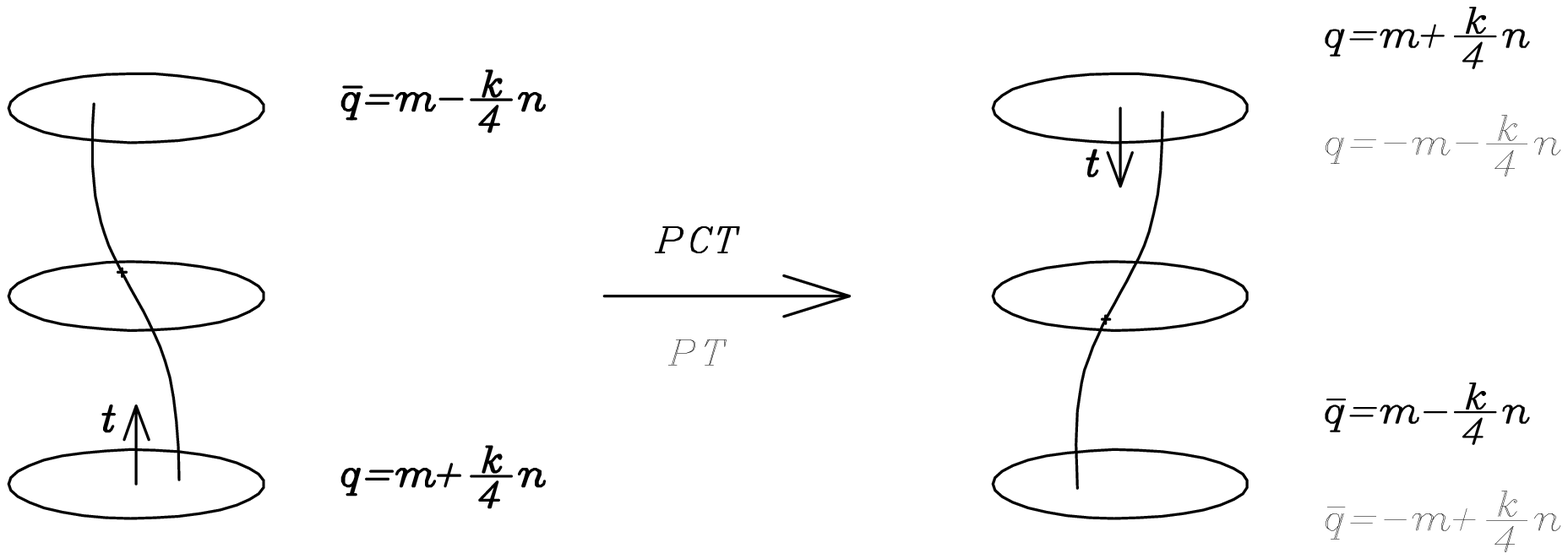}{Exchange of boundaries due to $PT$/$PCT$ transformation.}{fig.mirror}

As will be shown in detail there are important differences between the
two symmetries $CT$ and $T$, they will effectively gauge field
configurations corresponding to untwisted/twisted sectors of closed
strings and Neumann/Dirichlet boundary conditions of open strings.

Remembering that the final aim is to orbifold/quotient the theory
by gauging the discrete symmetries, let us proceed to check
compatibility with the desired symmetries in detail.  It is important
to stress that field configurations satisfying any $PT$/$PCT$
combinations of the previous symmetries exist, in principle, from the
start in the theory. One can either impose by hand that the physical
fields obey one of them (as is usual in QFT) or it can be assumed that
there is a wide theory with all of these field configurations and
obtain (self consistent) \textit{subtheories} by building suitable
projection operators that select some types of configurations.  It is
precisely this last construction that one must have in mind when
building several different theories out of one. In other words,
several different \textit{new} theories are going to be built by
gauging discrete symmetries of the type $PCT$ and $PT$.

It is important to stress what the orbifold means in terms of the
boundaries and bulk from the point of view of TM(GT).
It is splitting the manifold $M$ into two pieces creating one new
boundary at $t=1/2$. This boundary is going to feel only $CP$ or $P$
symmetries since it is located at the temporal fixed point of the
orbifold. Figure~\r{fig.orbi} shows this procedure.
\fig{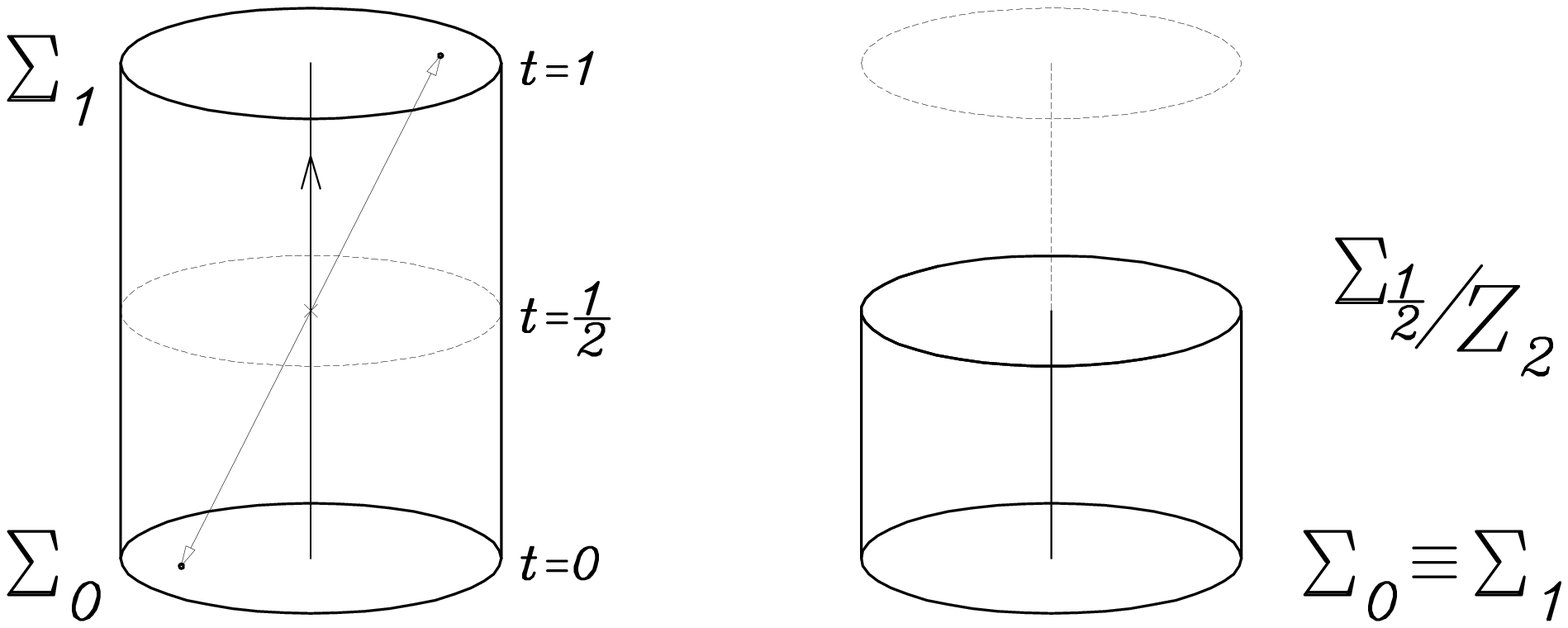}{Orbifolding of TM(GT). $\Sigma_{\frac{1}{2}}$ only
feels $PT$ or $P$ which are isomorphic to $Z_2$}{fig.orbi}
In this way this new boundary is going to constrain the \textit{new}
theory in such a way that the boundary theories will correspond 
to open and unoriented versions of the original full theory.

\subsection{Tree Level Amplitudes for\\ Open and Closed Unoriented Strings}

Let us start by considering tree level approximations to string amplitudes,
i.e. the Riemann surfaces are of genus $0$.  These surfaces are
the sphere (closed oriented strings) and its orbifolds: the disk (open
oriented) and the projective plane (closed unoriented) as was
discussed in section~\r{ch.opun:sec.riemman}.  From the point of view of
TM(GT), orbifolding means that the manifold $M$ is split into two
pieces that are identified.  As a result at $t=1/2$, the fixed point
of the orbifold, a new boundary is created.

For different orbifolds one will have different admissible field
configurations. In the following discussion we study
those field configurations which are compatible with $PT$ and $PCT$ type
orbifolds for the various parity operations already introduced.

\subsubsection{Disk}

Start from the simplest case - the disk. It is obtained by
the involution of the sphere under $P_1$ as given
by~(\r{P}). So consider the identifications under $P_1CT$ and $P_1T$.
For the first one the fields are transform as
\be
\ba{rrcl}
P_1CT:&\Lambda(t,z,\bz)      &=&\Lambda(1-t,\bz,z)\vspace{.1 cm}\\
            &\partial_iE^i(t,z,\bz)&=&-\partial_iE^i(1-t,\bz,z)\vspace{.1 cm}\\
            &B(t,z,\bz)            &=&-B(1-t,\bz,z)\vspace{.1 cm}\\
            &\displaystyle Q(t)=\int_{\Sigma(t)}\rho_0 &=&\displaystyle Q(1-t)=-\int_{\tilde{\Sigma}(1-t)}(-\rho_0)
\ea
\lb{PCT}
\ee
The orientations of $\Sigma$ and $\tilde{\Sigma}$ are opposite.  Under
these relations the Wilson lines have the property
\be
\exp{\left\{iQ\int_C dx^\mu A_\mu\right\}}=\exp{\left\{iQ\int_{-C} dx^\mu A_\mu\right\}}
\lb{PTW}
\ee
This means that for the configurations obeying the
relations~(\r{PCT}) the notion of time direction is completely lost!

Under the \textit{involution} of our $3D$ manifold, using the above
relations as geometrical identifications, the boundary is located at
$t=0$ and $t=1/2$.  For the moment let us check the compatibility of
the observables with the proposed orbifold constructions given by the
previous relations. In a very naive and straightforward way, when
$P_1CT$ is used as given by~(\r{PCT}) the charges should maintain
their sign ($q(t)\cong q(1-t)$).

\fig{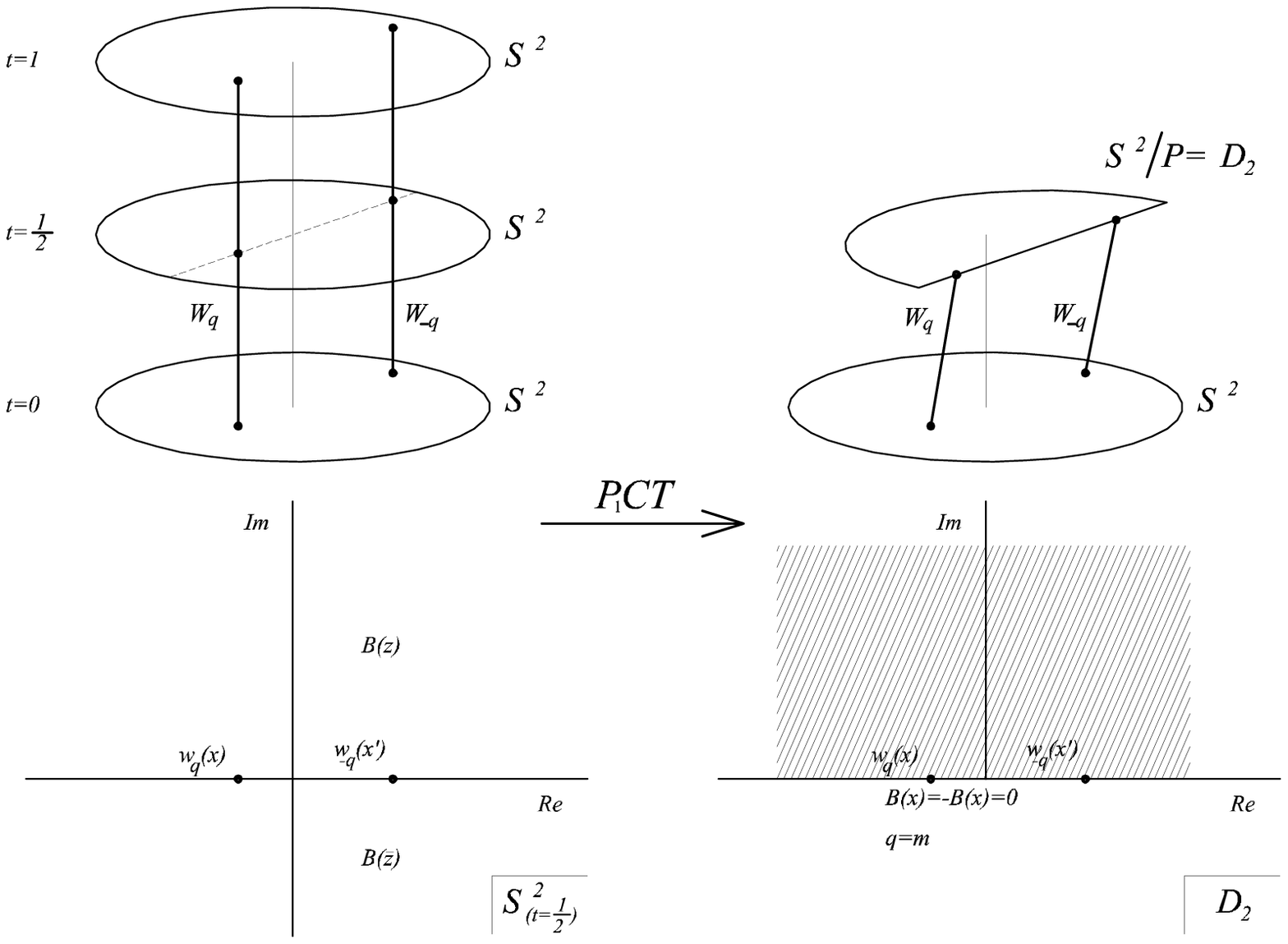}{Orbifold under $P_1CT$ in the presence of 2 Wilson
lines, $W_q$ and $W_{-q}$. They need to pierce
$\Sigma_{o\frac{1}{2}}=S^2/P_1=D_2$ on the real axis and the allowed
charges are $q=m$.}{fig.s2pct1a}\vspace{-.3cm}
\fig{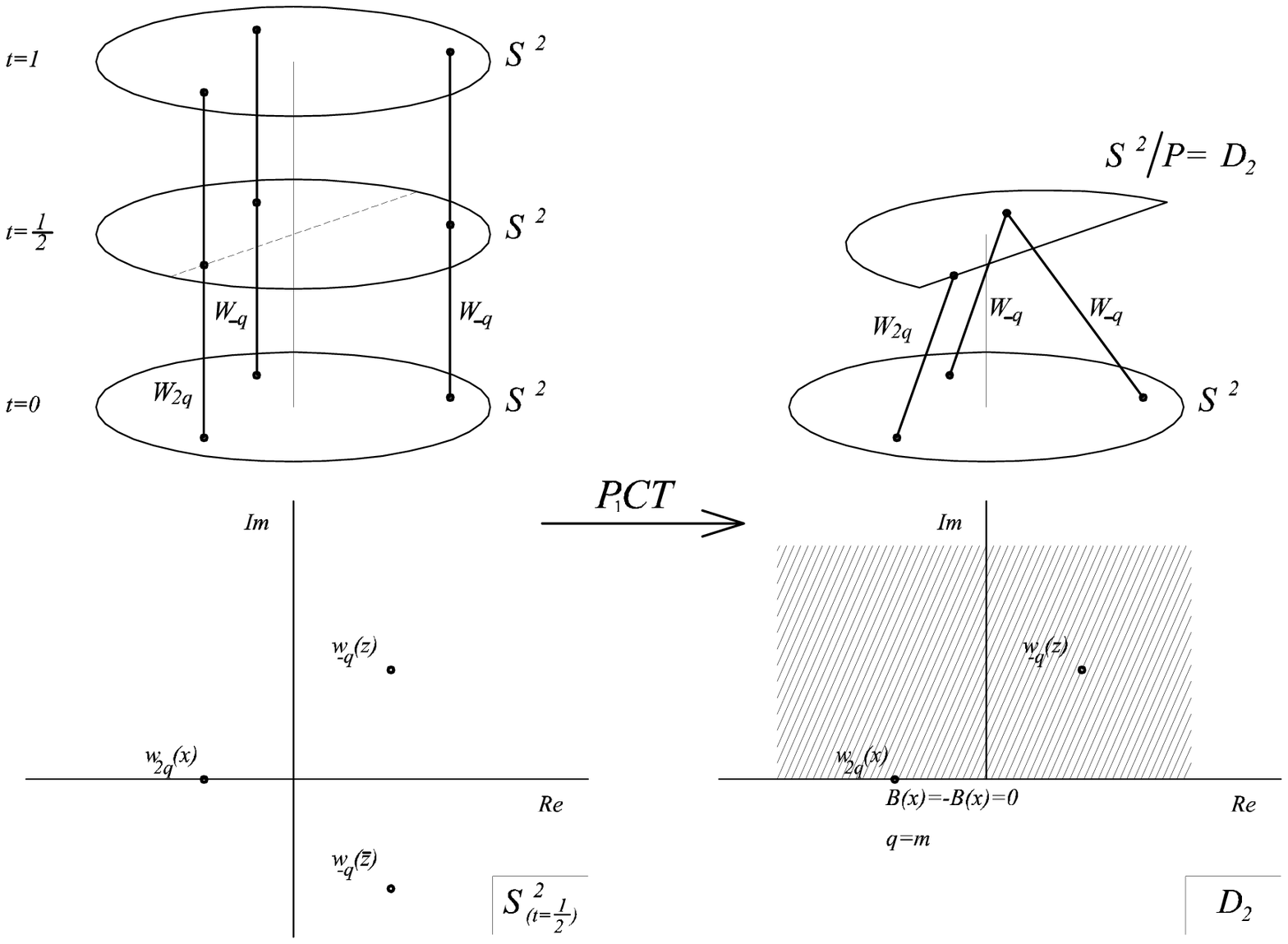}{Orbifold under $P_1CT$ in the presence of 3 Wilson
lines, $W_{2q}$ and two $W_{-q}$. $W_{2q}$ must pierce
$\Sigma_{o\frac{1}{2}}=S^2/P_1=D_2$ on the real axis and the allowed
charges are $q=m$.}{fig.s2pct1b}
\clearpage

Then, in order for the exchange of boundaries to be possible, it is
necessary to truncate the spectrum and set $q\cong{\bar{q}}=m$ for the
identification to make sense.  Let us check what happens at the
singular point of our orbifolded theory, $t=1/2$. The fields are
identified according to the previous rules but the manifold
$\Sigma(t=1/2)=S^2$ is only affected by $P_1$.

Take two Wilson lines that pierce the manifold in two distinct points,
$z$ and $z'$. Under the previous involution $P_1CT$, $z$ is identified
with $\bz$ for $t=1/2$.  Then, geometrically, the relation $z'=\bz$
must hold in order to have spatial identification of the
piercings. The problem is that when there are only two Wilson lines,
TM(GT) demands that they carry opposite charges. In order to implement
the desired identification $q=0$ is the only possibility.  For the
case where the Wilson lines pierce the manifold in the real axis,
$z=x$ and $z'=x'$, the involution is possible as pictured in
figure~\r{fig.s2pct1a} since the identification $x\cong x$ and
$x'\cong x'$ is considered.

In the presence of three Wilson lines, following the same line of
argument, there will necessarily exist one insertion on the boundary and
two in the bulk as pictured in figure~\r{fig.s2pct1b}. Only in the
presence of four Wilson lines, as pictured in figure~\r{fig.s2pct1c}
can any insertion on the boundary be avoided.

\fig{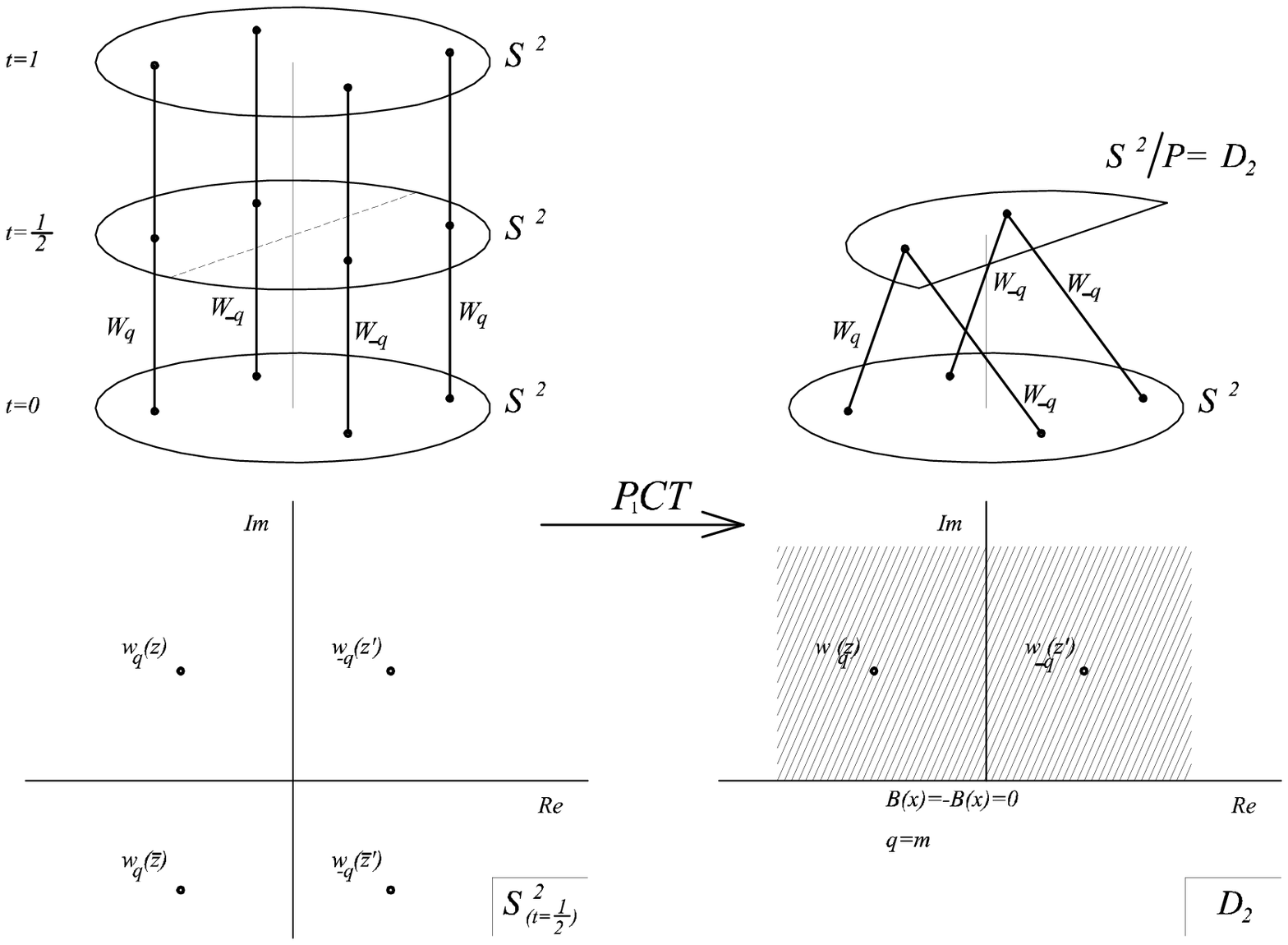}{Orbifold under $P_1CT$ in the presence of 4 Wilson
lines, two $W_q$ and two $W_{-q}$. The allowed
charges are $q=m$.}{fig.s2pct1c}

Note that the identification $B(z,\bz)\cong-B(\bz,z)$ on the real axis
implies necessarily $B(x,x)=0$. Remember that $2\pi n=\int B$
(see~\cite{TM_07,TM_15} for details). It is possible as well to have an
insertion on the boundary and one in the bulk.

This fact is simply the statement that by imposing $P_1CT$ symmetry on
the bulk fields one is actually imposing {\bf Neumann} boundary
conditions in the CFT. The charges of the theory become $q=m$, which
means that the string spectrum has only Kaluza-Klein
momenta. Furthermore, the monopole induced processes are suppressed.
Recall that they change the charge by an amount $kn/2$ which would
take the charges out of the spectrum allowed in these configurations.

Following our journey consider next $P_1T$. The fields now are related
in the following way
\be
\ba{rrcl}
P_1T:&\Lambda(t,z,\bz)       &=&-\Lambda(1-t,\bz,z)\vspace{.1 cm}\\
            &\partial_iE^i(t,z,\bz) &=&\partial_iE^i(1-t,\bz,z)\vspace{.1 cm}\\
            &B(t,z,\bz)             &=&B(1-t,\bz,z)\vspace{.1 cm}\\
            &\displaystyle Q=\int_{\Sigma(t)}\rho_0&=&-Q=-\int_{\tilde{\Sigma}(1-t)}(\rho_0)
\ea
\lb{PT}
\ee
The Wilson line has the same property~(\r{PTW}) as in the previous case.

Now the charges change sign under a $P_1T$ symmetry.
As before identifying the charges on opposite boundaries
truncates the spectrum, $q(t)\cong-q(1-t)$.
So, in order to have compatibility with the orbifold, $q\cong-{\bar{q}}=nk/4$.

In this case two piercings in the $2D$ bulk can be identified since
the charge identification $q\cong-q$ is now compatible with the
TM(GT) field configurations. But no other operator than the identity
$\phi_0$ can be inserted on the real axis since the corresponding
charge must be zero $q(x)=-q(x)=0$.  Therefore this kind of
orbifolding is only possible when there is an even number of Wilson
lines propagating in the $3D$ bulk.  The result for two Wilson lines is
pictured in figure~\r{fig.s2idpt1a} and for four in
figure~\r{fig.s2idpt1b}.

In terms of the full theory, a new $2D$ boundary has been defined. It
is a disk. The piercings of Wilson lines are none other than the
vertex operators (or fields) of a Conformal Field Theory defined on
that Disk.  In this case $B(z,\bz)=B(\bz,z)$, so that $B\neq 0$ on the
boundary.

\figh{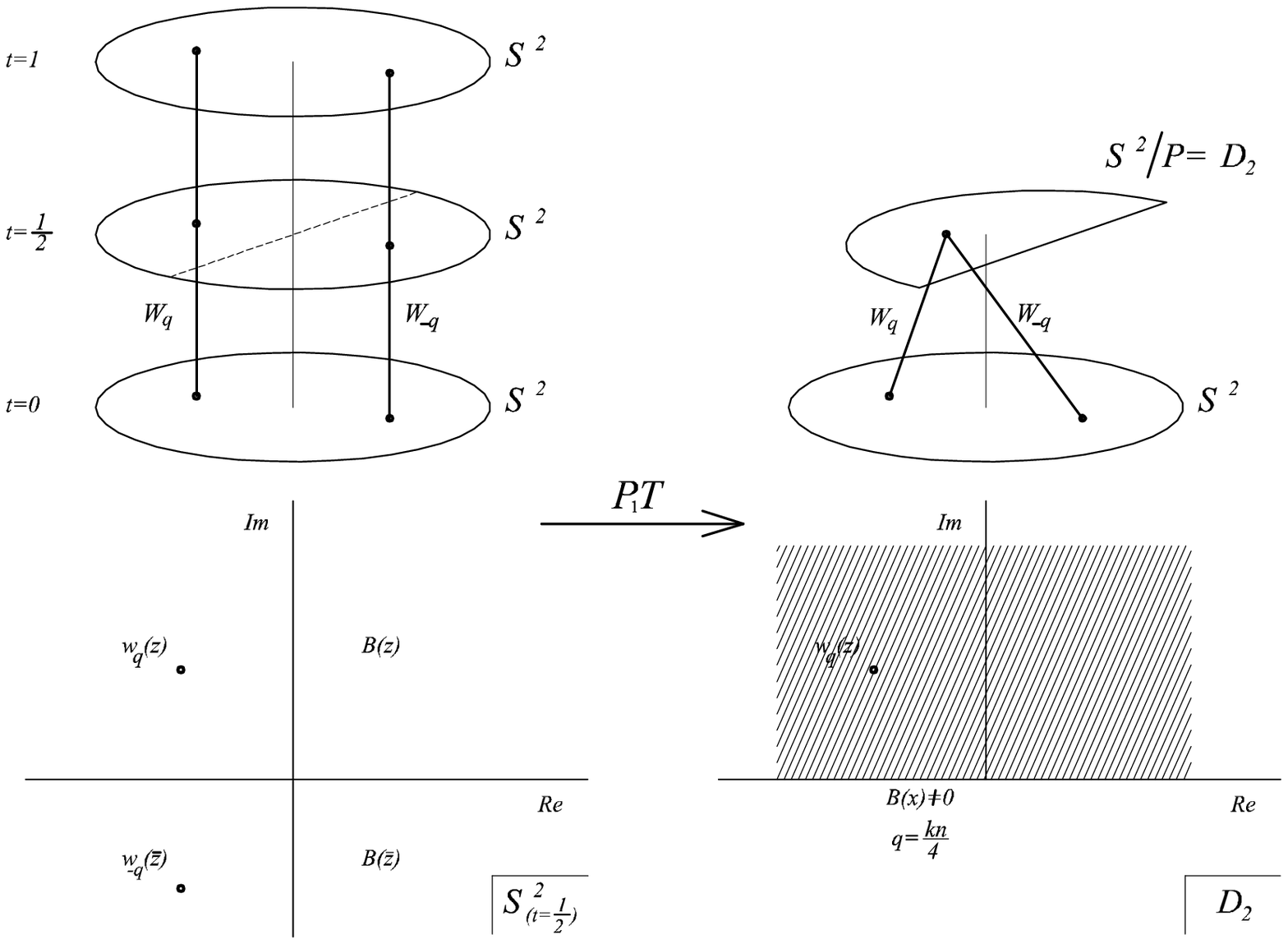}{Orbifold under $P_1T$ in the presence of 2 Wilson
lines, $W_q$ and $W_{-q}$. The new boundary is
$\Sigma_{o\frac{1}{2}}=S^2/P_1=D_2$. The allowed
charges are $q=kn/4$.}{fig.s2idpt1a}
\fig{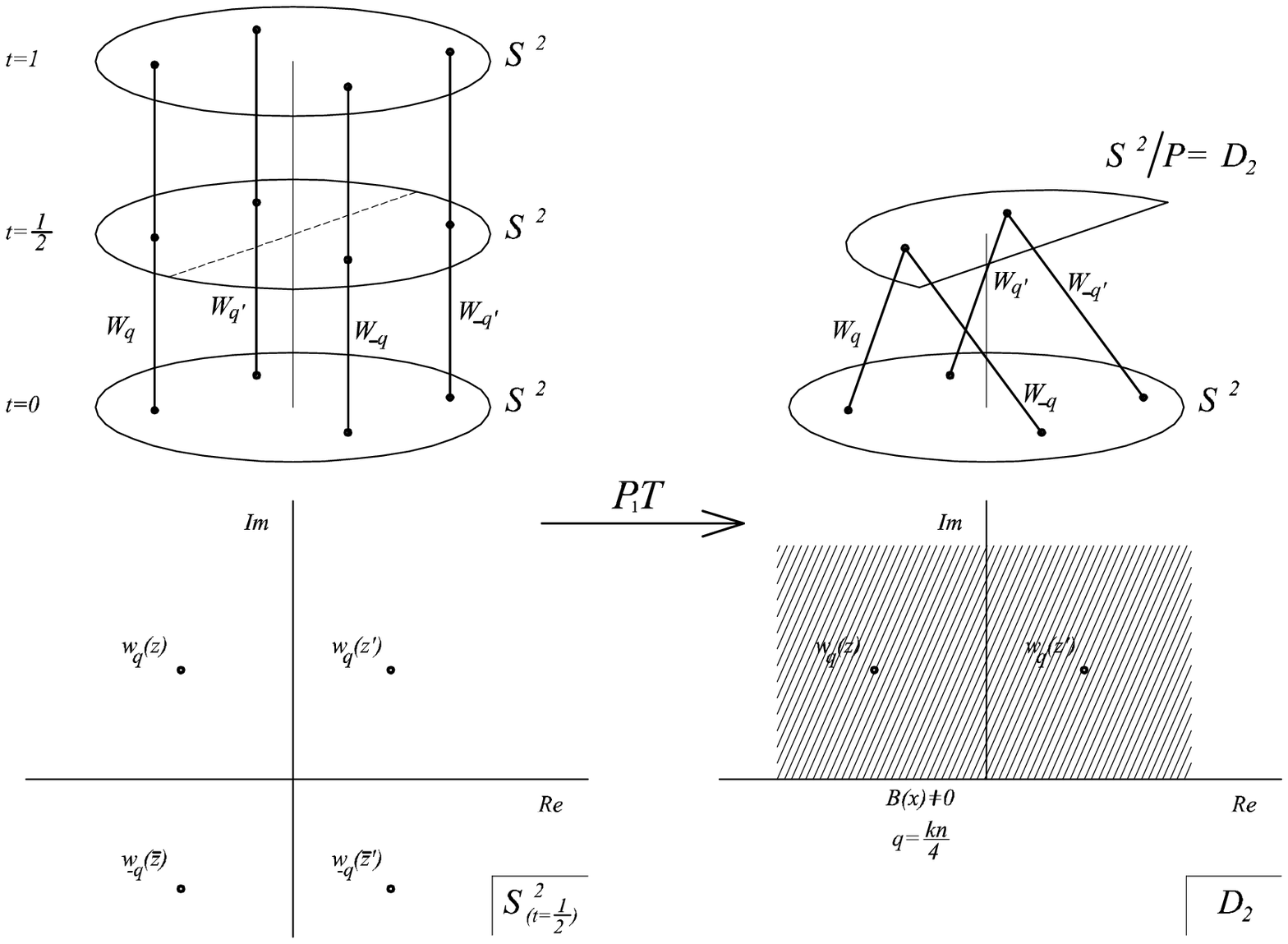}{Orbifold under $P_1T$ in the presence of 4 Wilson
lines, $W_q$, $W_q'$, $W_{-q}$ and $W_{-q'}$. The new boundary is
$\Sigma_{o\frac{1}{2}}=S^2/P_1=D_2$. The allowed
charges are $q=kn/4$.}{fig.s2idpt1b}
\clearpage

So these facts translate into {\bf Dirichlet}
boundary conditions for $P_1T$. The charges allowed, are $q=nk/4$,
the winding number of string theory. The monopole induced processes
are now allowed being crucial in this construction since they allow
the \textit{gluing} on this new boundary of two Wilson lines carrying
charges $q=nk/4$ and $\bar{q}=-nk/4$.

\subsubsection{Projective Plane}

Consider the parity operation as the antipodal identification
given in~(\r{PP}). The projective plane is obtained as the
\textit{new} $2D$ boundary of the orbifolded (under this symmetry) membrane.

The transformation is given by
\be
\ba{rrcl}
P_2: &z           &\rightarrow &\displaystyle-\frac{1}{\bz}\vspace{.1 cm}\\
   &\bz         &\rightarrow &\displaystyle-\frac{1}{z}\vspace{.1 cm}\\
   &\Lambda(z,\bz)  &\rightarrow &\displaystyle\Lambda(-\frac{1}{\bz},-\frac{1}{z})\vspace{.1 cm}\\
   &A_z(z,\bz)  &\rightarrow &\displaystyle\frac{1}{z^2} A_\bz(-\frac{1}{\bz},-\frac{1}{z})\vspace{.1 cm}\\
   &A_\bz(z,\bz)&\rightarrow &\displaystyle\frac{1}{\bz^2} A_z(-\frac{1}{\bz},-\frac{1}{z})\vspace{.1 cm}\\
   &B(z,\bz)&\rightarrow &\displaystyle\frac{1}{z^2\bz^2} B(-\frac{1}{\bz},-\frac{1}{z})\vspace{.1 cm}\\
   &E_z(z,\bz)&\rightarrow &\displaystyle\frac{1}{z^2} E_\bz(-\frac{1}{\bz},-\frac{1}{z})\vspace{.1 cm}\\
   &E_\bz(z,\bz)&\rightarrow &\displaystyle\frac{1}{\bz^2} E_z(-\frac{1}{\bz},-\frac{1}{z})\vspace{.1 cm}\\
\ea
\ee
Again the compatibility of the identifications
under this new discrete symmetry must be checked. Under $t'\cong 1-t$,
$z'\cong-1/\bz$ and $\bz'\cong-1/z$ we obtain, for $P_2CT$,
\be
\ba{rrcl}
P_2CT:&\Lambda(t,z,\bz)      &=&\Lambda(1-t,-\frac{1}{\bz},-\frac{1}{z})\vspace{.1 cm}\\
     &\partial_iE^i(t,z,\bz)&=&-\frac{1}{z^2\bz^2}\partial_iE^i(1-t,-\frac{1}{\bz},-\frac{1}{z})\vspace{.1 cm}\\
     &B(t,z,\bz)            &=&-\frac{1}{z^2\bz^2}B(1-t,-\frac{1}{\bz},-\frac{1}{z})\vspace{.1 cm}\\
     &\displaystyle Q(t)=\int_{\Sigma(t)}\rho_0 &=&\displaystyle Q(1-t)=-\int_{\tilde{\Sigma}(1-t)}(-\rho_0)
\ea
\lb{PPCT}
\ee
Note that the relation between the integrals
\be
\int_{\tilde{\Sigma}(t')}\frac{d^2z'}{z'^2\bz'^2}z'^2\bz'^2\left(B(t',z',\bz')+\partial_iE^i(t',z',\bz')\right)=\int_{\Sigma(t)}d^2z\left(B(t,z,\bz)+\partial_iE^i(t,z,\bz)\right)
\lb{Qzz}
\ee
follows from taking into account the second and third equalities of~(\r{PPCT}),
and the relations $dz=d\bz'/\bz'^2$, $d\bz=dz'/z'^2$, and consequently
$dz\wedge d\bz=-(1/z'^2\bz'^2)dz'\wedge d\bz'$.
$\Sigma$ and $\tilde{\Sigma}$ again have opposite orientations
and are mapped into each other by the given involution.
Under these relations and in a similar way to~(\r{Qzz})
the action transforms under $P_2$ as given in~(\r{PTS}) and any of the
combinations $P_2CT$ or $P_2T$ keep it invariant.
Also the Wilson lines have the same property given by~(\r{PTW}).

In the derivation of the previous identifications~(\r{PPCT})
it was necessary to demanding analyticity of the fields on the full
sphere. This translates into demanding that the transformation between the
two charts covering the sphere be well defined.
Since $\partial_u\Lambda=-z^2\partial_z\Lambda$  
and $\partial_{\bar{u}}\Lambda=-\bz^2\partial_\bz\Lambda$ the fields
must behave at infinity and zero like
\be
\ba{cc}
\Lambda\stackrel{\infty}{\to} z^{-1}\bz^{-1}&\Lambda\stackrel{0}{\to} z^3\bz^3\vspace{.1 cm}\\
A_z\stackrel{\infty}{\to} z^{-2}\bz^{-1}&A_z\stackrel{0}{\to} z^2\bz\vspace{.1 cm}\\
A_\bz\stackrel{\infty}{\to} z^{-1}\bz^{-2}&A_\bz\stackrel{0}{\to} z\bz^2
\ea
\ee

If naively one didn't care about these last limits the relations would
be plagued with Dirac delta-functions coming from the identity
$2\pi\delta^2(z,\bz)=\partial_z(1/\bz)=\partial_\bz(1/z)$.  Once the
previous behaviors are taken into account all these terms will vanish
upon integration. Another way to interpret these results is to note
that the points at infinity are not part of the chart (not physically
meaningful). To check the physical behavior at those points it is
necessary to compute it at zero in the other chart.

This time the charges compatible with $P_2CT$ are $q=m$ since
$q\cong\bar{q}$. Once there are no boundaries it is not possible to
have configurations with two Wilson lines which allow this kind of
orbifold.  In this way the smallest number of lines is four as pictured
in figure~\r{fig.s2idpct2}. Furthermore the total number of Wilson lines
must be even.
\fig{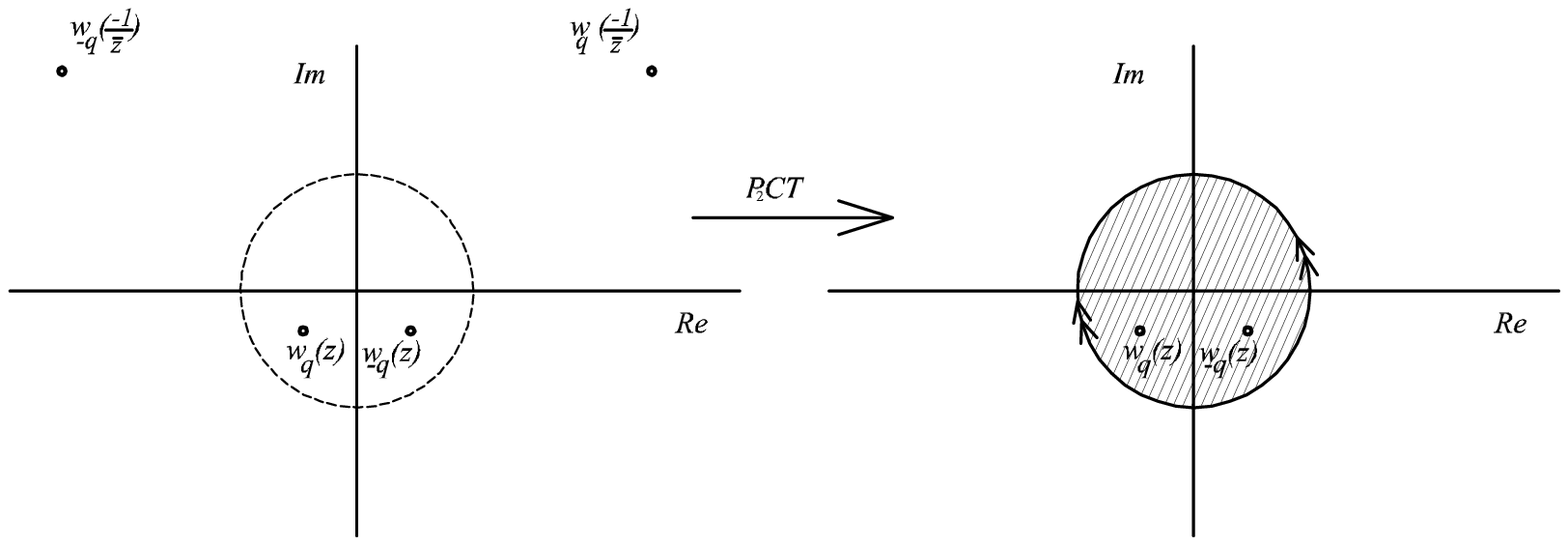}{Orbifold under $PCT$ in the presence of 4 Wilson
lines, two $W_q$ and two $W_{-q}$. The new $2D$ boundary is
$\Sigma_{o\frac{1}{2}}=S^2/P=RP_2$. The allowed
charges are $q=m$.}{fig.s2idpct2}

This configuration corresponds to {\bf untwisted} closed unoriented string
theories. Note that $\Lambda$, which is identified with string theory
target space, is not orbifolded by $P_2CT$. The charges allowed are $q=m$,
the Kaluza-Klein momenta of string theory. Once again the monopole processes
are suppressed.

For $P_2T$ the fields transform as
\be
\ba{cccc}
P_2T:&\Lambda(t,z,\bz)      &=&-\Lambda(1-t,-\frac{1}{\bz},-\frac{1}{z})\vspace{.1 cm}\\
     &\partial_iE^i(t,z,\bz)&=&\frac{1}{z^2\bz^2}\partial_iE^i(1-t,-\frac{1}{\bz},-\frac{1}{z})\vspace{.1 cm}\\
     &B(t,z,\bz)            &=&\frac{1}{z^2\bz^2}B(1-t,-\frac{1}{\bz},-\frac{1}{z})\vspace{.1 cm}\\
     &\displaystyle Q(t)=\int_{\Sigma(t)}\rho_0 &=&\displaystyle-Q(1-t)=-\int_{\tilde{\Sigma}(1-t)}(-\rho_0)
\ea
\lb{PPT}
\ee

In this case $q=kn/4$ since $q\cong-\bar{q}$, and furthermore
configurations with two Wilson lines are compatible with the orbifold
as pictured in figure~\r{fig.s2idpt2}.
\figh{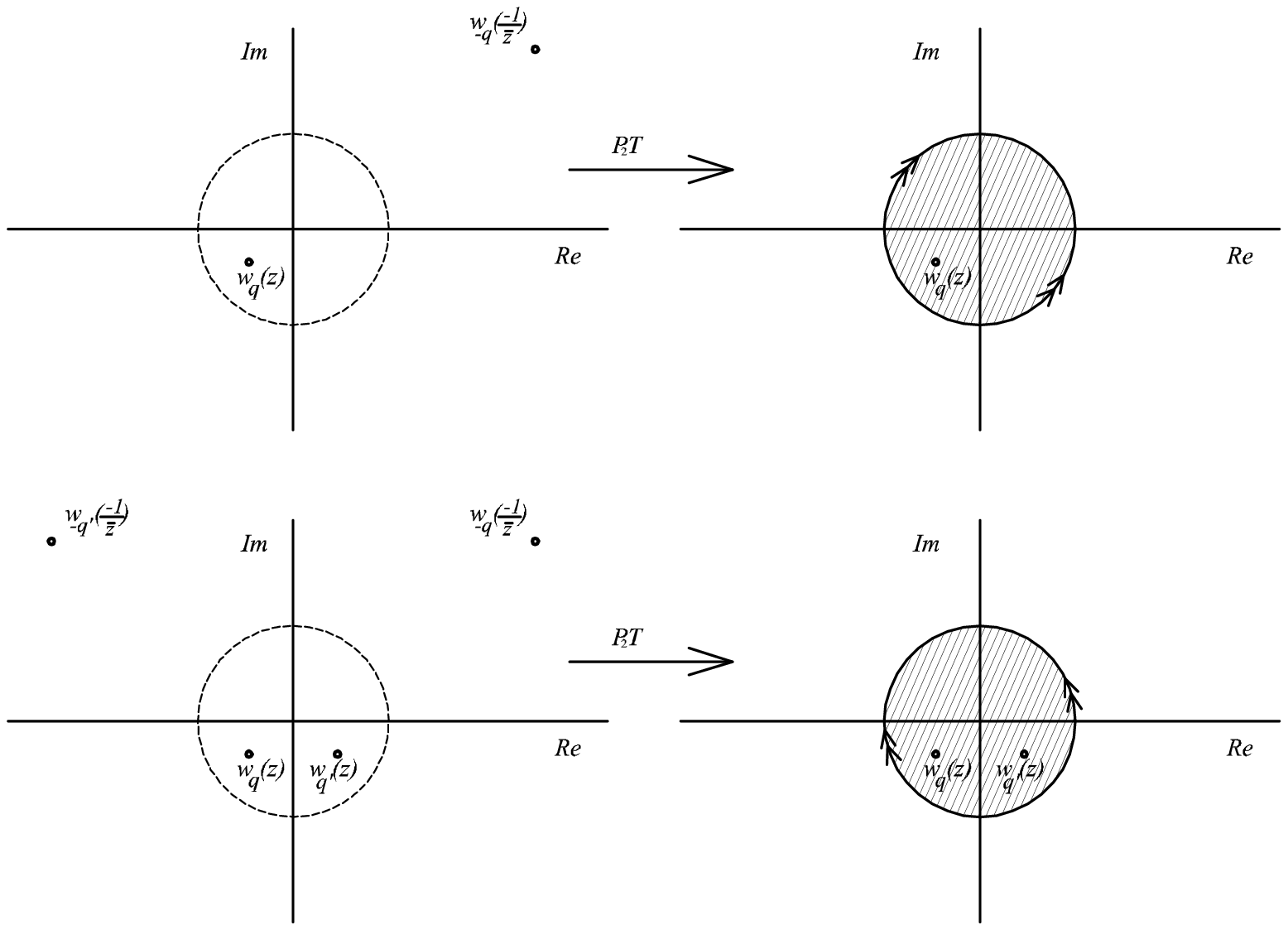}{Orbifold under $PT$ in the presence of 4 Wilson
lines, two $W_q$ and two $W_{-q}$. The new $2D$ boundary is
$\Sigma_{o\frac{1}{2}}=S^2/P=RP_2$. The allowed
charges are $q=kn/4$.}{fig.s2idpt2}

In this case one obtains {\bf twisted} unoriented closed
strings. Note that the orbifold identifies $\Lambda\cong-\Lambda$ such
that the target space of string theory is orbifolded. The full
construction, including the world-sheet parity, from the point of view
of string theory is called an orientifold. The allowed charges
$q=kn/4$ correspond to the winding number of string theory. The
monopole processes are again crucial since they allow, on the new boundary,
the \textit{gluing} of Wilson lines carrying opposite charges.
This discussion will be addressed later on.

\subsection{One Loop Amplitudes for\\ Open and Closed Unoriented Strings}

One loop amplitudes are computed for Riemann surfaces of genus $1$.
They correspond to the torus (closed oriented) and its orbifolds:
the annulus or cylinder (open oriented), the M\"{o}bius strip
(open unoriented) and the Klein bottle (closed unoriented).

\subsubsection{Annulus}
Let us start with the already studied parity transformation $\Omega$, as
given by~(\r{omega}).  There is nothing new to add to the field
transformations~(\r{PCT}) for $PCT$ and~(\r{PT}) for $PT$, this time under
the identifications $t'=1-t$, $z'=-\bz$ and $\bz'=-z$. The resulting
geometry is the annulus $C_2$ and has now two boundaries.

\figh{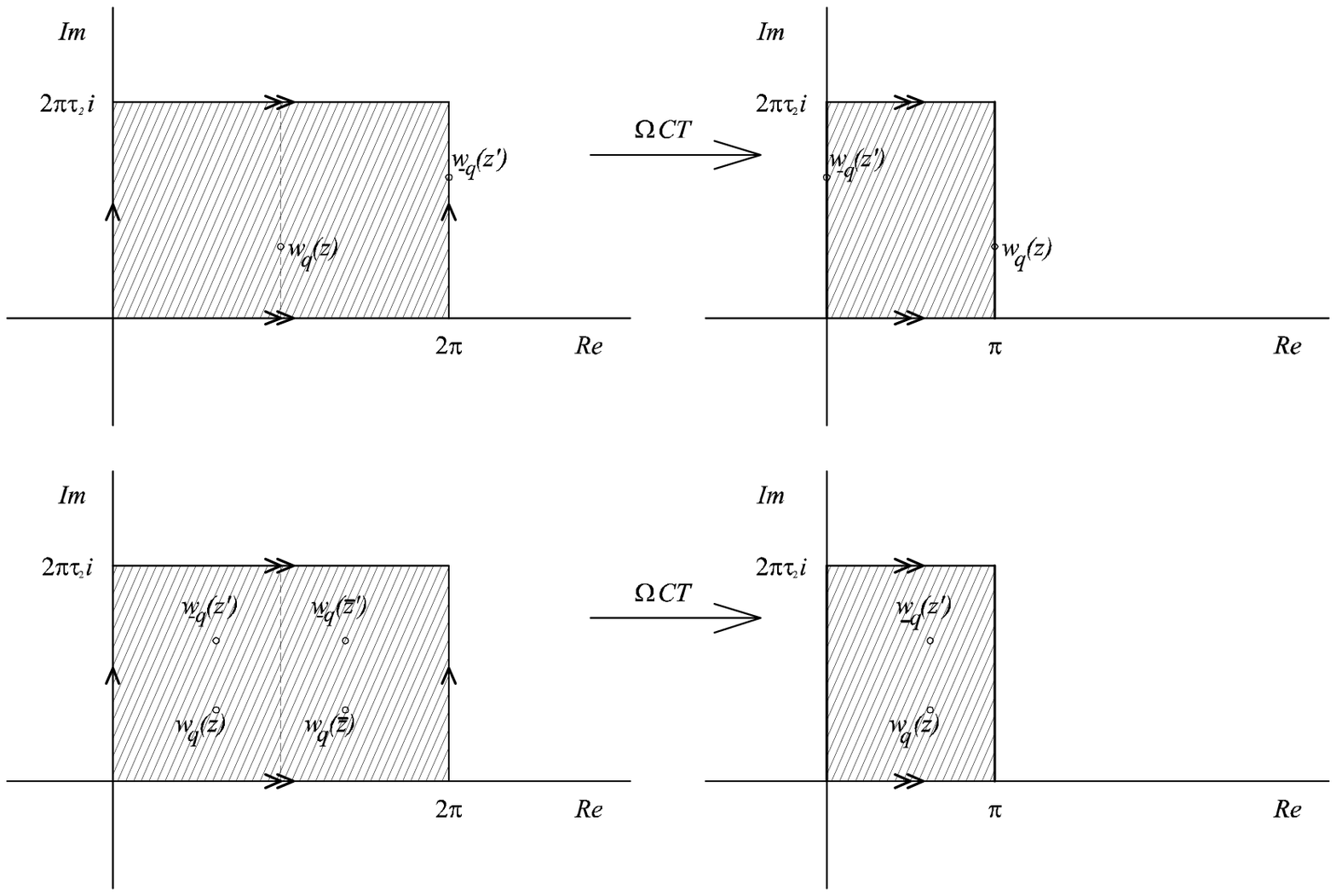}{Orbifold under $\Omega CT$ in
the presence of 2 and 4 Wilson lines. The new $2D$ boundary is
$\Sigma_{o\frac{1}{2}}=S^2/\Omega=C_2$. The allowed charges are
$q=m$.}{fig.t2idpct1}

For $\Omega CT$ the allowed charges are $q=m$ due to the identification
$q\cong\bar{q}$ and $B(x)=0$ at the boundaries.
There can exist two insertions on the boundaries of the
$2D$ CFT but not in its bulk due to the identifications of charges,
basically the argument is the same as that used for the disk.
As on the disk there cannot exist one single bulk insertion due
to the total charge being necessarily zero in the full plane.
Up to configurations with four Wilson lines one can have:
two insertions on the boundary; one insertion in the
bulk and one on the boundary corresponding to three Wilson lines;
three insertions on the boundaries (with $\sum q=0$); one insertion
in the bulk and two on the boundary corresponding to four Wilson
lines; and two insertions in the bulk corresponding to four Wilson
lines as pictured in figure~\r{fig.t2idpct1}.

This construction corresponds to open oriented strings with
{\bf Neumann} boundary conditions. The charge spectrum is $q=m$,
corresponding to Kaluza-Klein momenta in string theory and the monopole
induced processes are suppressed. It is Neumann because the gauged
symmetry is of $PCT$ type. Note that the definition of parity is
not important, as even for genus 1 surfaces the results hold similarly to 
the previous cases for $P_1$ and $P_2$ used in genus 0. What is
important is the inclusion of the discrete symmetry $C$!

For $\Omega T$ the allowed charges are $q=kn/4$
due to the identification $q\cong-\bar{q}$. There are no insertions in
the boundary. One insertion in the bulk corresponds to two Wilson
lines and two to four Wilson lines pictured in figure~\r{fig.t2idpt1}.
\figh{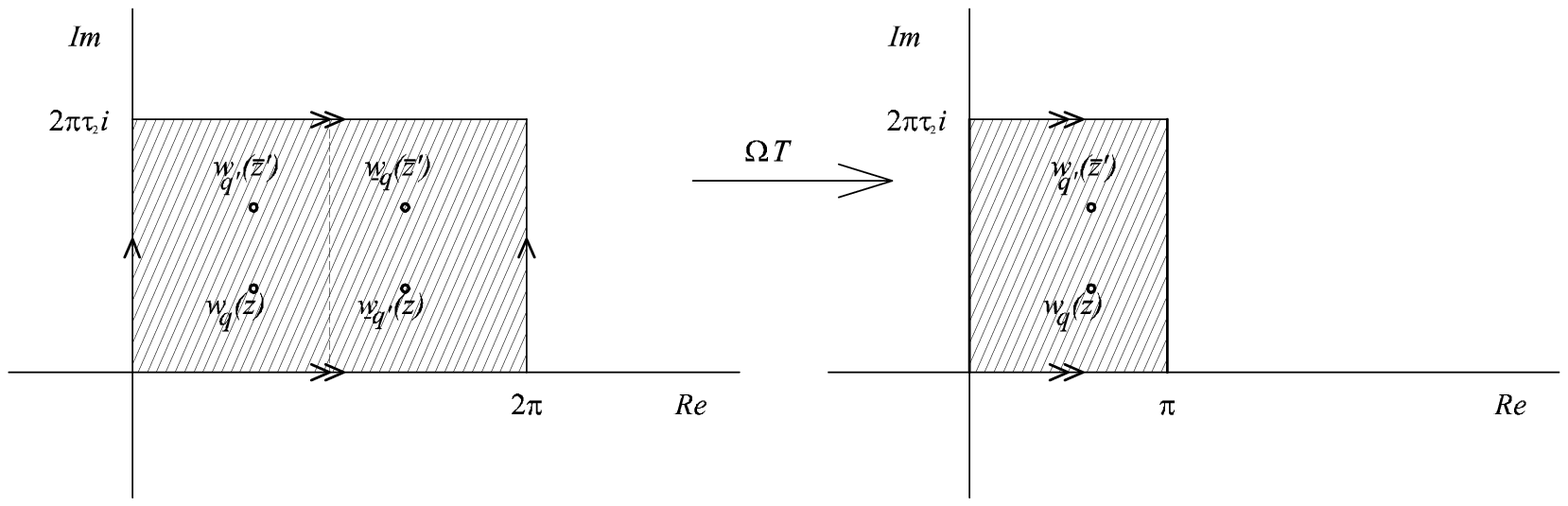}{Orbifold under $\Omega T$ in the presence of 4 Wilson
lines. The new $2D$ boundary is
$\Sigma_{o\frac{1}{2}}=S^2/\Omega=C_2$. The allowed
charges are $q=kn/4$.}{fig.t2idpt1}

This last construction corresponds to open oriented strings with {\bf
Dirichlet} boundary conditions.  The charge spectrum is $q=kn/4$,
corresponding to the winding number in string theory, and the monopole
induced processes are present allowing the \textit{gluing} of Wilson
lines with opposite charges.

\subsubsection{M\"{o}bius Strip}

Let us proceed to the parity $\tilde{\Omega}$ as given
by~(\r{omegat}).  The results are pictured in figure~\r{fig.t2idm2a}
and are fairly similar to the previous cases. Note that it corresponds
to two involutions of the torus with $\tau=2i\tau$, one given by
$\Omega$ resulting in the annulus, and $\tilde{a}$ which maps the
annulus onto the M\"obius strip. Then, for each insertion in the strip,
there exist four on the torus.

\figh{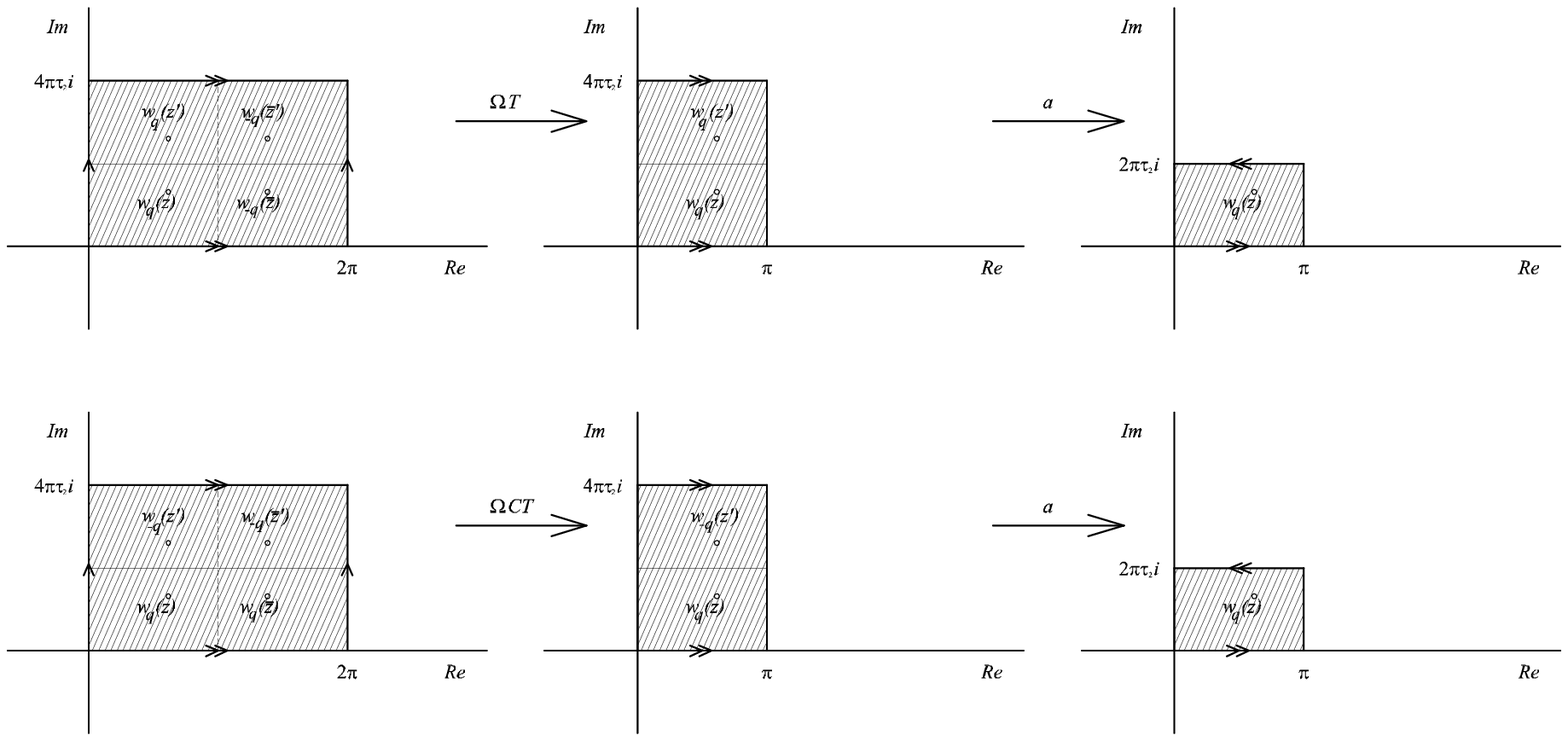}{Orbifold under $\tilde{\Omega}T$ and $\tilde{\Omega}CT$
of the torus with $\tau=2i\tau_2$ in the presence of four Wilson
lines.  For $\tilde{\Omega}CT$ the Wilson lines may pierce the
manifold on the real axis and the allowed charge is $q=m$. For
$\tilde{\Omega}T$, boundary insertions are not allowed and the
admissible charges are $q=kn/4$. The relation $z'=z+2\pi(1/2+i\tau_2)$
must hold.}{fig.t2idm2a}

Once more, for $\tilde{\Omega}CT$, one finds $B=-B=0$ on the
boundaries and $q$ is identified with $\bar{q}$ demanding the charges
to be $q=m$. This correspond to the Kaluza-Klein momenta of string
theory.  Due to this fact, the monopole processes are suppressed in the
configurations allowing this kind of orbifolding.  This corresponds to
{\bf Neumann} boundary conditions.

For the $\tilde{\Omega}T$ case the identification $q\cong-\bar{q}$
demands the charges to be $q=kn/4$, the winding number of string
theory. This time the monopole processes play a fundamental role and
the charges are purely \textit{magnetic}. This corresponds to
{\bf Dirichlet} boundary conditions.

As discussed in subsection~\r{sec:torus_inv} the involution of the
torus, with modulus $\tau=1/2+i\tau_2$ under $\Omega T$ or $\Omega CT$
can also be considered.  In this case four insertions on the torus
correspond to two insertions on the strip as presented in
figure~\r{fig.t2idm2b} for the $\Omega T$ case.
\figh{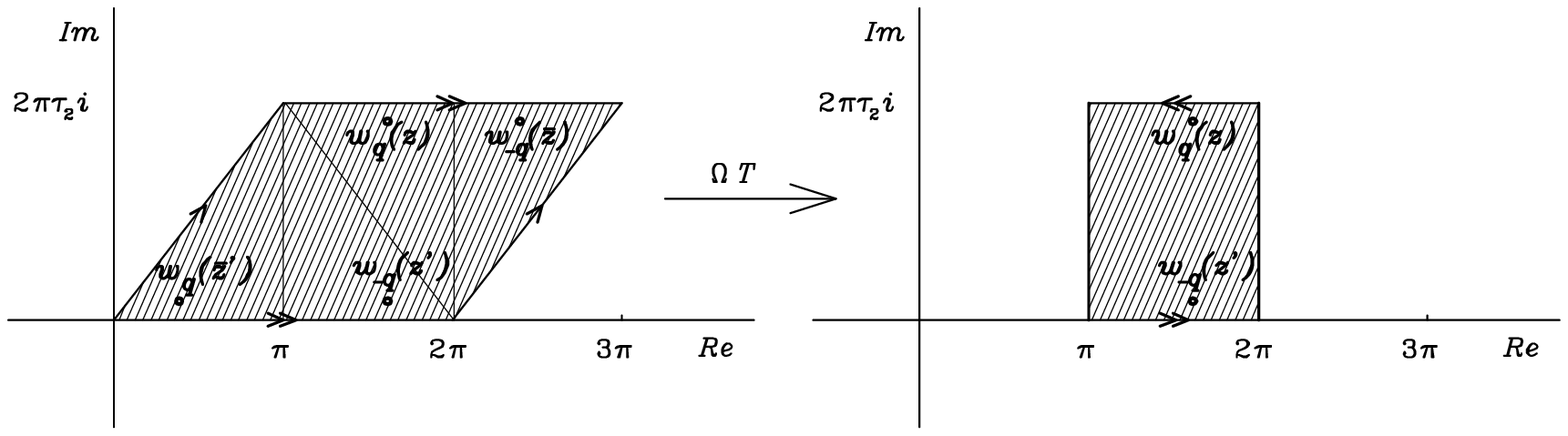}{Orbifold under $\Omega T$ of the torus with
$\tau=1/2+i\tau_2$ in the presence of four Wilson lines.}{fig.t2idm2b}

As previously explained both constructions result in the same region
of the complex plane. Note that the resulting area in both cases is
$2\pi^2\tau_2$ and that in both cases the region $[0,\pi]\times
i[0,2\pi\tau_2]$ is identified with the region $[\pi,2\pi]\times
i[0,2\pi\tau_2]$.

\subsubsection{Klein Bottle}

Finally using the parity $\Omega'$ as given by (\r{omega'}), the
points are identified under $t'=1-t$, $z'=-\bz+2\pi i\tau_2$ and
$\bz'=-z+2\pi i\tau_2$.  Upon orbifolding the new boundary of TM is a
Klein bottle.

Again, for $\Omega'CT$, $q=m$ is obtained because $q\cong\bar{q}$. The
minimum number of insertions is two corresponding to four Wilson lines 
in the bulk. This construction corresponds to {\bf untwisted}
unoriented closed strings with only Kaluza-Klein momenta in the spectrum.
The monopole processes are suppressed.

For $\Omega'T$ case the charge spectrum is truncated to $q=kn/4$ due
to the identification $q\cong-\bar{q}$.  One single insertion in the
$2D$ bulk can also be obtained corresponding to two Wilson lines or
two insertions corresponding to four Wilson lines. This construction
corresponds to {\bf twisted} unoriented closed strings with only
winding number. The monopole processes are present and are crucial in
the construction.

\figh{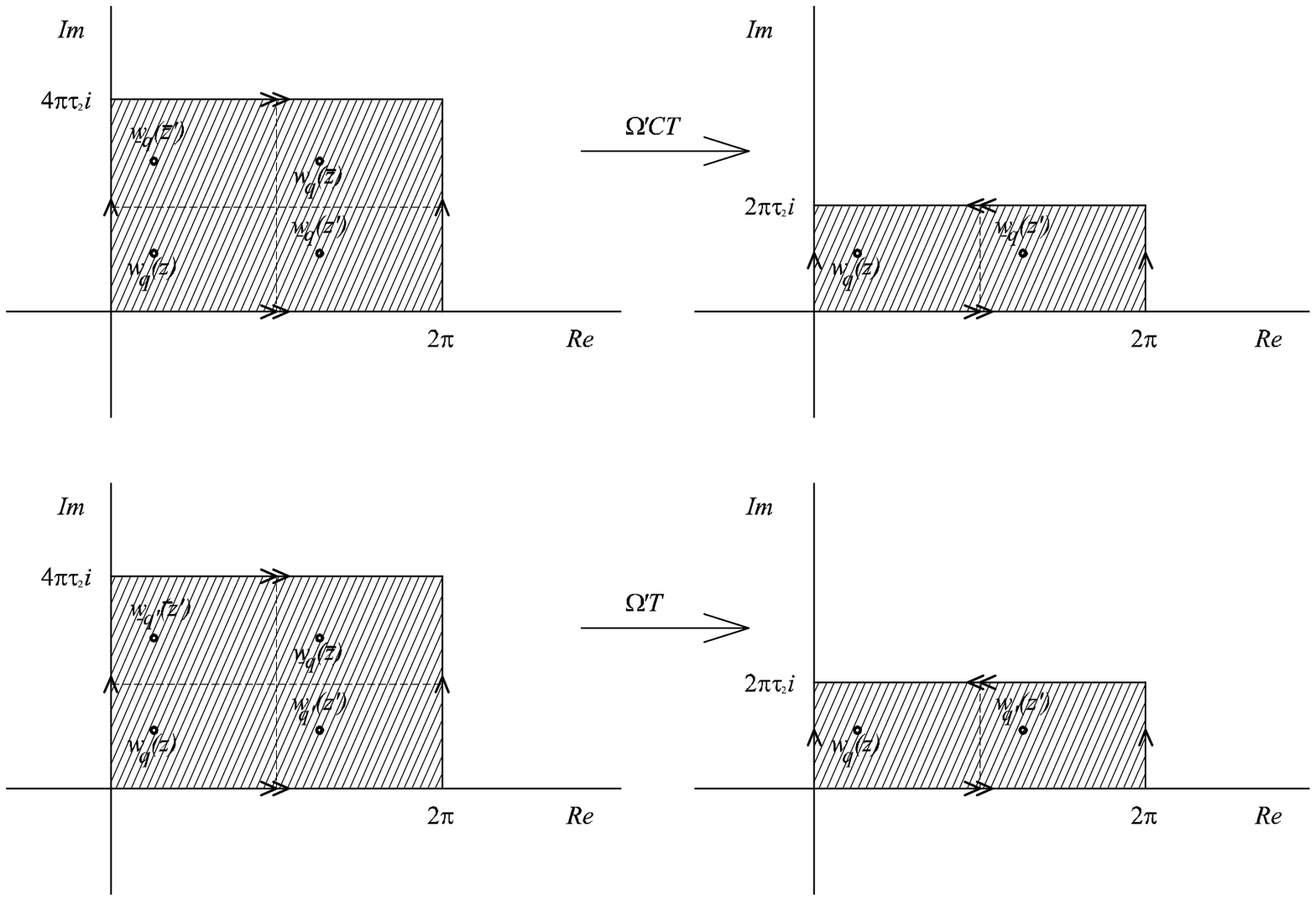}{Orbifold of $\Sigma_{\frac{1}{2}}=T^2$ under
$\Omega'CT$ and $\Omega'T$.}{fig.t2id3}

Two examples corresponding to four Wilson lines are pictured
in figure~\r{fig.t2id3}.

\subsection{Note on Modular Invariance and the Relative Modular Group\lb{ch.opun:sec.tmgt.mod}}

Modular invariance is a fundamental ingredient in string theory
which makes closed string theories UV finite. What about the
orbifolded theories? It is much more tricky. Each separated sector of
open and unoriented theories is clearly not invariant under a modular
transformation. The transformation $\tau\to-1/2\tau$ can be interpreted
as the exchange of the homology cycles $\alpha$ and $\beta$ of the
torus as represented in figure~\r{fig.mod}.  Equivalently it swaps
$PT$ and $PCT$ orbifold types. But then, if the first orbifold
corresponds to the twisted sector (closed unoriented strings) or to
Dirichlet boundary conditions (open strings), the second orbifold will
correspond to the untwisted sector (closed unoriented strings) or to
Dirichlet boundary conditions (open strings). Note the sign of the
charges in figure~\r{fig.mod} representing the Klein bottle
projection.
\figh{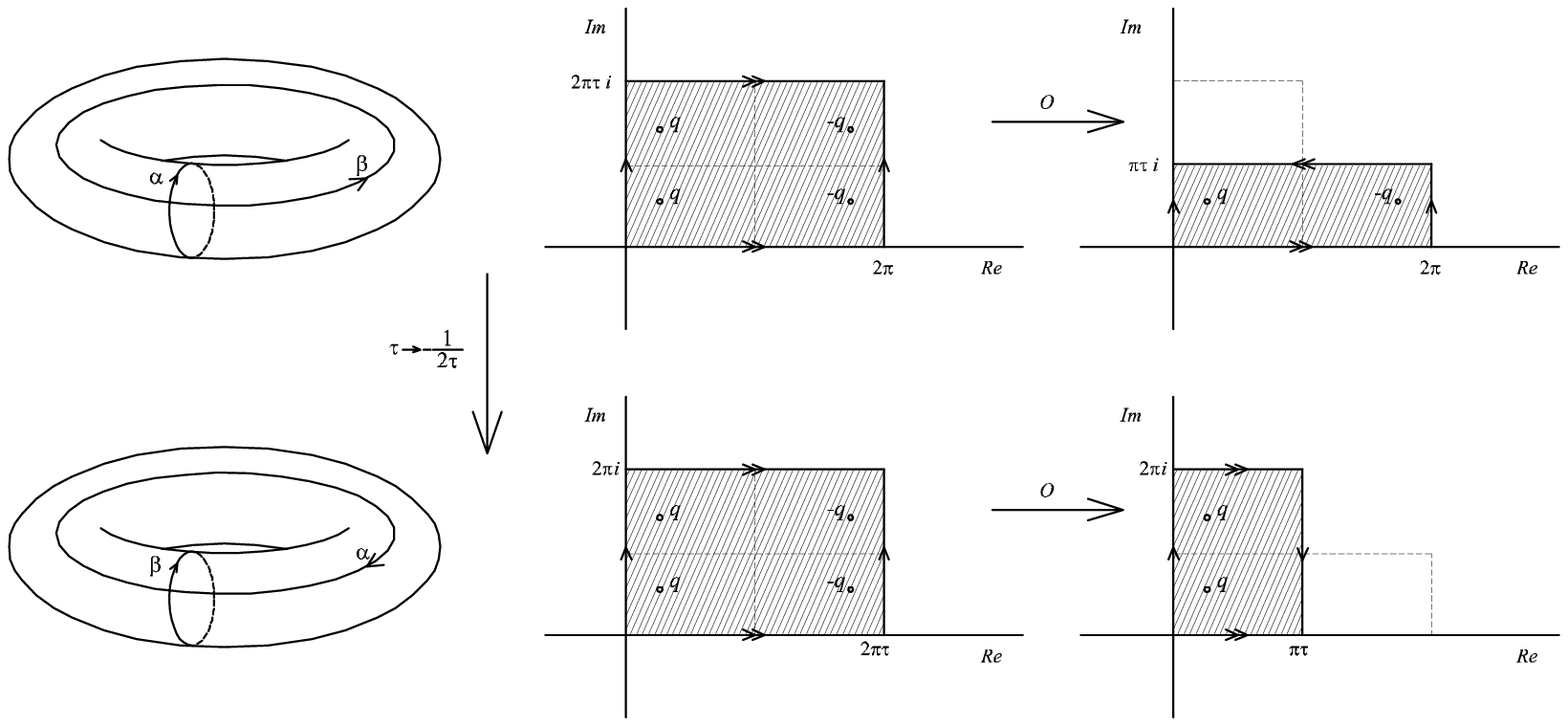}{A suitable modular transformation takes us from one
projection $PT$ (orbifold) to the other one $PCT$.}{fig.mod}

So if one actually wants to ensure modular invariance it is necessary
to build a projection operator which ensures it. A good choice would
be
\be
O=\frac{2+PT+PCT}{4}
\ee
such that the \textit{exchange} of orbifolds doesn't change
it. This fact is well known in string theory (see~\cite{POL_1} for details).

In the case that the orbifolded manifolds result in open
surfaces the modular transformation $\tau\to-1/\tau$, according to the
previous discussion, exchanges the boundary conditions
(Neumann/Dirichlet). Note that orbifolding the target space in string
theory (or equivalently the gauge group in TMGT) is effectively
creating an orientifold plane where the boundary conditions must be
Dirichlet (as for a D-brane). This is the equivalent of
\textit{twisting} for open strings. In terms of the bulk the modular
transformation is exchanging the projections $PCT\lra PT$.

Putting it in more exact terms, consider some discrete group $H$ of
symmetries of the target space (or equivalently the gauge group of
TMGT). Consider now the twist by the element $h=(h_1,h_2) \in H$,
where $h_1$ twists the states in the $x_1$ direction and $h_2$ in the
$x_2$ direction.  Then the modular transformation will change the
twist as
\be
\ba{lll}
{\mathcal{T}}:&\tau\to\tau+1&(h_1,h_2)\to(h_1,h_1h_2)\vspace{.1cm}\\
{\mathcal{S}}:&\displaystyle\tau\to-\frac{1}{\tau}&(h_1,h_2)\to(h_2,h_1^{-1})
\ea
\ee

Returning to the Horava picture of describing an open string as a
\textit{thickened surface} (or double cover), in the case of the
orbifold resulting in a new open boundary the picture is similar. In
this case a modular transformation takes an open string loop, with the
ends attached to the boundaries (direct-channel picture) to a closed
string propagating from boundary to boundary (transverse-channel
picture) as pictured in figure~\r{fig.mod2} for the annulus.
\fig{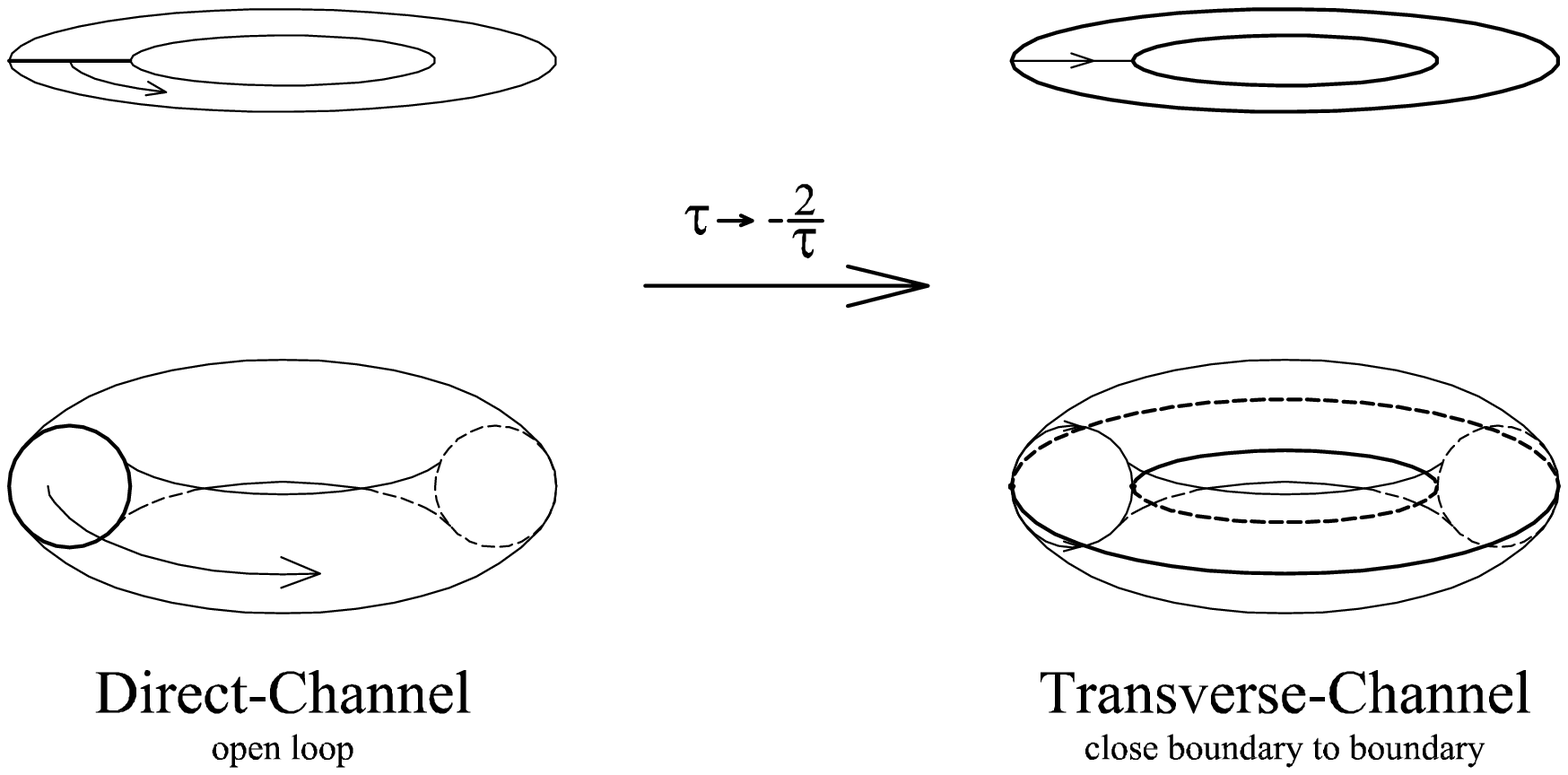}{A suitable modular transformation takes us from the
direct-channel picture to the transverse-channel picture.  Here this
construction is pictured in terms of the thickened string.}{fig.mod2}

The lower boundary of the membrane is a thickening of the string. In
the case of the open string loop one can imagine that the open string,
while propagating, splits into two parts.  The left modes propagate in
the top half of the torus while the right modes propagate in the
bottom half of the torus. In the case of the closed string, there is again
a splitting of the closed string exactly as before but the
propagation of the modes is transverse to the previous case as
pictured in figure~\r{fig.mod2}.

Basically this discussion explains relations~(\r{tr_BC}).
The direct channel-picture on the disk correspond to
$\Tr_{\rm open}\left(e^{-H_o\tau}\right)$ where the trace is
considered over the possible Chan-Paton factors carried by the
open string. The  transverse-channel picture corresponds to
$\left<B\left|e^{-H_o\tilde{\tau}}\right|B\right>$
where $\left|B\right>$ stands for the states of the closed string.

So far only one loop amplitudes were considered, i.e. genus 1
world-sheet surface orbifolds. For the pure bosonic case this is
sufficient, but once fermions and supersymmetry are introduced new
constraints emerge in two loop amplitudes. Specifically the modular
group of closed Riemann surfaces at genus $g$ is $SL(2g,\mathbb{Z})$,
upon orbifolding there is a residual conformal group, the so called
\textbf{Relative Modular Group}~\cite{SBH_06} (see
also~\cite{SBH_06a,SBH_06b,SBH_06c}). For genus 1 this group is
trivial but for higher genus it basically mixes neighboring tori. This
means it mixes holes and crosscaps (note that any surface of higher
genus can be obtained from sewing genus 1 surfaces). Furthermore, the
string amplitudes defined on these genus 2 open/unoriented surfaces
must factorize into products of genus 1 amplitudes. For instance a 2
torus amplitude can be thought as two 1 torus amplitudes connected
through an open string.  For a discussion of the same kind of
constraints for closed string amplitudes
see~\cite{MI_01,MI_02,MI_03,MI_04,MI_05}.

The factorization and modular invariance of open/unoriented superstring
theory amplitudes will induce generalized GSO projections ensuring the
consistency of the resulting string theories.

The correct Neveu-Schwarz (NS - antiperiodic conditions, target spacetime
fermions) and Ramond (R - periodic conditions, target spacetime bosons)
sectors were built from TMGT in~\cite{TM_11}. There the minimal
model given by the coset
$M_k=SU(2)_{k+2}\times SO(2)_2/U(1)_{k+2}$ with the CS action
\be
S^{N=2}[A,B,C]=kS_CS^{SU(2)}[A]+2S_CS^{SO(2)}[B]-(k+2)S_CS^{U(1)}[C]
\ee
was considered.
It induces, on the boundary, an $N=2$
Super Conformal Field Theory (see also~\cite{TM_08} for $N=1$ SCFT).
The boundary states of the $3D$
theory corresponding to the NS and R sectors are
obtained as quantum superpositions of the 4 possible ground states
(wave functions corresponding to the first Landau level
- the ground state is degenerate) 
of the gauge field $B$, that is to say it is necessary to choose the
correct basis of states.
The GSO projections emerge in this way as some particular
superposition of those 4 states at each boundary (for further
details see~\cite{TM_11}). It still remains to see how these
constraints emerge from genus 2 amplitudes from TM and its orbifolds
but this issue is not going to be addressed in this work.

\subsection{Neumann and Dirichlet World-Sheet Boundary Conditions,\\ Monopole Processes and Charge Conjugation}

It is clear by now that the operation of charge conjugation $C$ is
selecting important properties of the \textit{new} gauged theory. This
means the properties of the $2D$ boundary string theory.  Gauging
$PCT$ results in having an open CFT with Neumann boundary conditions,
while gauging $PT$ results in having Dirichlet boundary
conditions. So $C$ effectively selects the kind of boundary
conditions!  In the case that $PCT$ gives a closed unoriented
manifold, an untwisted theory is obtained, while for $PT$
a twisted theory is obtained (orientifold $X\cong-X$).  Again $C$ effectively
selects the theory to be twisted or not. These results are summarized
in table~\r{tabbc}.

\bt{|cccc|cccc|}
\hline  &$P_1$  &$P_2$          &\ \ \ \ &        &$\Omega$  &$\tilde{\Omega}$&$\Omega'$\\\hline
&&&&&&&\\
$S^2\to$&$D_2$     &$RP_2$       &        &$T^2\to$&$C_2$      &$M_2$                      &$K_2$\\
&O/O     &C/U          &        &     &O/O        &O/U&C/U\\
&&&&&&&\\
$CT$&$N$&Untwisted            & &&$N$&$N$&Untwisted\\
    &q=m&q=m                  & &&$q=m$&$q=m$&q=m\\
&&&&&&&\\
$T$ &$D$&Twisted              & &&$D$&$D$&Twisted\\
    &$q=kn/4$&$q=kn/4$        & &&$q=kn/4$&$q=kn/4$&$q=kn/4$\\
&&&&&&&\\\hline
\et{\lb{tabbc}Boundary conditions and twisted sectors.}

Although these facts are closely related with string T-duality,
the $C$~operation does not give us the dual spectrum. Upon
gauging the full theory it only selects the Kaluza-Klein
momenta or winding number as the spectrum of the configurations
being gauged.

From the point of view of the bulk theory the gauged configurations
corresponding to Neumann boundary conditions correspond to two Wilson
lines with one end attached to the $1D$ boundary of the new $2D$
\textit{boundary} of the membrane at $t=1/2$ and the other
end attached to the $2D$ boundary at $t=0$. For Dirichlet boundary
conditions there is one single Wilson line with both ends in the
$2D$ membrane boundary at $t=0$ and a monopole insertion in the
bulk of the $2D$ boundary at $t=1/2$. Note that the Wilson lines
no longer have a well defined direction in
time: time inversion has been gauged. These results are presented in
figure~\r{fig.monopoleDN}.
\fig{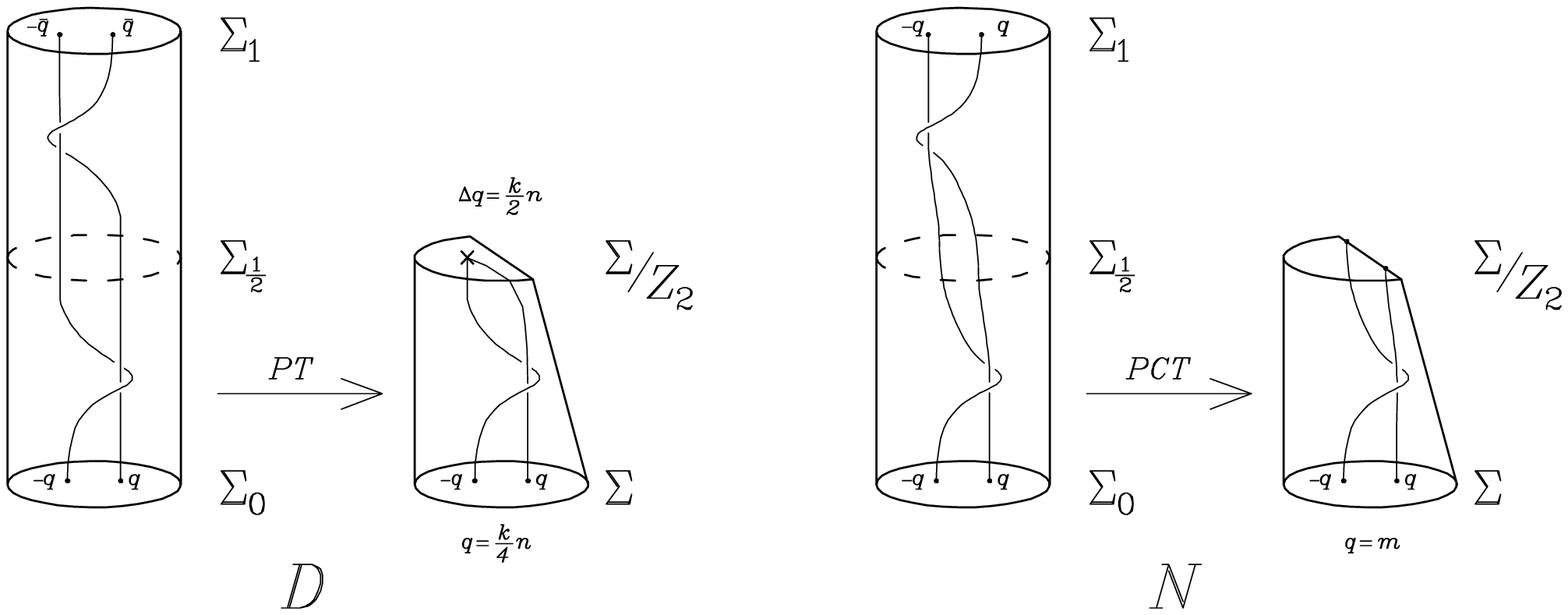}{For Dirichlet boundary conditions two Wilson
lines carrying charges $q$ and $-q$ meet in a monopole at the
orbifold singular point $t=1/2$. For Neumann boundary
conditions the two Wilson lines end on the boundary of
$\Sigma/Z_2$.}{fig.monopoleDN}

For the case where unoriented manifolds are obtained the picture is
quite similar. There are always an even number of bulk insertions.  In
the case of $PCT$ the Wilson lines which are identified have the same
charge, therefore there are no monopole processes involved. The two
Wilson lines are \textit{glued} at $t=1/2$ becoming in the orbifolded
theory one single line which has both ends attached to $\Sigma_0$ and
one point in the middle belonging to $\Sigma_{1/2}$. In the boundary
CFT it corresponds to two vertex insertions with opposite
momenta. This construction corresponds to untwisted string theories
since the target space coordinates (corresponding to the gauge
parameter $\Lambda$ in TM(GT)) are not orbifolded.

In the case of $PT$ the identification is done between charges of
opposite signs. Then two Wilson lines become one single line with its
ends attached to $\Sigma_0$, but at one end they have a charge $q$ and
at the other end they have a charge $-q$. In $\Sigma_{1/2}$ there is
a monopole insertion which changes the sign of the charge. This
construction corresponds to twisted string theories since the target
space coordinates are orbifolded ($\Lambda\cong-\Lambda$).

As a final consistency check, in $PCT$ type of orbifolds the charges
are always restricted to be $q=m$ due to compatibility with the
orbifold construction. By restricting the spectrum to this form one is
actually eliminating the monopole processes for these particular
configurations!

\subsection{T-Duality and Several U(1)'s}

The well know Target space or T-duality (for a review see~\cite{R_1})
of string theory is a combined symmetry of the background and the
spectrum of momenta and winding modes.  It interchanges winding modes
with Kaluza-Klein modes.  From the point of view of the orbifolded
TM(GT) corresponding to open and unoriented string theories the
projections $PT$ truncate the charge spectrum to $q=kn/4$ (due to
demanding $q=-\bar{q}$) which in string theory is the winding
number. The projections $PCT$ truncate the charge spectrum to $q=m$
(due to demanding $q=\bar{q}$) which corresponds in string theory to
the Kaluza-Klein momenta. Note that $PCT$ excludes all the monopole induced
processes while $PT$ singles out only monopole induced
processes~\cite{TM_07,TM_12,TM_15}.

T-duality is, from the point
of view of the $3D$ theory, effectively exchanging the two kinds of
projections
\be
\ba{lccc}
{\rm T-duality}:&PT&\lra&PCT\\
                &q=-\bar{q}&\lra&q=\bar{q}
\ea
\ee

This is precisely what it must do. The nature of duality in $3D$ terms
was discussed in some detail in~\cite{TM_10}. It was shown there that
it exchanges topologically non-trivial matter field configurations with
topologically non trivial gauge field configurations. Although charge
conjugation was not discussed there (only parity and time inversion),
this mechanism can be thought of as a charge conjugation operation. Note
that $C^2=1$.

It is also rather interesting that from the point of view of the
membrane both T-duality and modular transformations are playing the
same role. In some sense both phenomena are linked by the $3D$ bulk theory.

So far only a single $U(1)$ theory with gauge fields obeying
$\mathcal{B}$ boundary conditions was considered. But new phenomena
emerge in the more general case. The extra $\tilde{\mathcal{B}}$ gauge
sectors are necessary any how~\cite{TM_15}.

Take then the general action with gauge group $U(1)^d\times U(1)^D$
with $d$ type $\tilde{\mathcal{B}}$ gauge fields and the remaining $D$
gauge fields of type $\mathcal{B}$,
\be
S_{d+D}=\int_Mdx^3\left[-\frac{\sqrt{-g}}{\gamma}\left(F^{M}_{\mu\nu}F_{M}^{\mu\nu} 
+F^{I}_{\mu\nu}F_{I}^{\mu\nu}\right)+\frac{\epsilon^{\mu\nu\lambda}}{8\pi}\left( 
K_{MN}A^{M}_\mu\partial_\nu A^{N}_{\lambda}+K_{IJ}A^{I}_\mu\partial_\nu A^{J}_{\lambda}\right)\right]
\label{SIJ}
\ee
where $M,N=1,\ldots,d$ correspond to type $\tilde{\mathcal{B}}$
gauge fields and $I,J=d+1,\ldots,d+D$ to type $\mathcal{B}$ fields.

For a given parity $P$, one can build an operator $O$ that acts on
every $A$ field through $PT$ and only on some of them through $C$
\be
O=PT\left(\sum_{I'} C\delta_{I' I}+\sum_{I''}\delta_{I'' I}\right)
\ee
Due to the charges not being quantized and the non-existence of
monopole-induced processes in the $\tilde{\mathcal{B}}$ gauge sector,
the mechanism is slightly different (see
section~\r{ch.opun:sec.cft.bc}). But this operator can act as well
over the $\tilde{\mathcal{B}}$ sector.

For the case of open manifolds $M/PT$, $I'$ runs over the indices for
which one wants to impose Neumann boundary conditions
(on $\Lambda^{I'}$) and $I''$ over
the indices corresponding to Dirichlet boundary conditions.
For the case of closed manifolds $M/PT$ the picture is similar but $I'$
runs over the indices for which  one wants $\Lambda^{I'}$ to be orbifolded
(obtaining an orientifold or twisted sector).

In the case of several $U(1)$'s more general symmetries (and therefore
orbifold groups) can be considered (for instance $Z_N$).  Those
symmetries are encoded in the Chern-Simons coefficient $K_{IJ}$.

\chapter{Final Remarks}
\lb{ch.conc}

The work presented here is one further step towards the description of
string theory from TM.  Most of the constraints on string theories
exist because of their intimate relations to $2D$ geometry, $2D$
conformal field theories and their symmetries. One of these
constraints is modular invariance which puts severe restrictions
on the spectrum of the theories.  Modular invariance, as employed by
Narain in the context of toroidal compactification, reduces the
physical problem of finding the allowed momenta and winding modes in
string theory to the problem of constructing lattices on a
pseudo-Euclidean space ${\mathbb{R}}^{d_R,d_L}$ on which physical
invariants are modular invariant functions. These lattices turn out to
be
\textit{self-dual}, \textit{integer} and
\textit{even} lattices.

In Chapter~\r{ch.tor} the results of Narain compactification were
reproduced starting from $3D$ gauge dynamics without referring to
modular invariance at all.  The fact that $3D$ topologically massive
gauge theory carries full information about the lattices of string
theory is quite fascinating. The approach presented here and the
modular invariance construction are compatible and moreover they are
logically connected. But there is still an element of surprise here
because of the crucial difference between the ideology of these two
approaches.  To see this let us summarize what has been obtained so
far.  The starting point is a TMGT defined on a three manifold with
two disconnected boundaries. Gauge degrees of freedom became dynamical
on the boundaries generating chiral CFT's. Each boundary of the three
manifold is interpreted as a ``chiral worldsheet'' of string theory,
meaning that the left and right modes are physically separated. This
is the framework of Topological Membrane theory.  The Abelian theory
with fields of type $\mathcal{B}$ which has a discrete charge spectrum
of the form $\vb{Q} =\vb{m} + k \vb{n}/4$ was considered. A particle
with charge $\vb{Q}$ is inserted at one boundary and it travels
through the bulk, interacts with all possible charge violating
instantons on its path and links with other charges and emerges as a
new charge $\bar{\vb{Q}} = \vb{m} - k \vb{n}/4$ on the other boundary.
The path of the charged particle is represented by a Wilson line in
the bulk theory. After taking care of all the linkings and instanton
interactions it is quite an amusing result to obtain that the emerging
charge in the other boundary is of the $\bar{Q}$ form and nothing
else.  The Aharonov-Bohm phases of the linkings interfere in such a
way that for a particle which does not have the charge $\pm \bar{Q}$
there is zero probability to emerge at the other boundary.  Moreover a
natural self-dual Lorentzian lattice structure emerges from the
linkings of the Wilson lines as was shown in Chapter~\r{ch.tor}.  The
connection between this approach and the modular invariance arguments
was not worked out here, but naively one can see that modular
transformations at the boundaries will yield linkings of the Wilson
lines in the bulk.  This is why these results were obtained.

Chapter~\r{ch.opun} has also shown how one can obtain open and closed
unoriented string theories from the Topological Membrane. There were
two major ingredients: one is the Horava idea about orbifolding, the
second is that the orbifold symmetry was a discrete symmetry of TMGT.
The orbifold works from the point of view of the membrane as a
projection of field configurations obeying either $PT$ or $PCT$
symmetries (the only two kinds of discrete symmetries compatible with
TMGT). For $PCT$ type projections, Neumann boundary conditions were
obtained for open strings, and untwisted sectors for closed unoriented
strings. For $PT$ type projections Dirichlet boundary conditions were
obtained for open strings, and twisted sectors for closed unoriented
strings.  For $PCT$, $q=\bar{q}=m$, so only the string Kaluza Klein
modes survive. In this case the monopole induced processes are
completely suppressed. For $PT$, $q=-\bar{q}=kn/4$, so only the string
winding modes survive. In this case only monopole induced processes
are present, the charges being purely magnetic. Charge conjugation $C$
plays an important role in all these processes, playing the role of a
$Z_2$ symmetry of the string theory target space. These results can be
generalized to symmetries of the target space encoded in the tensor
$K_{IJ}$ and are closely connected. Both to modular transformations
and T-duality which exchange $PT\lra PCT$.

As a final remark note that the string photon Wilson line has been
left out. TM(GT) can take account of it as well: for any closed
$\Sigma$ there is a symmetry of the gauge group coupling tensor
$K_{IJ}\to K_{IJ}+\delta_I\chi_J-\delta_J\chi_I$ where each
$\chi_I=\chi_I[A]$ is taken to be some function of the $A^I$'s.  This
transformation affects only $B_{IJ}$ and the induced terms vanish upon
integration by parts. Once the orbifold of the theory is considered,
the new orbifolded $\Sigma_o$ has a boundary and the induced terms
will no longer vanish. Instead they will induce a new action on the
boundary $\partial\Sigma_o$. This will be precisely the \textit{new}
gauge photon action of open string theories. As is well known the
choice of the gauge group of string theory, i.e. the Chan-Paton
factor structure carried by this photon Wilson line, will be
determined by the cancellation of the open string theory gauge
anomalies (see~\cite{POL_1} and references therein).

\ \\

There are a couple of important things which need to be addressed in the
future. The proper treatment of the string photon Wilson line~\cite{II}
will allow the introduction of D-branes in the context of TM constituting a
promising field of research connected with this last subject.

Another work would be the bulk interpretation of A-D-E
classification~\cite{Z_1} of modular invariant partition functions on
the boundary~\cite{BN_1,LR_1,O_1,AFC_1,W_1b}. Also, related to these
modular invariants, one would like to derive the orbifolded partition
functions of the boundary CFT from the bulk TMGT~\cite{II}.

Although it was shown that the charge conjugation symmetry $C$ plays
an important role both in modular invariance and in T-duality, neither of
them have been properly studied. To properly describe modular
invariance it is necessary to take the gravitational
diffeomorphisms due to the TMG sector into account.
Also, so far the full duality group~\cite{R_1} has not been properly
described in terms of symmetries of the $3D$ QFT, although a study
has been done in~\cite{TM_09,TM_10}.

Concerning the generalized GSO projections of
subsection~\r{ch.opun:sec.tmgt.mod} it still remains to study how they
emerge from genus 2 amplitudes from the point of view of TM and its orbifolds.

Also an issue to address in future work will be to
generalize the orbifold constructions presented here to non-trivial
boundary CFT's, for example WZWN models and different coset models
which can be obtained from the TM with a non-Abelian TMGT.

Some further interesting directions of research are to obtain boundary
CFT logarithmic theory~\cite{log_1} and deriving the Sen
mechanism~\cite{sen_1} (tachyon condensation) from the TM
(see~\cite{TM_13} for an initial study). These aspects are vital if
one wants to build consistent effective non supersymmetric string
theories.

Finally the framework presented in this work can be applied to other
quantum theories which include the CS term such as the effective
Quantum Hall Effect. For example in~\cite{jain} is studied
the Jain hierarchy using such a theory.

\end{document}